\documentclass[twocolumn,twocolappendix,hypersetup]{openjournal}
\usepackage{amsmath}
\usepackage{amssymb} 
\usepackage{upgreek}
\usepackage{iondefs}

\usepackage[breaklinks,colorlinks,allcolors=blue,citecolor=blue,urlcolor=blue]{hyperref}
\usepackage{orcidlink}

\newcommand{\subB}{_{\hbox{\tiny B}}}
\newcommand{\subH}{_{\hbox{\tiny H}}}

\newcommand{\subHI}{_{\hbox{\tiny HI}}}

\newcommand{\subE}{_{\hbox{\tiny E}}}
\newcommand{\supT}{^{\hbox{\tiny T}}}

\newcommand{\subR}{_{\hbox{\tiny R}}}
\newcommand{\subLOS}{_{\hbox{\tiny LOS}}}

\newcommand{\subcalC}{_{\cal C}}
\newcommand{\subCGM}{_{\hbox{\tiny CGM}}}

\newcommand{\ihat}{\hat{\hbox{\bf e}}_x}
\newcommand{\jhat}{\hat{\hbox{\bf e}}_y}
\newcommand{\khat}{\hat{\hbox{\bf e}}_z}
\newcommand{\ihato}{\hat{\hbox{\bf e}}_{x_o}}
\newcommand{\jhato}{\hat{\hbox{\bf e}}_{y_o}}
\newcommand{\khato}{\hat{\hbox{\bf e}}_{z_o}}
\newcommand{\shat}{\hat{\hbox{\bf s}}}
\newcommand{\qhat}{\hat{\hbox{\bf e}}_k}
\newcommand{\rvec}{\hbox{\bf r}}

\newcommand{\rhat}{\hat{\hbox{\bf e}}_r}
\newcommand{\phihat}{\hat{\hbox{\bf e}}_\upphi}
\newcommand{\thetahat}{\hat{\hbox{\bf e}}_\uptheta}
\newcommand{\rhohat}{\hat{\hbox{\bf e}}_\uprho}
\newcommand{\Gsys}{$G$}
\newcommand{\Osys}{$O$}
\newcommand{\subG}{_{\hbox{\tiny G}}}
\newcommand{\subO}{_{\hbox{\tiny O}}}

\shorttitle{\sc SKAM II}
\shortauthors{\sc Churchill}

\defcitealias{churchill25-skamI}{Paper~I}

\begin{document}

\title{Spatial-Kinematic Absorption Models of the Circumgalactic Medium. \\ II. Ionized Gas Phases and Absorption Lines\vspace{-15mm}}




\author{Christopher W. Churchill\,\orcidlink{0000-0002-9125-8159}}

\affiliation{New Mexico State University\\
Department of Astronomy \\
1320 Frenger Mall \\
Las Cruces, NM 88011, USA}
\thanks{email: \href{mailto:cwc@nmsu.edu}{cwc@nmsu.edu}}

\begin{abstract}
In this two-paper series, we present a straightforward mathematical model for synthesizing quasar absorption line profiles from sight lines through idealized, spatial-kinematic models of the circumgalactic medium (CGM) and their host galaxies. In \citetalias{churchill25-skamI}, we developed the spatial components of the galaxy/CGM structures (disk, halo, wind, accretion) and their 3D velocity fields. We derived the formalism for arbitrary observed orientation of the galaxy/CGM model and quasar line of sight positioning. In this paper, following a brief review of \citetalias{churchill25-skamI}, we present the formalism for populating the galaxy/CGM structures with multiphase photoionized and collisionally ionized gas and for generating {\HI} and metal-line absorption profiles. Example absorption line systems through a fiducial galaxy/CGM model are presented. These flexible spatial-kinematic absorption models (SKAMs) can be directly applied to and/or easily modified/expanded for studying individual or ensembles of observed absorption line systems, for exploring various competing theoretical scenarios of the baryon cycle as studied through quasar absorption line systems, and/or serving as pedagogical tools for developing physical intuition. We briefly describe a SKAM GUI that is in early stages of development.
\end{abstract}


\section{Introduction}
\label{sec:introduction}

Insights into the formation and evolution of galaxies have been slowly developed through observational efforts and theoretical studies aimed at understanding the spatial distribution, kinematics, multiple gas phases, chemical enrichment, and ionization conditions of the baryons populating the circumgalactic medium \citep[CGM, e.g.,][]{tpw-araa17, faucher-giguere23}. The CGM, an extended region around galaxies that harbors prominent components of the baryon cycle, is highly accessible for empirical study \citep[][]{peroux-howk20}. Functionally, this important medium serves as the interface between the interstellar medium (ISM), where stars form, live and die, and the intergalactic medium (IGM) in which the galaxy is embedded. 

Throughout the CGM, baryons in the gaseous phase are all at once flowing into their host galaxies from the IGM and being propelled back out to the IGM by stellar feedback in the ISM of their host galaxies \citep[e.g.,][]{keres05, dekel06, vandevoort11, lilly13, nelson18, peroux20}.  Those baryons that remain gravitationally bound within the dark matter halo may recycle back into the host galaxy or mix and remain trapped in the CGM \citep[e.g.,][]{oppenheimer08, ford14}.  The resulting dynamic multiphase CGM environment is a product of an array of kinetic interactions driving radiative and collisional ionization processes, photo-heating, shock heating, radiative cooling, turbulent mixing, advection and/or diffusion \citep[e.g.,][]{maller04, ceverino09, dave11,  faerman17, stewart-proc17, nelson19, oppenheimer18, rupke18, zhang18, faerman20, fielding20, esmerian21, hafen22, nguyen22, trapp22}. 

In recent decades, the technique of quasar absorption lines has been instrumental in establishing the statistical properties of the CGM \citep[see][and references therein]{churchill25-book1}. Constrained by the growing body of quasar absorption line studies, ever more powerful hydrodynamic simulations have refined the theoretical picture of the inflowing and outflowing components of the baryon cycle in the CGM while also providing observational tests to discriminate between various theoretical predictions and feedback scenarios \citep[e.g.,][]{wright24}.  

\begin{figure}[h!bt]
\centering
\vglue 0.05in
\includegraphics[width=\linewidth]{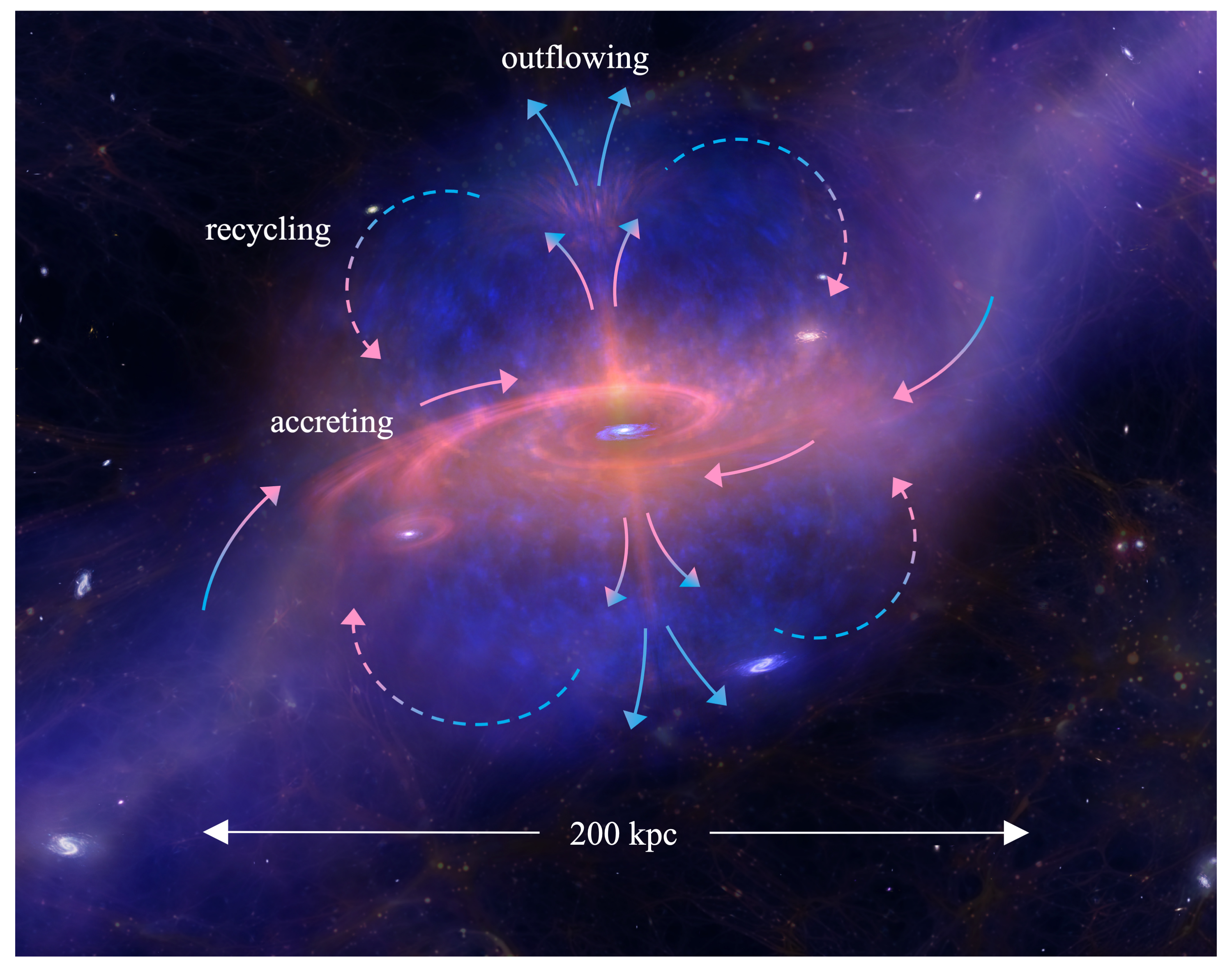}
\vglue 0.05in
\caption{\small  Schematic of the baryon cycle in the CGM of a typical spiral galaxy.  The prominent processes occurring throughout the CGM are outflowing winds directed along the galactic poles and accreting IGM/filaments converging into an extended planar region. Some fraction of the outflowing stellar-driven wind, a result of stellar feedback, can escape the galaxy, whereas some fraction can be recycling back into the accreting inflow. (original image credit: Zheng Cai, Tsinghua University; annotations added)}
\label{fig:baryoncycle}
\end{figure}

Observations and theory are converging toward a cohesive picture of the baryon cycle in the CGM, which we illustate in Figure~\ref{fig:baryoncycle}. The CGM appears to be influenced by stellar-driven outflows that are propelled outward perpendicular to the galaxy plane \citep[e.g.,][]{nelson19, peroux20, faucher-giguere23} and generally form a bi-polar conical geometry \citep[see][]{rupke18, zhang18, fielding20, nguyen22}. On the other hand, infalling gas conserves angular momentum and tends to spiral inward and concentrate in an extended co-planar geometry \citep[e.g.,][]{keres05, stewart11, stewart17, stewart-proc17, hafen22, trapp22, stern23}. However, we are not so naive that we would fail to recognize the CGM is a galactic zone of organized chaos.

In efforts to organize both the common trends and complexity learned from observations and hydrodynamic simulations, the community of researchers has converged on a simplified heuristic composite spatial-kinematic model of the CGM. There are four main spatial-kinematic components: (1) a kinematically static or turbulent spherical halo, (2) an outflowing stellar-driven wind represented by a conical geometry, (3) a rotating ISM and extra-planar gas in a disk-like geometry, and (4) inward spiraling IGM accretion confined to  a ``flared" extended planar structure 

The use of simple spatial-kinematic models of the CGM to explore and interpret quasar absorption line profiles has a rich history \citep[e.g.][]{weisheit78, lanzetta92, charlton98, prochaska98, steidel02, kacprzak10, kacprzak11, bouche12, gauthier12, chen14, kacprzak19, bordoloi14, diamond16, ho17, lan18, martin19, schroetter19, french20, ho20, zabl20, nateghi21, beckett22, beckett23, carr22, casavecchia23}. Running mock lines of sight through CGM structures with known 3D geometries, baryon density fields, and kinematics can untwist the tangled complexity of observations and hydrodynamic simulations. As such, simple intuitive spatial-kinematic absorption models (SKAMs) of the CGM hold great potential for enhancing our interpretation of LOS kinematics of selected ionic transitions captured in quasar absorption line spectra.  

In \citet[][hereafter \citetalias{churchill25-skamI}]{churchill25-skamI}, we developed a flexible and generalized formalism for constructing the geometric structures of SKAMs that simultaneously incorporate all four of the galaxy/CGM components (see Figure~\ref{fig:LOSskyview}). We further showed how to populate these structures with arbitrary kinematic velocity fields and presented kinematic models consistent with those inferred from observations and simulations. The principal axes of these SKAMs can be arbitrarily oriented with respect to the observer's viewing angle using a simple two-rotation coordinate transformation.  The line of sight (LOS) to the background quasar is defined by its impact parameter and position angle on the sky. In the reference frames of both the SKAM and the observer, the LOS is described by a parametric equation of a vector.  For the individual principal velocity components of 3D velocity fields (in both Cartesian and spherical coordinates systems), simple generalized projection functions for computing LOS velocities were derived that can be evaluated at any point along the LOS. 

For a given galaxy-quasar impact parameter and observer/LOS orientation with respect to a galaxy/CGM structure, these spatial-kinematic models can predict only the ``allowed" range of LOS velocities over which each galaxy/CGM structure can {\it potentially\/} give rise to absorption. This LOS velocity information provides a degree of insight into the CGM and baryon cycle, but how are we to interpret cases in which two or more of the galaxy/CGM structures probed by a LOS have overlapping LOS velocity ranges? There is no information as to what limited intervals within these ``allowed" LOS velocity ranges the absorption profiles will actually arise, nor their relative line strengths. Neither is there any predictive power as to which ionic transition(s) will arise in which structures-- critical for insights into the different chemical enrichment and ionization conditions of the individual galaxy/CGM structures.  In order to have full predictive power for interpreting observations, SKAMs must be capable of predicting absorption lines. As such, the spatial-kinematic models presented in \citetalias{churchill25-skamI} are not yet complete SKAMs.

In this paper, we develop the additional formalism required for the SKAMs to predict absorption lines. Section~\ref{sec:thesetup} through Section~\ref{sec:kinematicmodels} provides a brief summary of  \citetalias{churchill25-skamI}. In Section~\ref{sec:thesetup}, we review the geometric and mathematical formalism of the model, including the relationship between the observer, background quasar, galaxy, LOS, and LOS velocities.  In Section~\ref{sec:spatialmodels}, we review the idealized geometric spatial components of the galaxy/CGM model. The kinematics and LOS velocities for each spatial galaxy/CGM structure are reviewed in Section~\ref{sec:kinematicmodels}.  Section~\ref{sec:gasphysics} describes how the SKAM is populated by baryons, including their density and temperature distributions, and their metal abundances and ionization conditions. Section~\ref{sec:abslines} describes how absorption line spectra are generated. In Section~\ref{sec:discussion}, we discuss some insights into the SKAM formalism, including its methods, flexibilities, and oversimplifications.  Examples of synthesized quasar absorption lines are presented in Section~\ref{sec:examples}. Concluding remarks and discussion of potential future work are reserved for Section~\ref{sec:conclusion}.  The full SKAM with a graphical interface is available as a Fortran 95 code at \href{https://github.com/CGM-World}{github.com/CGM-World}.

\section{The Set Up}
\label{sec:thesetup}

In this section, we review the spatial-kinematic aspects of a SKAM. This review comprises material that was motivated and developed in \citetalias{churchill25-skamI}. Much of that work includes derivations and accompanying illustrations and the reader should refer to \citetalias{churchill25-skamI} if seeking a detailed treatment.  Here, we do not repeat observational and/or theoretical motivations for geometric and kinematic representations described below; our aim is to present the foundational context for further development of the SKAM.

\subsection{The Dark Matter Halo}

The host galaxy is characterized by its virial mass
\begin{equation}
    M_{\rm vir} = \frac{4\pi}{3}
    R_{\rm vir}^3 (\Delta_c \rho_c) \, ,
\label{eq:virialmass-rvir}
\end{equation}
where $R_{\rm vir}$ is the virial radius. The virial radius is defined such that the average mass density within this radius is $\Delta_c\rho_c$, where $\rho_c$ is the critical density of the universe.  Typically, an overdensity factor of $\Delta_c \sim 200$ is adopted. Employing an ``NFW" dark matter halo \citep{NFW96}, we derived an expression for $R_{\rm vir}$ in terms of the halo maximum circular velocity $V_c$.  We obtained,
\begin{equation}
 R_{\rm vir} = V_c 
  \left\{ \frac{3}{4\pi G} 
  \frac{\xi/\mathcal{C}}{\Delta_c \rho_c}    \frac{A\subcalC}{A_\xi}  
 \right\} ^{1/2} \, ,
\label{eq:virialradius}
\end{equation}
where $\mathcal{C}$ is  the concentration, $\xi=2.16258$ is the positive root that solves $ \ln(1+\xi) = \xi (1+ 2\xi)/(1+\xi)^2$, and $A_x = \ln(1+x) - x/(1+x)$.  Since $\xi$ is fixed, $A_\xi = 1.83519$. Adopting a typical value of $\mathcal{C} \sim 10$, we have $A\subcalC\simeq 1.48$, yielding $(\xi/\mathcal{C})(A\subcalC/A_\xi) \simeq 1.24$.  Thus, the galaxy virial mass and radius are defined by three free parameters, $V_c$, $\Delta_c$, and $\mathcal{C}$. For example, $V_c \simeq 220$~{\kms}, $\Delta_c = 200$, and $\mathcal{C}=10$, yields ${R_{\rm vir} \simeq 200}$~kpc and $\log (M_{\rm vir}/{\rm M}_\odot) \simeq 12$.

\subsection{Coordinates and Projections}
\label{sec:coordinates}

\begin{figure}[thb]
\centering
\vglue 0.15in
\includegraphics[width=0.95\linewidth]{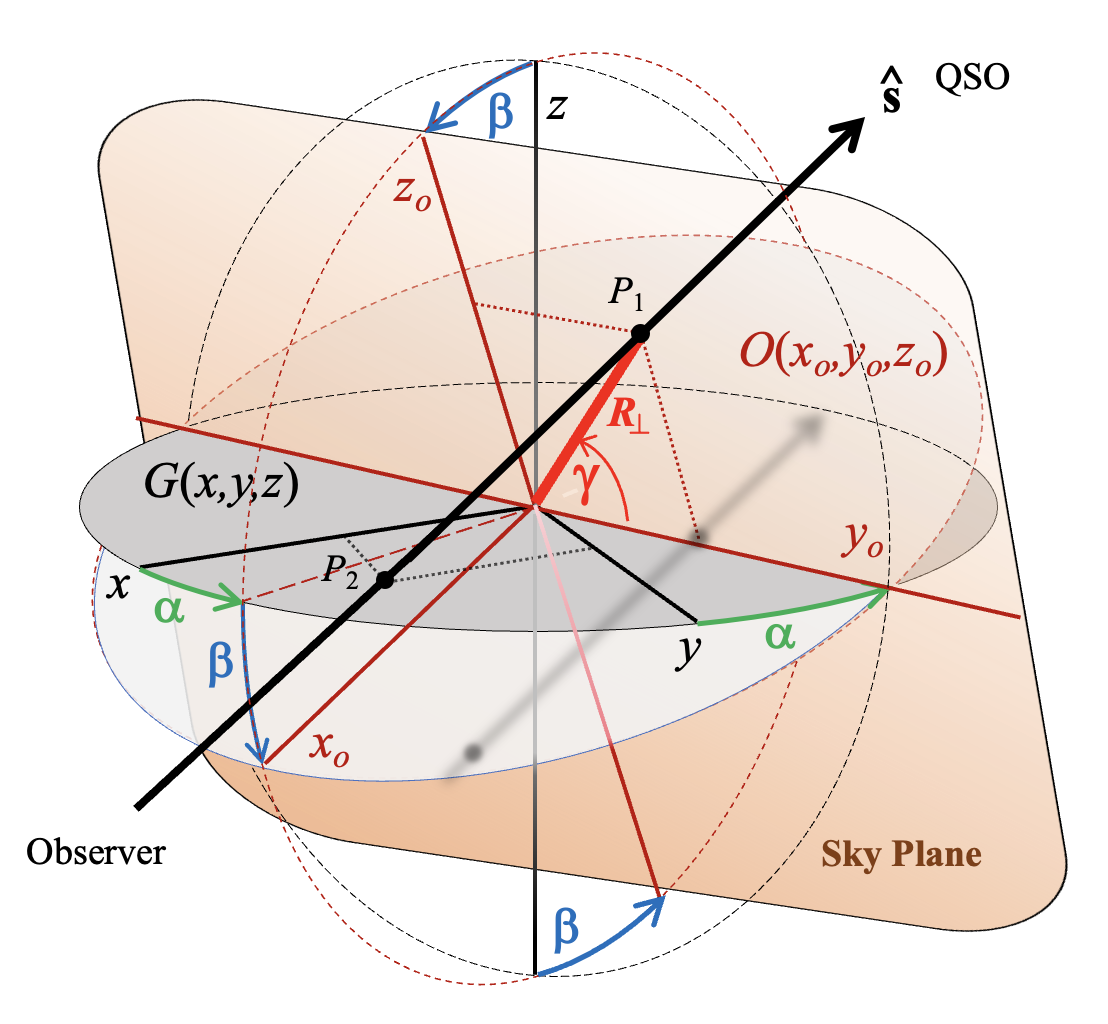}
\caption{\small  The galaxy coordinate system {\Gsys}$(x,y,z)$ and the observer coordinate system {\Osys}$(x_o,y_o,z_o)$.  The LOS is parallel to the $x_o$ axis and has unit vector  ${\shat}= - {\ihato}$. In {\Osys}, the LOS intersects the sky plane at $P_1 = P\subO(0,R_\perp \cos \gamma, R_\perp \sin \gamma)= P\subG(X_0,Y_0,Z_0)$, where $R_\perp$ is the impact parameter and $\gamma$ is the sky position angle. The observer is positioned at $P\subO(+\infty, R_\perp \cos \gamma, R_\perp \sin \gamma)$, while the quasar is at $P\subO(-\infty, R_\perp \cos \gamma, R_\perp \sin \gamma)$. }
\label{fig:godseyeview}
\end{figure}

As illustrated in Figure~\ref{fig:godseyeview}, the galaxy/CGM is fixed at the origin of coordinate system {\Gsys} with standard Cartesian coordinates $P\subG(x,y,z)$ with unit vectors  vector $\ihat$, $\jhat$, and $\khat$ and spherical coordinates $P\subG(r, \theta, \phi)$ with corresponding unit vectors 
\begin{equation}
\begin{array}{rcl}
{\rhat}\!\! &=& \! (x/r) \, {\ihat} + (y/r)\, {\jhat} + (z/r) \, {\khat} \, , \\[2pt]
{\thetahat}\!\! &=&  (x/\rho)(z/r) \, {\ihat} + (y/\rho)(z/r) \, {\jhat} - (\rho/r) \, {\khat} \, , \\[2pt]
{\phihat}\!\! &=& \!  - (y/\rho) \, {\ihat} + (x/\rho) \, {\jhat} \, ,
\end{array}
\label{eq:sphereunitvecs}
\end{equation}
where the axial distance from the $z$ axis is $\rho^2 = x^2 + y^2$. The galaxy rotation (polar) axis is the $z$ axis and the galactic plane is the $xy$ plane.

The observer's frame of reference is coordinate system {\Osys}, which has coincident origin with {\Gsys} and has standard Cartesian coordinates $P\subO(x_o,y_o,z_o)$ with unit vectors ${\ihato}$, ${\jhato}$, and ${\khato}$ and standard spherical coordinates and unit vectors.  
For a general viewing angle of objects in {\Gsys}, two rotations of {\Osys} are required. The first is a counterclockwise rotation about the $z$ axis of {\Gsys} through angle $\alpha \in (0^\circ,90^\circ]$.  The second is a counterclockwise rotation about the now rotated $y$ axis by angle $\beta \in (-90^\circ,+90^\circ)$.  The rotations are implemented through the operation $[x,y,z]\supT\subG = R_y(\beta)R_z(\alpha) [x_o,y_o,z_o]\supT\subO$, where $R_y(\beta)R_z(\alpha)$ is the rotation matrix for the two rotations $\alpha$ and $\beta$, applied in that order.

The position of the quasar on the plane of the observer's sky (the $y_o$$z_o$ plane) is defined using the impact parameter, $R_\perp$, and position angle, $\gamma$.  In {\Osys}, the observer is at $P\subO(+\infty, R_\perp \cos \gamma, R_\perp \sin \gamma)$, while the quasar is located $P\subO(-\infty, R_\perp \cos \gamma, R_\perp \sin \gamma)$. Thus, in {\Osys} the LOS is a line parallel to the $x_o$ axis passing through the sky plane at the point $P\subO(0, R_\perp \cos \gamma, R_\perp \sin \gamma)$ and has unit vector ${\shat} = -{\ihato}$. We define the unit vector of the LOS in {\Gsys} to be $\shat = \sigma_x \ihat + \sigma_y \jhat + \sigma_x \khat$,  where $\sigma_x  =\! -({\ihato} \!\!\cdot {\ihat})$,  $\sigma_y =\! -({\ihato} \!\!\cdot {\jhat})$, and $\sigma_z =\! -({\ihato} \!\!\cdot {\khat})$ are the direction cosines. As shown in Appendix~A of \citetalias{churchill25-skamI},  
\begin{equation}
\begin{array}{rcl}
\sigma_x \!\! &=& \! -\cos\beta\cos\alpha \, ,\\[2pt]
\sigma_y \!\! &=& \!  + \sin\alpha \, ,\\[2pt]
\sigma_z \!\! &=& \!  - \sin \beta \cos \alpha \, .
\end{array}
\label{eq:shat-derived}
\end{equation}  

Position along the LOS in {\Gsys} is mapped using the parametric equation 
\begin{equation}
x(t) =  \sigma_x t \!+\! X_0 \, , \,
y(t) = \sigma_y t \!+\! Y_0 \, , \,
z(t) = \sigma_z t \!+\! Z_0 \, , 
\label{eq:defineLOSPOS}
\end{equation}
where parameter $t$ is the 1D coordinate position along the LOS measured in physical units. We define $t$ such that it increases from the observer ($t=-\infty$) to the quasar ($t=+\infty$).  The LOS is perpendicular to and intersects the sky plane and at $t=0$. 
As shown in Appendix~A of \citetalias{churchill25-skamI}, 
\begin{equation}
\begin{array}{rcl}
    X_{0} \!\!&=&\! R_\perp \cos\gamma \cos\beta \sin\alpha - R_\perp \sin\gamma \sin\beta  \, , \\[2pt] 
    Y_{0} \!\!&=&\! R_\perp \cos\gamma \cos\alpha \, ,  
    \\[2pt] 
    Z_{0} \!\!&=&\! R_\perp \cos\gamma \sin\beta \sin\alpha + R_\perp \sin\gamma \cos\beta\, .
\end{array}
\label{eq:skypos-derived}
\end{equation}
Note that the galactocentric distance corresponding to position $t$ on the LOS is $r_o^2(t) = r^2(t) =  R^2_\perp + t^2 $ and is invariant in {\Gsys} and {\Osys}.

\subsection{The Galaxy-Quasar Orientation}

The galaxy-quasar orientation is defined by the observed galaxy inclination ($i\in [0^\circ,90^\circ]$) and the quasar-galaxy azimuthal angle ($\Phi \in [0^\circ,90^\circ]$). Using the convention that $i=0^\circ$ is face-on and $i=90^\circ$ is edge-on orientation, the observed inclination angle is the angle between the $z$-axis of the galaxy and the LOS, 
\begin{equation}
  i = \cos^{-1} \left( \sin|\beta| \cos\alpha \right) \, .
\label{eq:inclination}
\end{equation}
 The observed azimuthal angle between the {\it sky projected} major axis of the galaxy and a line connecting the galaxy center to the position of the quasar on the sky plane is 
\begin{equation}
\Phi =  \gamma + \cot ^{-1}
\left( \frac{\cos\beta}{\sin\beta \sin\alpha}  \right) \, ,
\label{eq:phi}
\end{equation}
which is always converted to a primary angle.

\subsection{Kinematics and LOS Velocities}
\label{sec:losvelocity}

A velocity vector field at spatial location ${\rvec}$ in coordinate system {\Gsys} is written ${\hbox{\bf V}}({\rvec})$.  
In terms of vector components, the velocity field is written as
\begin{equation}
 {\bf V}({\rvec}) =  \sum_k V_k({\rvec}) \, {\qhat} ({\rvec})\, ,
 \label{eq:V-general}
\end{equation}
where $k$ represents a principal component and ${\qhat}({\rvec})$ represents its unit vector direction at ${\rvec}$. At LOS location ${\rvec}(t) = x(t) {\ihat} + y(t) {\jhat} + z(t) {\khat}$, the LOS projected velocity is
\begin{equation}
\begin{array}{rcl}
 V\subLOS(t) = \displaystyle  \sum _k V_k(t) ({\qhat} (t) \! \cdot {\shat}) = \sum _k V_k(t) \mathcal{P}_k(t) \, ,
\end{array}
  \label{eq:Vlos-general}
\end{equation}
where  ${\shat} = \sigma_x {\ihat} + \sigma_y {\jhat} + \sigma_z {\khat}$ is the LOS unit vector. The  $\mathcal{P}_k(t) = {\qhat} (t) \! \cdot {\shat}$ are LOS projection functions for component $k$ given by 
\begin{equation}
\begin{array}{rcl}
{\cal P}_z(t) &\!\!=\!\!& \sigma_z \, , \\[2pt]
{\cal P}_r(t) &\!\!=\!\!& \{ \sigma_x  x(t) + \sigma_y y(t) + \sigma_z z(t) \} / r(t) \, , \\[2pt]
{\cal P}_\upphi(t) &\!\!=\!\!& \{ \sigma_y x(t) - \sigma_x y(t) \} / \rho(t) \, , \\[2pt]
{\cal P}_\uprho(t) &\!\!=\!\!& \{ \sigma_x x(t) + \sigma_y y(t) \} / \rho(t) \, , \\[2pt]
{\cal P}_\uptheta(t) &\!\!=\!\!& \{ z(t){\cal P}_\uprho(t) - \sigma_z \rho(t) \} / r(t) \, ,
\end{array}
\label{eq:vlosprojections}
\end{equation}
where the components are vertical (${\khat}$), radial (${\rhat}$), azimuthal (${\phihat}$) axial (${\rhohat}$), and polar (${\thetahat}$), respectively. Note that $V\subLOS(t) < 0$ is defined to be toward the observer and $V\subLOS(t) > 0$ is away from the observer.

\section{The Spatial Models}
\label{sec:spatialmodels}

In Table~\ref{tab:downtocases}, we list the geometric solids representing the spatial components of the galaxy/CGM and the geometric parameters governing their morphologies. For complete motivations, derivations, and accompanying illustrations the reader should refer to \citetalias{churchill25-skamI}. Here, we summarize the parametric equations mapping the locus of points $(X,Y,Z)$ defining the surface of each geometric solid in {\Gsys}.  

\begin{table}[thb]
\centering
\caption{Spatial Models \label{tab:downtocases}}
\begin{tabular}{lll}
\hline\\[-8pt]
Component &  Structure  &  Parameters \\
\hline\\[-8pt]
Halo             & Sphere      & $R\subCGM$  \\
Disk             & Cylinder    & $\rho_d; h_d$ \\
Wind    & Hyperboloid & $\Theta_w; \rho_{w,0}; R_w$ \\
Accretion & Hyperboloid & $\Theta_a; \rho_{a,0}; R_a$  \\[2pt]
\hline \\[-12pt]
\end{tabular}
\end{table}

\begin{figure*}[tb]
\centering
\includegraphics[width=0.98\textwidth]{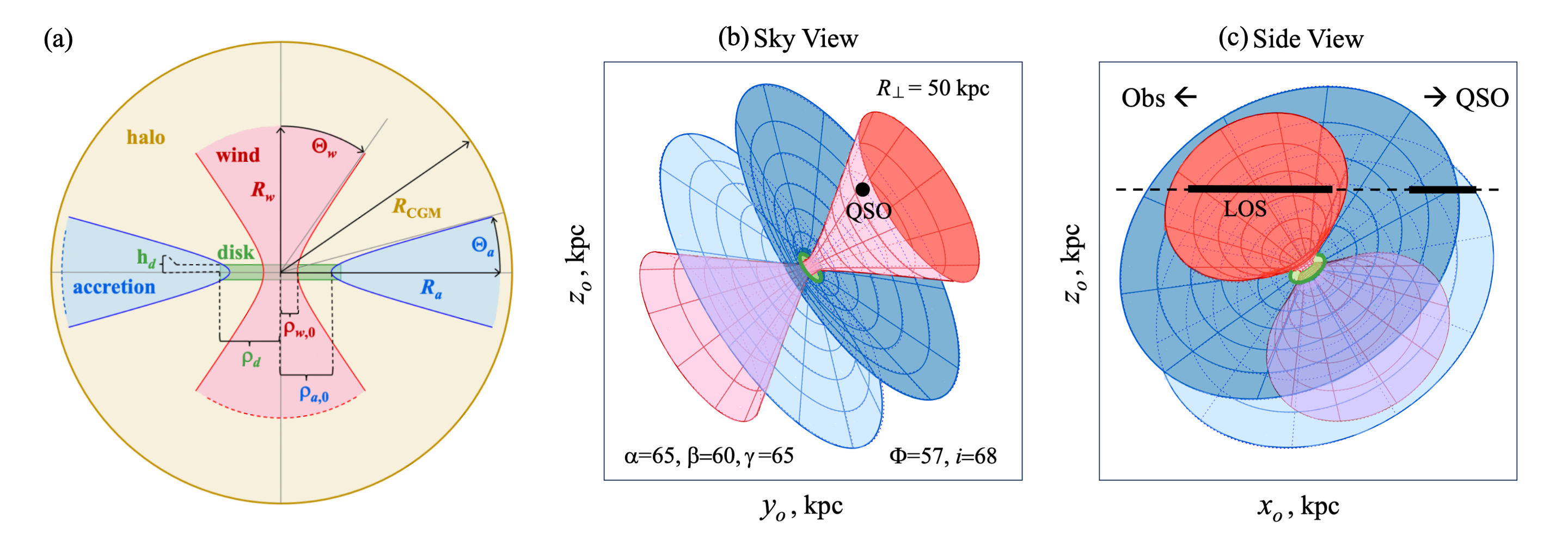}
\vglue -0.05in
\caption{\small  (a) A cross-sectional schematic (not to scale) of the galaxy/CGM structures in the galaxy frame, {\Gsys}. The spherical ``halo" has radius $R\subCGM$. The hyperboloidal wind has opening angle $\Theta_w$, base radius $\rho_{w,0}$, and maximum extent $R_w$. The cylindrical disk has axial radius $\rho_d$ and height $h_d$.  The planar accretion fills the void of the hyperbola with accretion radius $\rho_{a,0}$, maximum extent $R_a$, opening angle complement $\Theta_a$. (b, c) A schematic of a spatial-kinematic model and LOS in the observer frame, {\Osys}. The galaxy is obliquely inclined with $\alpha,\beta = 65^\circ, 60^\circ$, which yields inclination $i=68^\circ$. The quasar is placed at $R_\perp = 50$~kpc  with position angle $\gamma = 65^\circ$, which yields $\Phi = 57^\circ$.  The wind parameters are $\rho_{w,0}=10$~kpc, $\Theta_w=40^\circ$, and $R_w=R_{\rm vir}$, where $R_{\rm vir}= 200$~kpc. The accretion parameters are $\rho_{a,0}= 23$~kpc, $\Theta_w=20^\circ$, and $R_a=R_{\rm vir}$. The disk parameters are $\rho_d=25$~kpc, and $h_d=5$~kpc. (left) The sky plane showing the observer perspective. (right) The side view showing where LOS is probing the wind and accretion structure (thicker portion of the LOS).}
\label{fig:LOSskyview}
\vglue 0.2in
\end{figure*}

\subsection{The Halo}

The gas halo is approximated as a sphere of radius $R_h$ and is given by 
\begin{equation}
    X_h^2 + Y_h^2 + Z_h^2 = R^2_h \, .
    \label{eq:sphere}
\end{equation}
We define $R_h \!=\! R\subCGM$, and substitute $R\subCGM \!=\! \eta\subCGM R_{\rm vir}$ into Eq.~\ref{eq:sphere} where $\eta\subCGM \in (0,\infty)$ is a free scaling parameter.

\subsection{The Galactic Disk}

For the galaxy disk, we adopt a cylindrical structure with axial radius $\rho_d$ and height $h_d$. The parametric equations for the surface of this cylinder are
\begin{equation}
    X_d = \rho_d \cos \phi \, , \quad
    Y_d = \rho_d \sin \phi \, , \quad
    Z_d = z \, ,
\label{eq:diskstructure}
\end{equation}
for $z \in (0,h_d]$. The disk component is appropriate for the ISM, rather than the CGM.

\subsection{The Wind}

The volume occupied by the galactic wind is approximated as a hyperboloid of one sheet with base radius\footnote{The proper geometric term is skirt radius.} $\rho_{w,0} < \rho_d$ and opening angle $\Theta_w \in (0^\circ,90^\circ)$.  The parametric equations for the hyperboloid surface are written
\begin{equation}
\begin{array}{rcl}
X_w \!\!&=&\! \left(\rho_{w,0}^2 + Z_w^2 \tan^2 \Theta_w \right) ^{1/2} \cos \phi \, , \\[4pt]
Y_w \!\!&=&\! \left(\rho_{w,0}^2 + Z_w^2 \tan^2 \Theta_w \right) ^{1/2} \sin \phi \, ,
\end{array}
\label{eq:hypersurface}
\end{equation}
where $Z_w = z$. For $\rho_{w,0}=0$, Eq.~\ref{eq:hypersurface} describes a cone. Eq.~\ref{eq:hypersurface} describes a hyperboloid of infinite height above and below the galactic plane. The wind can be ``capped'' at a radius, $R_w = \eta_w R_{\rm vir}$, where $\eta_w \in (0,\infty)$ is a free scaling parameter.

\subsection{The Extended Planar Accretion}

The volume occupied by extended planar accretion is modeled as the void space of a hyperboloid of one sheet. By replacing the opening angle in Eq.~\ref{eq:hypersurface} by its complementary angle, we can write
\begin{equation}
\begin{array}{rcl}
X_a \!\!&=&\!  \left(\rho_{a,0}^2 + Z_a^2 \cot^2 \Theta_a \right) ^{1/2} \cos \phi \, , \\[4pt]
Y_a \!\!&=&\! \left(\rho_{a,0}^2 + Z_a^2 \cot^2 \Theta_a \right) ^{1/2}  \sin \phi \, , 
\end{array}
\label{eq:planarsurface}
\end{equation}
where $Z_a = z$. We refer to $\rho_{a,0}$ as the accretion radius and $\Theta_a \in (0^\circ,90^\circ)$ as the accretion opening angle. If one adopts $\rho_{a,0} \simeq \rho_d$, the structure interfaces with the disk edge. Alternatively an accretion zone overlapping the disk edge can be specified. If we apply the condition that the height of the hyperboloid structure matches the disk height, $h_d$, at the disk edge, we obtain $\rho^2_{a,0} = \rho_d^2 - h_d^2 \cot^2 \Theta_a$. This results in a torus around the disk where the structures overlap.

\subsection{Composite Model and LOS Probing}

A schematic showing the cross sections of the four geometric solids representing a galaxy/CGM SKAM is presented in Figure~\ref{fig:LOSskyview}(a). A sphere of radius $R\subCGM$ represents the overall extent of the gaseous CGM. A cylinder with axial radius $\rho_d$ and vertical height $h_d$ represents a galaxy disk. The wind fills the solid volume of a hyperboloid of one sheet with base radius $\rho_{w,0}$, opening angle $\Theta_w$, and radial extent $R_w$.  The extended planar accretion resides throughout the void of the solid volume of a hyperboloid of one sheet with accretion radius $\rho_{w,0}$, accretion opening angle $\Theta_{a}$, and radial extent $R_w$. Note that $R_w$ and/ or $R_a$ can extend beyond $R\subCGM$.  Each of these geometric parameters is defined using a multiplicative factor of the virial radius, i.e, as $R_x = \eta_x R_{\rm vir}$. 

In Figure~\ref{fig:LOSskyview}(b), we present an example of a LOS through an obliquely inclined galaxy ($\alpha,\beta = 65^\circ, 60^\circ$) in the observer frame of reference (sky view).  This galaxy has inclination $i=68^\circ$ and the quasar is located at $R_\perp=50$~kpc with azimuthal angle $\Phi = 57^\circ$.  The $z>0$ surface of the wind hyperboloid is located on the observer side of the sky plane and the $z<0$ surface is on the far side of the galaxy. 

The LOS positions $t$ where the LOS probes a geometric solid satisfies a simple quadratic inequality of the form $At^2 + Bt + C > 0$, where the coefficients depend on the geometric parameters defining the morphology of the geometric solid, the LOS zero point $P\subG(X_0,Y_0,Z_0)$ and direction cosines ($\sigma_x ,\sigma_y, \sigma_z$).  The coefficients and inequality conditions are presented in \citetalias{churchill25-skamI}. The most straight-forward strategy is to scan along the LOS from some $t_{\rm min} \ll 0 $ to some $t_{\rm max} \gg 0$ and track the inequality condition as a function of $t$ for each structure. In Figure~\ref{fig:LOSskyview}(c), we show that for this example SKAM, the wind is probed by the quasar LOS (thick line portions) on the near (observer) side of the galaxy.  Also probed is the accretion (in the void between the surfaces) on the far side of the galaxy.

\section{Kinematic Models}
\label{sec:kinematicmodels}

Idealized spatial-kinematic models for the four components of the galaxy/CGM are listed in Table~\ref{tab:kinematics-downtocases}.   For complete motivations, derivations, and accompanying illustrations the reader should refer to \citetalias{churchill25-skamI}. Here we summarize the kinematic models.

\begin{table}[h!bt]
\centering
\caption{Spatial-Kinematic Models$^{\rm a}$ \label{tab:kinematics-downtocases}}
\begin{tabular}{ll}
\hline\\[-8pt]
Model & Parameters  \\
\hline\\[-8pt]
Halo \\
-- Static & $V_r=0$  \\
-- Radial Outflow & $V_r>0$  \\
-- Radial Infall  & $V_r<0$ \\
-- Polar Orbit  & $V_\uptheta$\\[2pt]
Disk/EPG \\ 
-- Flat Rotation & $V_c$ \\
-- Lagging Halo$^{\rm b}$ & $V_\upphi (\rho,0)$, $H_v$  \\
-- Lagging Halo$^{\rm c}$ & $V_c$, $\rho_{d,0}$, $dV_\upphi/dz$, $dV_\upphi/d\rho$ \\[2pt]
Wind \\
-- Radial (Cone) & $V_w$   \\
-- Hyperbolic & $V_w$, $\rho_{w,0}$, $\Theta_w$ \\[2pt]
Planar Accretion \\
-- Keplerian  ($e\geq1$) & $\rho_{a,0}$, $V_\upphi(\rho_{a,0},0)$\\[2pt]
\hline \\[-8pt]
\multicolumn{2}{l}{(a) These models do not include stall, breaking,} \\
\multicolumn{2}{l}{\phantom{(a)} or enhancement functions (see \citetalias{churchill25-skamI})} \\
\multicolumn{2}{l}{(b) \citet[][]{steidel02}} \\
\multicolumn{2}{l}{(c) \citet[][]{bizyaev17}} 
\end{tabular}
\end{table}

\subsection{The Halo}

The spherical halo is not a spatial-kinematic structure per se, as much as it is a volume within which the spatial-kinematic structures are embedded. Simple kinematics can be assumed, such as radial infall or outflow, with ${\bf V}({\bf r}) = V_r\, {\ihat}({\bf r})$, where $V_r<0$ for the former and $V_r>0$ for the latter.  Alternatively, one could assume a static halo with $V_r=0$. Polar orbits emulating tidal streams (such as the Magellanic Stream of the Milky Way) could be described as great circles at fixed radii, i.e., ${\bf V}({\bf r}) = V_\uptheta\, {\thetahat}({\bf r})$. The LOS velocities are simply $V\subLOS(t) = V_r {\cal P}_r(t)$ and $V\subLOS(t) = V_\uptheta {\cal P}_\uptheta(t)$, respectively.

\subsection{The Disk/EPG}

The velocity vector field for a flat azimuthal rotation curve is ${\bf V}({\rvec}) = V_c\, {\phihat} ({\rvec})$, which has LOS velocity $V\subLOS(t) = V_c {\cal P}_\upphi(t)$. The free parameter $V_c$ scales the rotation speed and can be equated to the dark matter halo maximum circular velocity. Two models incorporating halo lag of extra-planar gas (EPG) have been explored. 

\citet{steidel02} introduced an exponential decrease with height above the disk plane as governed by scale height, $H_v$, 
\begin{equation}
{\bf V} (\rho,z) = V_\upphi(\rho,0) \exp \left\{ - |z|/H_v \right\} {\phihat} ({\rvec})\, .
\label{eq:vsteidel02}
\end{equation}
The quantity $V_\upphi(\rho,0)$ is the rotation speed in the galaxy plane ($z=0)$ at axial radius $\rho$ on the galaxy disk. This formalism allows one to specify a non-flat rotation curve. 

\citet{bizyaev17} characterized rotation lags with height above the plane using the expression
\begin{equation}
{\bf V} (\rho,z) = \left[ V_\upphi(\rho,0) + |z| \frac{dV_\upphi}{dz} \right] {\phihat} ({\rvec}) \, , 
\label{eq:vrbizaev17}
\end{equation}
where $dV_\upphi / dz$ is the negative velocity gradient with height above the plane, and the rotation curve is
\begin{equation}
 V_\upphi(\rho,0) = \left\{
\begin{array}{lcl}
\displaystyle V_c (\rho/\rho_{d,0}) && (\rho \leq \rho_{d,0}) \\[5pt]
\displaystyle V_c + (\rho - \rho_{d,0}) \frac{dV_\upphi}{d\rho}  && (\rho> \rho_{d,0}) \, .
\end{array}
\right.
\label{eq:vphibizaev17}
\end{equation}
The parameter ${\rho_{d,0} \simeq 0.1 \rho_d}$ is the axial radius out to which the rotation curve increases linearly; for $\rho > \rho_{d,0}$, the rotation curve decreases or increases linearly with $\rho$ depending on the value of $dV_\upphi / d\rho$. The rotation cut-off height is ${z(\rho) = \pm V_\upphi(\rho,0)/|dV_\upphi / dz|}$.  

\begin{figure*}[tb]
\centering
\includegraphics[width=0.9\linewidth]{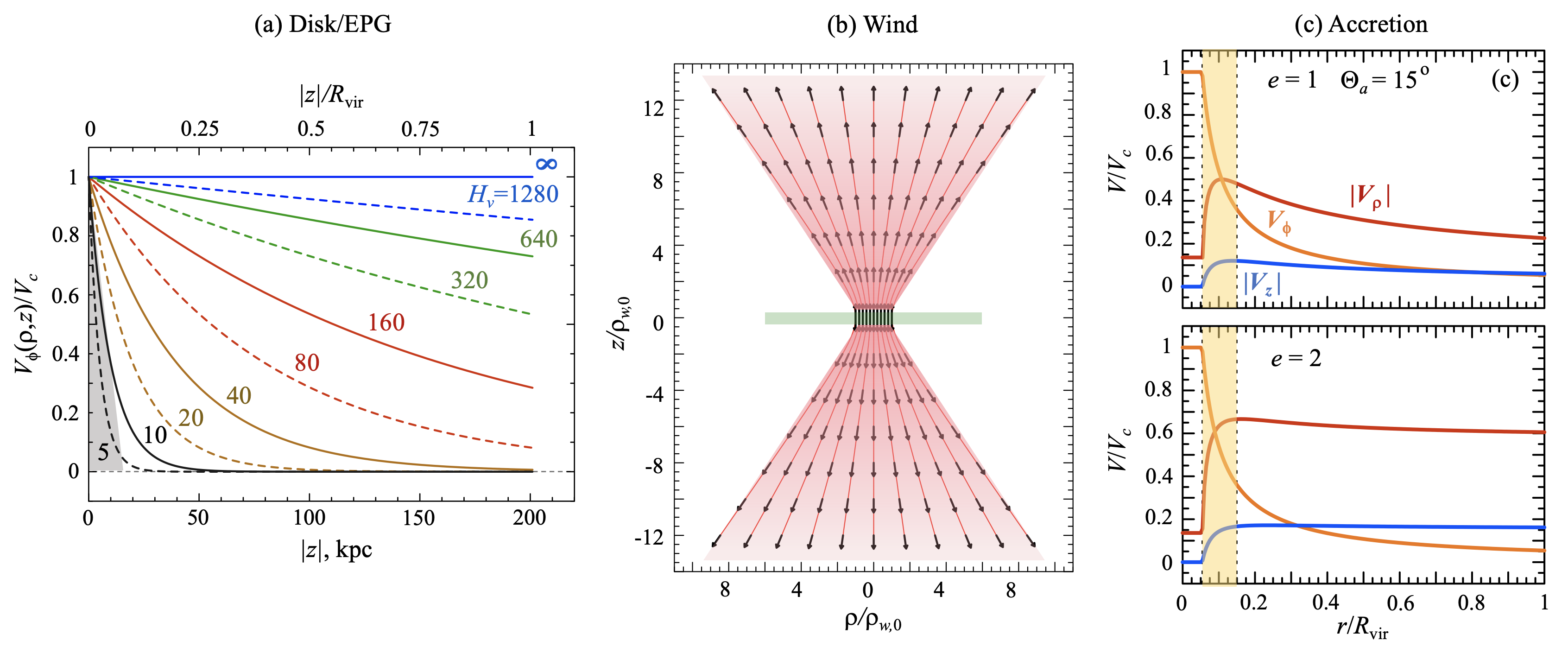}
\caption{\small  (a) The Disk/EPG kinematic model with exponential scale height governing lagging halo kinematics. The gray shading corresponds to the range of observed $dV_\upphi/dz$. (b) The cross section of a wind kinematic model for hyperboloid velocity trajectories,  illustrate as arrows.  (c) The accretion kinematic model of Keplerian infall from infinity. The illustrated $z$ component is the maximum, $|V_z(r,z_{\rm max})|$. (top) For $e=1$ describing the parabolic trajectory. (bottom) For $e=2$ describing a hyperbolic trajectory.}
\label{fig:myskam}
\vglue 0.15in
\end{figure*}

In Figure~\ref{fig:myskam}(a), we show $ V_\upphi(\rho,z)/V_c$ as a function of $z$ above the disk plane for arbitrary $\rho$ for the exponential halo-lag model of \citet{steidel02}. Several selected values of the scale height, $H_v$, are illustrated.  
The LOS velocity for the \citet{steidel02} model is given by
\begin{equation}
V\subLOS (t) = V_\upphi(t) \exp \left\{ - |z(t)|/H_v \right\} {\cal P}_\upphi(t) \, ,
\label{eq:VLOS-diskexp}
\end{equation}
and for the \citet{bizyaev17} model is
\begin{equation}
V\subLOS (t) = \left[ V_\upphi(t) - |z(t)| \frac{dV_\upphi}{dz} \right]  {\cal P}_\upphi(t) \, .
\label{eq:VLOS-bizaev}
\end{equation}

\subsection{The Wind}

At the base of the wind, the outflow velocities should be nearly perpendicular to the galaxy plane.  At height above the plane the wind fans outward from its central axis and asymptotically approaches radial outflow. We developed a model of the form,
\begin{equation}
   {\bf V}(\rho,z) = V_\uprho(\rho,z)\, {\rhohat}({\rvec}) + V_z (\rho,z)\, {\khat} \, ,
\label{eq:thehyperveltotal}
\end{equation}
where,
\begin{equation}
\begin{array}{rcl}
V_\uprho(\rho,z) \!\!\!&=&\!\!\! \displaystyle \frac{\Upsilon(\rho,z) V_w}{\sqrt{1+ \Upsilon^2(\rho,z)}} \\[15pt]
V_z(\rho,z) \!\!\!&=&\!\!\! \displaystyle \frac{{\rm sgn}(z)V_w}{\sqrt{1+ \Upsilon^2(\rho,z)}}  \, , 
\end{array}
\label{eq:mywindmodel}
\end{equation}
where $V_w$ is a free parameter (the wind launch speed), and where 
\begin{equation}
    \Upsilon(\rho,z) =\frac{|z|\rho \tan^2 \Theta_w}
    {\rho^2_{w,0} + z^2 \tan^2 \Theta_w} \, ,
\label{eq:upsilon4wind}
\end{equation}
flares the wind on hyperbolic trajectories. The cross-sectional view of the wind model is presented in Figure~\ref{fig:myskam}(b). Note that when $z= 0$ (galactic plane) or $\rho\!=\!0$ (wind axis), $\Upsilon(\rho,z) =0$ yielding $V_\uprho(\rho,0)=V_\uprho(0,z)= 0$ and $|V_z(\rho,{z\!\rightarrow\!0})| \simeq V_w$, all desired features of bi-polar wind kinematics. 
The LOS wind velocity is then
\begin{equation}
    V\subLOS(t) = 
    \frac{V_w\left[ \Upsilon(t) P_\uprho(t) + 
    {\rm sgn}(z(t)) P_z \right]}
    {\sqrt{1+ \Upsilon^2(t)}} 
     \, .
\end{equation}

\subsection{The Extended Planar Accretion}

Allowing for a velocity field with axial, azimuthal, and vertical components, we have
\begin{equation}
 {\bf V}({\bf r}) = 
V_\uprho (r) \, {\rhohat}({\rvec}) +
V_\upphi(r)\, {\phihat} ({\rvec}) 
+ V_z(r) \, {\khat} \, .
\label{eq:vaccretion}
\end{equation}
Keplerian infall from infinity can be parameterized by orbital eccentricity. For originally unbound material infalling  on planar trajectories we have $e \geq 1$. Enforcing that the azimuthal velocity component matches the rotation velocity at the disk/accretion interface, $V_\upphi(\rho_{a,0},0)$, we developed the kinematic model
\begin{equation}
\begin{array}{rcl}
V_\uprho(r)  \!\!&=&\!\! \displaystyle 
  {\cal E}_e(r) \, V_\upphi(\rho_{a,0},0) \\[12pt]
V_\upphi(r) \!\!&=&\!\!  \displaystyle
  \frac{\rho_{a,0}}{r} \,  V_\upphi(\rho_{a,0},0)  \\[12pt]
V_z(r) \!\!&=&\!\! \displaystyle 
  \frac{z}{\rho} \, {\cal E}_e(r) \, V_\upphi(\rho_{a,0},0)  \, ,
\end{array}
\label{eq:keplervels}
\end{equation}
for $r \geq \rho_{a,0}$, where
\begin{equation}
{\cal E}_e(r) = \frac{1}{1\!+\!e}
\left[ e^2 - \left\{ 1\!-\! (e\!+\!1) \frac{\rho_{a,0}}{r}\right\}^2 \, \right] ^{1/2} \, .
\label{eq:vaccretion-rhomag}
\end{equation} 

In Figure~\ref{fig:myskam}(c), we illustrate the radial behavior of the three velocity components for parabolic ($e=1$) and a hyperbolic ($e=2$) infall trajectories for a representative accretion model with wind opening angle $\Theta_a = 15^\circ$ and accretion zone (shaded region) spanning $\rho_{a,0}/R_{\rm vir} \leq r/R_{\rm vir} \leq r_d/R_{\rm vir}$, with $\rho_{a,0}/R_{\rm vir} = 0.05$ and $r_d/R_{\rm vir} = 0.15$.  Features of the accretion model are: as $r$ decreases, (1) the infall speed steadily increases following conservation of energy and momentum, (2) the axial component of the velocity increases, peaks in the accretion zone, and then rapidly decelerates so as to match the disk azimuthal velocity (which can be nonzero), while (3) the azimuthal component monotonically increases to match the disk rotation as $r \rightarrow \rho_{a,0}$, and (4) the vertical component of the velocity ensures that the accreting material decelerates to $V_z =0$ as it reaches the galactic plane and joins the disk at the accretion interface. 

In Figure~\ref{fig:myskam}(c), we plotted the maximum of $|V_z(\rho,z)|$, which is computed for the planes of maximum obliquity (the surface of the accretion hyperbola), for which $z_{\rm max} = \pm (r^2 - \rho_{a,0}^2)/\cot \Theta_a$.  These inflow trajectories (which are shown in \citetalias{churchill25-skamI}) emulate those shown in Fig.~2 of \citet{hafen22}, Fig.~7 of \citet{trapp22}, and Fig.~1 of \citet{stern23}.  

For the LOS accretion velocity, we have 
\begin{equation}
  V\subLOS(t) = \displaystyle V_\uprho(t)  {\cal P}_\uprho(t) + V_\upphi(t) {\cal P}_\upphi(t)  +
   V_z(t) {\cal P}_z(t) \, . 
\label{eq:VLOS-planaraccretion}
\end{equation}

\subsection{Enhancement Functions}
\label{sec:enhancefuncs}

As written, the free parameter for the wind launch velocity, $V_w$, is a constant. Spatial variation in the wind velocity can be modeled using an ``enhancement function." For example, we can write the wind launch velocity as $V_w + \tilde{V}\subE f\subE(x)$, where $\tilde{V}\subE$ is the amplitude of the variation of $V_w$ as a function of $x$ with $f\subE(x) \in [0,1]$ normalized such that $f\subE(0) = 1$. Several enhancement functions were described in Appendix~D of \citetalias{churchill25-skamI}.

\begin{figure}[h!tb]
\centering
\vglue 0.1in
\includegraphics[width=1.0\linewidth]{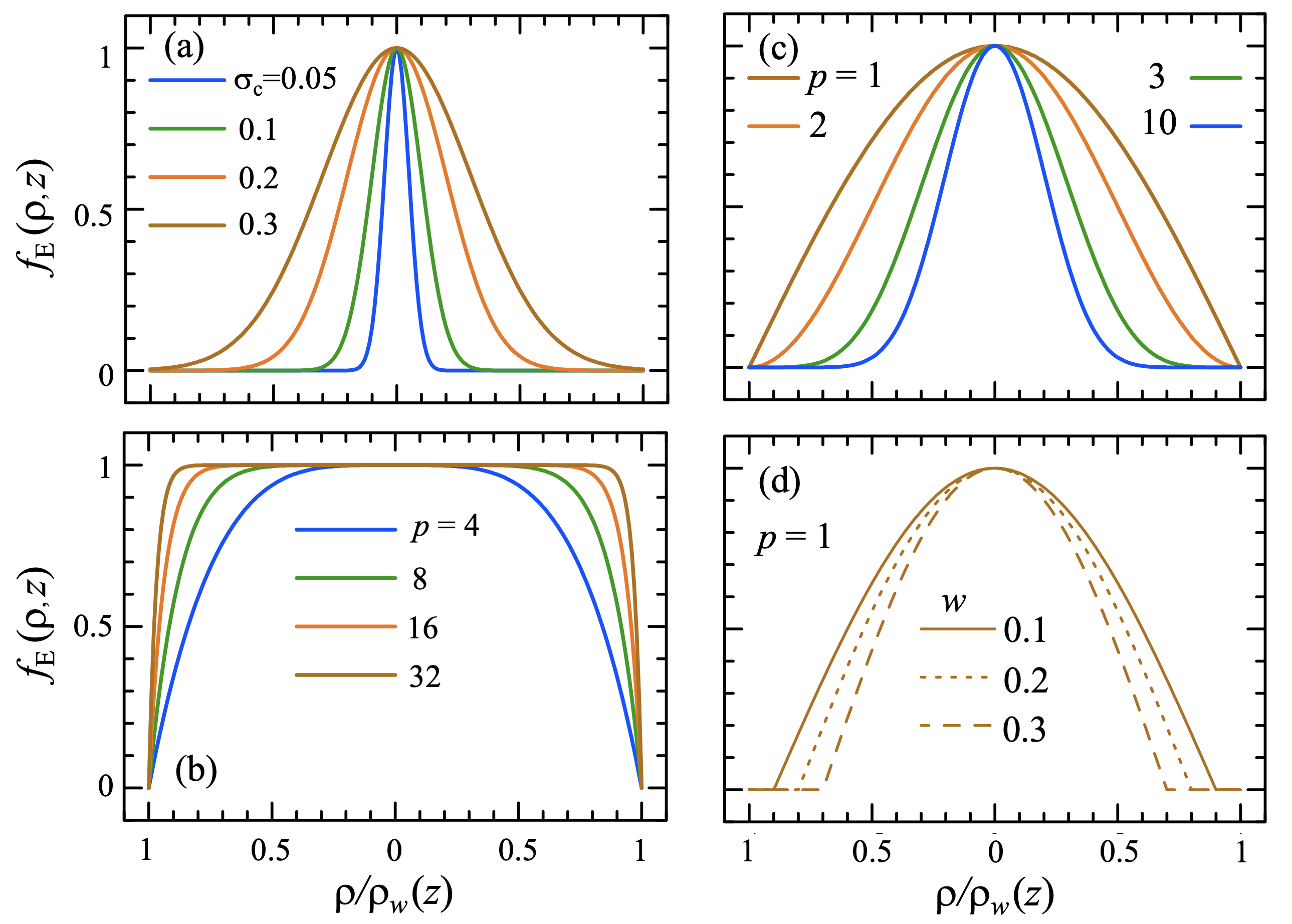}
\caption{\small  Examples of enhanced wind models with the highest wind speeds along the wind axis ($\rho/\rho_w(z)=0$) and the slowest speeds on the wind surface ($\rho/\rho_w(z)=1$). (a) Gaussian enhanced wind profiles. (b) Even powered polynomial wind profiles. (c) Truncated cosine wind profiles (that become Gaussian-like as $p$ is increased. (d) A truncated cosine model with $p=1$ and scaling $w$ that emulates a wind wall of thickness $w\rho_w(z)$}
\label{fig:enhance}
\end{figure}

In Figure~\ref{fig:enhance}, we show example enhanced wind models as a function of $x = \rho/\rho_w(z)$ with $0 \leq x \leq 1$, where $\rho_w(z)$ is the axial extent of the wind at height $z$ above the disk, given by $\rho_w(z) = [\rho^2_{w,0} + z^2\tan^2 \Theta_w]^{1/2}$.  Figure~\ref{fig:enhance}(a) illustrates a Gaussian $ f\subE(x) = \exp\{ - x^2 / {2\sigma^2_c} \}$ for $\sigma_c = 0.05$, $0.1$, $0.2$, and $0.3$. In Figure~\ref{fig:enhance}(b), we
illustrate the polynomial $f\subE(x) = 1 - x^p $, where $p$ is an even integer for $p=4$, 8, 16, and 32.  Figure~\ref{fig:enhance}(c) illustrates a truncated cosine function $f\subE(x) = \cos^p (\pi x /2 )$, where $p\geq 1$ is an integer for $p=1$, 2, 3, and 10. For $\tilde{V}\subE >0$, these enhancement functions all yield $V_w+\tilde{V}\subE$ in the wind core ($x=0$) and $V_w$ on the wind surface ($x=1$) at all heights $z$ above the disk plane.

The wind surface/wall can be given a thickness $w$ by replacing $x$ with $x/(1-w)$ in the enhancement functions, where $w$ is specified as a fraction of $\rho_w(z)$. Figure~\ref{fig:enhance}(d) illustrates wall thickness of $w=0.1$ 0.2, and 0.3 for a truncated cosine function with $p=1$. As described here, an enhanced kinematic wind model would include choice of functional form and two free parameters, $\tilde{V}\subE $ and $w$.

\section{Modeling Baryonic Gas}
\label{sec:gasphysics}

In order to synthesize absorption profiles of atomic transitions for various ions, we need to develop formalism for predicting the optical depths of various ion transitions as a function of LOS position and velocity. As optical depth is proportional to the product of the number density of the absorbing ion and the total absorption cross section, we must specify the densities of all ions and the line broadening physics of the gas at all locations within the CGM structures.  In this section, we develop the formalism for quantifying the gas phase properties (density and temperature) and the chemical and ionization conditions.  We also describe how we populate the individual spatial-kinematic structures with the gas. 

We adopt a simplified approach of quantifying a gas phase at spatial location ${\rvec}$ in the galaxy frame by its hydrogen number density, $n\subH({\rvec})$, and gas temperature $T({\rvec})$. To specify the gas chemical abundances, we adopt the abundance fraction, $\epsilon_k({\rvec})$, of each atomic species $k$. The number density of species $k$ is then $n_k({\rvec}) = \epsilon_k({\rvec}) n\subH({\rvec})$.  Obtaining the number density of each ion requires ionization modeling of the gas.  The ionization model provides the ionization fraction, $f_{kj}({\rvec}) = n_{kj}({\rvec}) /n_k({\rvec}) $, for each ionization stage $j$ of each atomic species $k$.  We adopt the convention that $j=0$ is the neutral atom so that the notation X$^{+j}$ applies for $j$ electrons having been ionized from species X (for example Mg$^+$, C$^{+3}$, and O$^{+5}$).  At each location in the gas structure, we determined the number density of each ion using the relation
\begin{equation}
    n_{kj}({\rvec}) = f_{kj}({\rvec}) \, n_k({\rvec}) = 
    f_{kj}({\rvec})  \, \epsilon_k({\rvec})  \, n\subH({\rvec}) \, .
\label{eq:nkj}
\end{equation}
The ionization fraction depends on the ionization thresholds of the ionization stages of the element. It also depends on gas density and temperature, and the spectral energy distribution (SED) of the local ionizing radiation field (hereafter, ``ionizing SED") within the gas. Thus, $f_{kj} = f_{kj}(n\subH,T,J_\nu)$, where $J_\nu$ represents the mean intensity of the ionizing SED.

For the ionization corrections, we use the ionization code Cloudy \citep[version C22.02][]{ferland98, ferland13, ferland17}, which is commonly employed for chemical-ionization modeling of the ISM, CGM, and IGM over a large range of gas densities and temperatures.  We built a grid of cloud models as a function of galaxy redshift, $z_{\rm gal}$, hydrogen number density, $n\subH$, and neutral hydrogen column density $N\subHI$ over the range $z_{\rm gal} \in (0.0,4.0)$ in steps of $\Delta z_{\rm gal}=0.2$, $\log (n\subH/{\rm cm}^{-3}) \in (-5.0,+1.0)$ in steps of 0.25 dex and $\log (N\subHI/{\rm cm}^{-2}) \in (14.0,21.5)$ in 0.5 dex steps.  All models are run using 10\% solar metallicity scaled from the ``recommended abundances" listed in Table~4 of \citet{lodders19} for the Sun.  A cosmic ultraviolet background (UVB) can be selected from one of various models \citep[e.g.,][]{haardt96, haardt12, finke10, finke22, ks19, faucher-giguere20}. For the adopted UVB model, the adopted redshift of the SKAM provides the cosmic epoch of the UVB.  In addition, stellar ionizing SEDs \citep[e.g.,][]{leitherer99} can be added to the UVB for assumed stellar populations (characteristic mass, age, metallicity) in the galaxy.  If stellar ionizing SEDs are to be included, the distance between the LOS position and the stellar population must be taken into account. For the illustrative examples shown in this work (see Section~\ref{sec:examples}), we employ the ionizing SEDs of \citet{haardt96, haardt12}; we do not incorporate stellar radiation.  

For a given point, $P\subG(x,y,x)$ in the SKAM frame of reference at redshift $z$, we specify $n\subH({\bf r})$, $T({\bf r})$, $\epsilon_k({\bf r})$ and $J_\nu({\bf r},z)$.  For these physical condition, we then use the Cloudy models to approximate the ionization fraction, $f_{kj}({\bf r})$,  from which we obtain the ion density $n_{kj}({\bf r})$ using Eq.~\ref{eq:nkj}. We further describe application of the ionization models in Section~\ref{sec:abslines} when we discuss the generation of spectral absorption lines.

\subsection{Spherical Halo}

Several researchers have developed models of the density, temperature, and pressure profiles of the CGM for extended halos \citep[e.g.,][]{mo96, maller04, marasco11, fang13, sharma14, faerman17, faerman20,  stern16, stern18, stern19, mathews17, pezzulli17, sormani18, faerman23, oppenheimer25}. Many of these models represent higher sophistication of earlier isothermal hot halo models with cooling and condensation of a cool/warm phase.  The observational constraints provided by X-ray and high- and low-ionization UV absorption lines are continually improving and better constraining these models. 

We begin with the single temperature isothermal sphere model.  \citep{maller04} showed that,
\begin{equation}
    T_{\rm h} \simeq 10^6 \left( \frac{V_c}{163~{\kms}}\right) ^2 ~{\rm K} \, ,
\end{equation}
which is consistent with the $T=2\times 10^6$ K temperature of the hot Milky Way halo as detected in X-ray emission and in {\OVII} absorption using the best fit circular velocity $V_c = 183\pm 41$~{\kms} for an extended halo model
\citep{hodges-kluck16}.  For an NFW halo \citep{NFW96}, the gravitational potential is  
\begin{equation}
    \Phi({\rvec}) = -\frac{GM_{\rm vir}}{x_sR_s}
    \frac{\ln( 1+x_s)}{A\subcalC} \, ,
\end{equation}
with $x_s=r/R_s$, where $R_s = R_{\rm vir}/{\cal C}$. The NFW isothermal density profile for a highly ionized gas is \citep{sharma14},
\begin{equation}
    n({\rvec}) = n_0 \exp
\left\{-
\frac{\mu m\subH \left( \Phi(r)-\Phi(r_b)\right) }{kT}
\right\} \, ,
\end{equation}
for $r\geq r_b$, 
where $\mu \simeq 0.6$ is the mean molecular weight, $m\subH$ is the hydrogen mass, and $n_0$ is the density at $r=r_b$. The central density is $n_0 \simeq 10^{-3}$~cm$^{-3}$ \citep{maller04,fang13,faerman17}. The inner boundary, $r_b$, is the galactocentric distance of the transition from the ISM to the CGM. This transition is observed to occur around $r_b \simeq 0.035R_{\rm vir}$ \citep[][]{nielsen24} which corresponds to $r_b\simeq 7$~kpc for $R_{\rm vir}= 200$~kpc. 

\begin{figure*}[tb]
\centering
\includegraphics[width=0.75\linewidth]{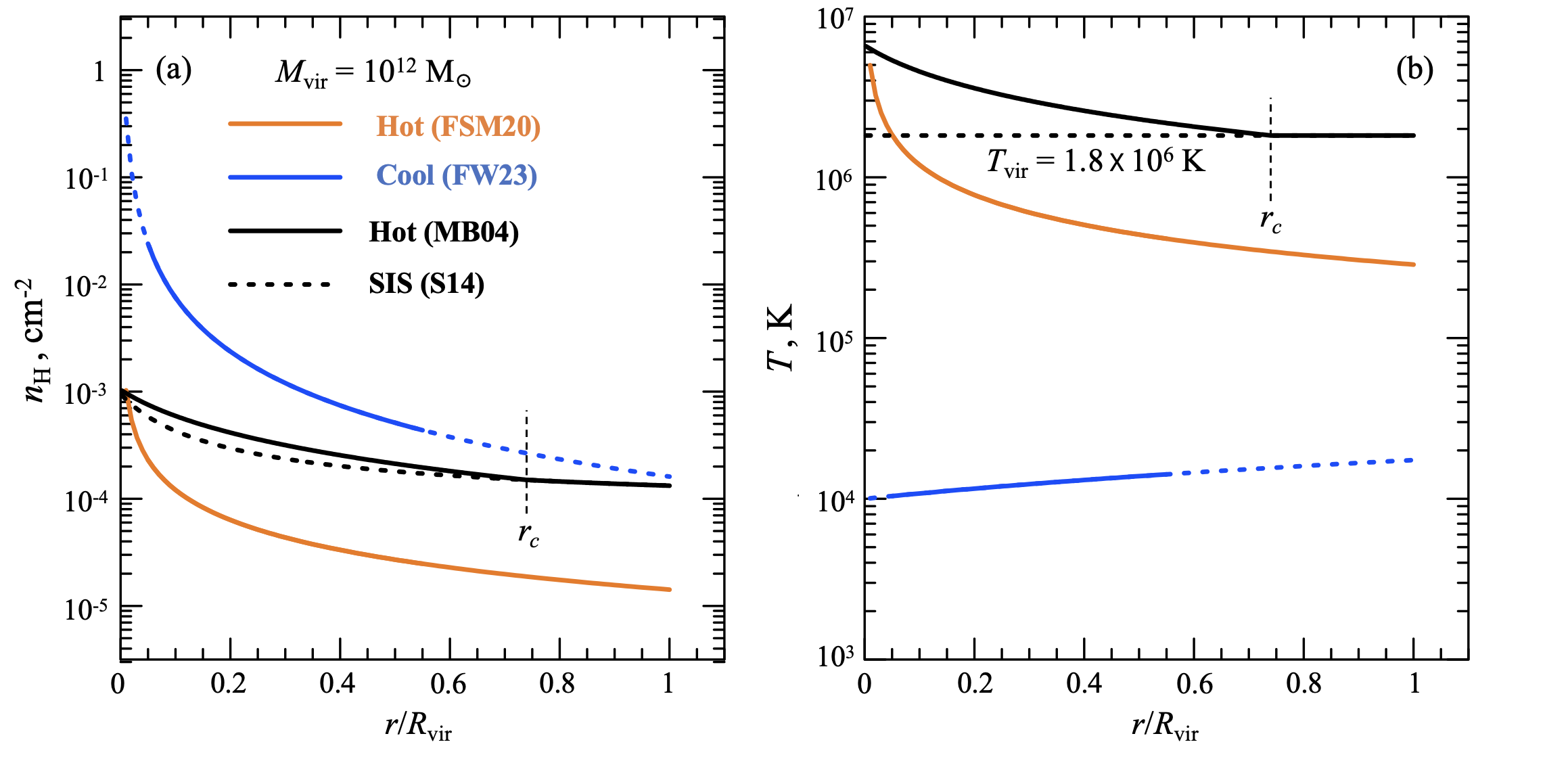}
\caption{\small  Various model hydrogen density and temperature profiles for a $M_{\rm vir} = 10^{12}$ M$_\odot$ halo. (a) The hot (orange) and cool (blue) gaseous components from the works or \citet{faerman20} and \citet[][]{faerman23}. The hot residual component (solid black) is from \citet[][]{maller04} and the single-isothermal sphere (SIS) model (dashed black) is from \citet{sharma14}.}
\label{fig:hothalo}
\vglue 0.1in
\end{figure*}

A two-phase model \citep{maller04} that is anchored to the cooling density, $n_c$, at the cooling radius, $r_c$, yields modified density and temperature profiles of the ``residual" hot phase,
\begin{equation}
    n({\rvec}) = n_c \, [g(x_s,x_c)]^{3/2} \, ,
    \quad
    T_{\rm h}({\rvec}) = T_{\rm h} \, g(x_s,x_c) \, ,
\end{equation}
where $x_s=r/R_s$ and $x_c = r_c/R_s$, and where the cooling density and radius for a Milky Way sized halo\footnote{The cooling density and radius are shown in Figs.~2 and 3 of \citet{maller04} for a broad range of halo masses (as a function of $V_c$) and metallicity. Representative values of $\Lambda_z$ are listed in their Table~A1.} are 
\begin{equation}
\begin{array}{rcll}
 n_c & \!\!\simeq\!\! & 6 \,
    T^2_6 / (\Lambda_z t_8 ) \times10^{-5} & {\rm cm}^{-3} \, , \\[5pt]
    r_c & \!\!\simeq\!\! & 160 \, 
    T^{-1/8}_6 (\Lambda_z t_8 )^{1/3} & {\rm kpc} \, ,  
\end{array}
\end{equation}
with $T_6 = T_{\rm h}/(10^6~{\rm K})$, where $\Lambda_z$ is a temperature and metallicity dependent normalization constant to the cooling function over the range $0.2$--$4.4$, and $t_8=t_{\rm f}/(10^8~{\rm yr})$ is the look-back time to the formation of the halo. The radial dependence for an adiabatic equation of state is given by the function
\begin{equation}
    g(x_s,x_c) = 1 + \frac{3.7}{x_s} \ln (1\!+\!x_s) - \frac{3.7}{x_c} \ln(1\!+\!x_c) \, .
\end{equation}
This hot phase provides the pressure support for the cool phase.  Note that the integrated baryon mass density of the halo gas, $M_b$, is normalized to the virial mass via $M_b=f_b M_{\rm vir}$, where $f_b \simeq 0.16$ is the cosmic baryon fraction.

In Figure~\ref{fig:hothalo}, we illustrate the radial density and temperature profiles for similar halo models. We show the single-isothermal sphere model of \citet[][S14]{sharma14} and the residual hot halo model of \citet[][MB04]{maller04}.  These density and temperature profiles are for a halo with $M_{\rm vir} = 10^{12}$~M$_{\odot}$. We have assumed $t_8=1$ and 10\% solar metallicity ($\Lambda_z = 1$). The densities have been converted to hydrogen density by multiplying by the hydrogen abundance fraction, $n\subH(r) = 0.91n(r)$.  Outside the cooling radius, the model of \citet{maller04} is equated to the isothermal sphere model, reflecting the lack of cooling for $r>r_c$. 

\citet{faerman23} extended the work of \citet{maller04} and \citet{faerman17,faerman20} by estimating the density  and temperature profiles for both the hot and the cool phase. They introduce the free parameter $\eta = \alpha_{\rm h}/\alpha_{\rm c}$, where 
${\alpha_{\rm h} = P_{\rm hot,tot}/P_{\rm hot,th}}$ and ${\alpha_{\rm c} = P_{\rm cool,tot}/P_{\rm cool,th}}$, where $P$ is the pressure, ``tot" denotes total pressure and ``th" denotes thermally supported pressure.  The hot component hydrogen density\footnote{For the hot density component of \citet{faerman20} a correction for $r\subCGM = 1.1R_{\rm vir}$ had been applied and for the cool density component of \citet{faerman23} the total density has been scaled by the abundance of hydrogen, 0.91.} is $n_{\hbox{\tiny H},\rm h}(r) = 1.4\times10^{-5} (r/R_{\rm vir})^{-0.93}$~cm$^{-3}$ and temperature is ${T_{\rm h}(r) = 2.9\times10^{5} (r/R_{\rm vir})^{-0.62}}$~K. For the cool component, the density is parameterized $n_{\hbox{\tiny H},\rm c}(r) = 2.1\times10^{-4} (r/R_{\rm vir})^{-1.67}$~cm$^{-3}$ over the range $r/R_{\rm vir} \in (0.05,0.55)$ where the filling factor of the cool gas in their model is non-zero. Note that a wide variation in $n_{\hbox{\tiny H},\rm c}(r)$ was found based on a range of thermal support scenario, such that the normalization density ranges from $0.6$--$8.4\times 10^{-4}$~cm$^{-3}$ and the power law index ranges from $1.5$--$1.8$. No parameterization was provided for the cool temperature, but we find that the function $\log T_{\rm c}(r) = 4.0 + \tau_{\rm c}\log(1+r/R_{\rm vir})$ with $\tau_{\rm c} =0.80^{+0.30}_{-0.24}$ is in reasonable agreement for the range of thermal support scenarios \citep[see Fig.~1 of][]{faerman23}. 

In Figure~\ref{fig:hothalo}, we show both the hot (orange curve) and cool component (blue curve) models of \citet[][]{faerman20} and \citet[][]{faerman23} for a halo with mass $M_{\rm vir} = 10^{12}$~M$_{\odot}$.  Unlike the \citet{maller04} model, it is not perfectly clear how the \citet{faerman23} density and temperature profiles for an $M_{\rm vir} = 10^{12}$~M$_{\odot}$ halo scale with halo mass and concentration parameter. But, what is clear is that \citet{faerman23} propose a halo populated by a cooler but less dense hot component than do \citet{maller04}.  The differences between these predictions suggest that models of the two-phase CGM/halo provide roughly an order of magnitude of freedom in the adopted density and temperature profiles for spatial-kinematic modeling.

Overall, galactic halos are complex; the density and temperature profiles discussed here are smoothed versions that omit sub-halo size dynamical structures.  For example, the Magellanic System, which resides at a galactocentric distance of  $\sim\! 50$--100 kpc from the center of the Milky Way, is a complex, cool, moderate-density structure that interacts with the hot halo \citep[see][]{d-onghia116}.  Some external galaxies also exhibit polar rings \citep[e.g.,][]{maccio06, brook08, spavone10}. These streaming structures serve as examples of how satellite galaxies are stripped of their gas out to large radii as they move through a diffuse hot halo medium \citep[e.g.,][]{putman12}.

If the spherical halo is to be modeled as a unique component of the CGM, then the adopted density and temperature profiles for the halo should be applied only at LOS positions where the LOS is not probing the bi-polar wind nor the extended planar accretion. Thus, at LOS position $t$, the criterion $r_b \leq r(t)\leq R\subCGM$ would need to hold while also ensuring that the LOS is not probing either the wind structure or accretion structure.  In addition, there would need to be a transition from the extra planar gas of the galaxy disk (which we discuss below).

\subsection{Bi-Polar Wind}
\label{eq:wind-nHT}

In general, the hot wind component can be modeled as thermalized ejecta driven by massive stars \citep[e.g.,][]{chevalier85, zhang18, veilleux20}. As concluded by \citet{rupke18} in their review of stellar driven galactic winds, ``The simple model of a modest-velocity, biconical flow of multiphase gas and dust perpendicular to galaxy disks continues to be a robust descriptor of these flows."  Indeed, an idealized cone model for the spatial-kinematics of bi-polar winds has been explored for interpreting quasar absorption line data and is supported by such studies \citep[e.g.,][]{bouche12, chen14, schroetter19, zabl20, beckett23}.

The highly dynamic multiphase density and temperature distributions of galactic winds are complex and challenging to model. Understanding stellar driven winds requires adequate treatment of the dynamics on large spatial scales as well as detailed treatment of the rapid interactions between the ISM and CGM on small spatial scales at the base of the wind \citep[e.g.,][]{naab17, huang22}. The exchange of mass, momentum, and energy between the hot and the cool/warm phases requires diligent treatment of turbulent radiative mixing layers \citep[e.g.,][]{gronke20, tan21}. The sophistication of multiphase models has advanced such that both a hot component and a cold/warm component in the form of embedded clouds entrained by the hot wind are included and these provide guidance to the gas phase distributions within wind structure \citep[e.g.,][]{bustard16, huang20, fielding22, nguyen22, ghosh25}. 

Here, we denote $n\subH(\rho,z)$ and $T(\rho,z)$ as the hydrogen number density and gas temperature distributions throughout the wind structure.
A zeroth-order model for the baryons confined in the wind is a single isodensity-isothermal gas phase with $n\subH(\rho,z)=n_w$ and temperature $T(\rho,z)=T_w$.  A first-order variation might yield a density that declines with height above the wind base but is constant for each cross sectional area at height $z$; for example, a wind cross-sectional area dependence can be modeled as, $n\subH(z) = n_w\, [\rho_{w,0}/\rho_w(z)]^2$, where $n_w$ is the hydrogen density at the wind base and $\rho_w(z) = [\rho^2_{w,0} + z^2\tan^2 \Theta_w ]^{1/2}$ is the wind cross sectional radius at height $z$. 

To capture an axis-symmetric gradient in the distribution of wind gas phases, we can use enhancement functions (see Section~\ref{sec:enhancefuncs}), and write 
\begin{equation}
\begin{array}{rcl}
n\subH (\rho,z) &\!\!=\!\!& \displaystyle 10^{\nu(\rho,z)} \, , \\[7pt]
T (\rho,z) &\!\!=\!\!&  \displaystyle 10^{\tau(\rho,z)} \, ,
\label{eq:temperature-wind}
\end{array}
\end{equation}
where 
$\nu(\rho,z) = \nu_1 +\nu_2 f\subE(\rho/\rho_{w}(z))$ 
is a spatially varying exponent specifying the hydrogen number density, and $\tau(\rho,z) = \tau_1 + \tau_2 f\subE(\rho/\rho_{w}(z))$ is the same but for the temperature. The parameters $\nu_1$ and $\tau_1$ are the exponents giving the density and temperature along the wind walls and $\nu_2$ and $\tau_2$ give the amplitude of the variations in the wind core relative to the wind walls. The enhancement functions, $f\subE (\rho/\rho_{w,0}(z))$, provide the functional variation. The density enhancement function and the temperature enhancement function can be unique to one another. Enhancement functions were briefly described in Section~\ref{sec:enhancefuncs} with selected examples illustrated in Figure~\ref{fig:enhance}. They are discussed with further detail in Appendix~D of \citetalias{churchill25-skamI}. The essential criteria of enhancement functions is that $0 \leq f\subE(x) \leq 1$ with $f\subE(0) =1$ and that the function is axisymmetric about the wind axis.

Since the adopted enhancement functions are arbitrary and can be unique to the density and temperature, Eq.~\ref{eq:temperature-wind} provides a great deal of flexibility in the gas phase profiles populating the wind structure (clearly, these distribution functions are not
unique and other forms can be imagined). Insights into the chosen $\nu_1$, $\nu_2$, $\tau_1$, $\tau_2$ and the form of the enhancement functions can be garnered from models such as those of \citet{bustard16}, \citet{fielding22}, \citet{nguyen22}, and \citet{ghosh25}. Empirical insights can be gleaned from the work of \citet{xu23-classy6}, who developed methods for measuring wind densities, pressures, and temperatures from absorption profiles.  

Enhancement functions can approximate a spatially segregated multiphase medium.  For example, a hot, low density wind core surrounded by a thin low-ionization surface layer that is cooler and denser could be modeled by truncated cosines with wall thickness $w=0.1$ (see Figure~\ref{fig:enhance}(d)). The models of \citet{nguyen22} suggest $\nu_1 \simeq -2$ with $\nu_2 \simeq -3$ yielding a density along the wind wall surface of ${n\subH \simeq 10^{-2}}$~cm$^{-3}$ and ${n\subH \simeq 10^{-5}}$~cm$^{-3}$ in the core, and ${\tau_1 \sim 4}$ with ${\tau_2\simeq 3}$ so that temperature along the wind wall surface is $ T \simeq 10^4$~K rising to $T \simeq 10^7$~K in the core.  These conditions would be expected to yield X-ray absorption from ions such as {\OVII} and {\OVIII} in the wind core and UV absorption from ions such as {\OVI}, {\CIV} and {\MgII} nearer to and in wind walls.

Mapping $z(t)$ and $\rho^2(t) = x^2(t)+y^2(t)$ as a function of  LOS position $t$, we write
\begin{equation}
\begin{array}{rcl}
\log n_{\hbox{\tiny H}}(t) &\!=\!& \nu_1 + \nu_2 f\subE (t) \\[7pt]
 \log T(t) &\!=\!& \tau_1 + \tau_2 f\subE (t) \, .
\end{array}
\label{eq:LOSdensity-wind}
\end{equation}

\subsection{Galactic Disk}

For purpose of illustration in this work, we will discuss disk galaxies. The vertical distribution of gas in our Galaxy has been studied for several decades \citep[e.g.,][]{dickey-lockman90, padoan01, berkhuijsen06, kalberla08,  savage09, marasco11, langer14, qu19, qu20}. Observationally constrained models of the density structure of gaseous disks are commonly expressed as the product of two independent functions, i.e., $n(\rho,z) = f_\uprho(\rho;L_d)f_z(z;H_d)$ where the most common fitted parameters are a disk scale length, $L_d$, and a vertical scale height, $H_d$. 

For Milky Way studies, the deprojection of the data can depend on the assumed rotation curve, which can be derived assuming gravitational dynamics in a Galactic dark matter halo  \citep[e.g.,][]{kalberla08}.  The shape of the disk influences how the scale height changes with axial distance \citep[see][]{deGrijs97}; though it is generally found that the {\HI} scale height is approximately constant over a large range of axial radii \citep[][also see \citet{putman12} and references therein]{dickey-lockman90, marasco11}.  It is commonly assumed that the {\HI} gas is supported in the Galactic gravitational potential primarily by radiatively  driven turbulence \citep{ferrara96}, with contributions from magnetic and cosmic-ray pressure. 

\citet{marasco11} modeled the {\HI} halo of the Milky Way assuming a density distribution $n\subHI(\rho,z)$ analogous to that used by \citet{oosterloo07} for the external galaxy NGC 891,
\begin{equation}
n\subHI(\rho,z) = n_0 \left[ 1\!+\!\frac{\rho}{L_d} \right]^{\gamma_\rho} \!\! 
e^{-\rho/L_d} \, {\rm sech}^2 ({z}/{H_d}) 
\, ,
\label{eq:nHImarasco11}
\end{equation}
where $n_0 = n\subHI(0,0)$ and with best-fitted parameters $\gamma_\rho \simeq 5$, $L_d \simeq  1.6$~kpc, and $H_d \simeq  1.6$~kpc.  
If we adopt this functional form for $n\subHI$, we can determine $n_0$ by normalizing to $N\subHI(0)$, the path-length integrated hydrogen column density perpendicular to the disk at $\rho = 0$.  We obtain $n_0 = N\subHI(0)/2H_d$. Defining $N_{20} = N\subHI(0)/10^{20}$~cm$^{-2}$, we obtain $n_0 \simeq 0.016(N_{20}/H_d)$~cm$^{-3}$, where $H_d$ is expressed in kpc.  For $N_{20} = 2$ and $H_z=1.6$~kpc, we have $n_0 \simeq 0.02$~cm$^{-3}$.

Because of the unknown ionization structure of the disk, the {\HI} density profile does not necessarily reflect the total hydrogen density profile. It is well known that {\HI} absorption traces ``cold" neutral hydrogen clouds, whereas absorption from ions such as {\CII}, {\NII}, {\SiIV}, {\CIV}, and {\OVI} trace the warm-ionized medium in which hydrogen is partially ionized \citep[e.g.][]{haffner09, savage09, qu19, qu20, langer21}. Pulsar dispersion measures also probe the warm-ionized medium,  yielding the distribution of electron density in ionized clouds throughout the Galaxy \citep[e.g.,][]{berkhuijsen06, gaensler08}.

From {\SiIV} and {\OVI} absorption, the warm gas distribution in the Galaxy has been modeled as a
1D plane-parallel slab with distribution $n(z) \propto \exp\{-z/H_d\}$, where $H_d$ is the scale height perpendicular to the disk.  \citet{savage09} found $H_d\simeq3.2$, 3.6, and 2.6~kpc for {\SiIV}, {\CIV} and {\OVI} respectively, whereas \citet{qu19} reported $H_d\simeq 2.6$~kpc for both {\SiIV} and {\OVI}.
\citet{qu19} also measured the disk scale length of $L_d \simeq 6.1$~kpc assuming $n(\rho) \propto \exp\{-\rho/L_d\}$.

\citet{qu20} estimated the ``warm gas density" in the Milky Way follows the parameterization
\begin{equation}
n(\rho,z) = n_0 \,
e ^{-(\rho/L_d)^{\alpha_\rho}} e ^{ - (|z|/H_d)^{\alpha_z}}
\, ,
\label{eq:diskHden}
\end{equation}
with best-fitted parameters $\alpha_z \simeq 0.8$, $H_d \simeq 3$--$6$ kpc, $\alpha_\rho \simeq 3.4$, and $L_d \simeq 12$--13 kpc. In the axial radial direction, this density distribution is much flatter than the {\HI} distribution model of \citet[][]{marasco11} given by Eq.~\ref{eq:nHImarasco11}; it also shows a ``sharp edge" at 15--20 kpc (not found in the {\HI} model), just beyond the disk scale length owing to the large value of $\alpha_\rho$. 

Care must be taken in adopting 1D plane-parallel models of the absorption line data.  \citet{zheng19}
propose that low-to-intermediate velocity  {\SiIV} absorption in the Milky Way extends well beyond the inner Galactic halo and into the CGM and is best described by a spherical geometry. They argue that this ``global component" is not likely to arise from either the Local Bubble, Galactic superbubbles, the Fermi Bubble, the CGM of M31, or the Local Group medium. Instead, they suggest there is likely to be a large amount of ionized gas moving at low-intermediate velocities that is not part of the extra-planar gas (EPG) associated with the disk and lower halo of the Milky Way.  This would imply that the scale heights of fitted 1D plane-parallel disk models are overestimated from the data.

We aim to model the total hydrogen density, $n\subH(\rho,z)$.  We adopt a model that combines elements of the {\HI} model of \citet{marasco11} and of the warm ionized medium of \citet{qu20},
\begin{equation}
    n\subH(\rho,z) = n_0\,
    e^{-(\rho/L_d)^{\alpha_\rho}} \,
    {\rm sech}^2(z/H_d) \, ,
\label{eq:diskdensity}
\end{equation}
valid for $\rho \leq \rho_d$
with free parameter $n_0 = n\subH(0,0)$. We adopt $\alpha_\rho = 3.5$ based on the fit of \citet{qu20} in order to induce a ``sharp edge'' just beyond the adopted disk scale length.

We aim to smoothly tie this density model to the density model for the extended planar accretion (see Section~\ref{sec:xpa-density}).  The boundary condition at the disk edge, $\rho=\rho_d$, is where the models will be required to yield the same hydrogen density in the plane, $n\subH(\rho_d,0)$, and the same column density perpendicular to the structures, $N\subH(\rho_d)$. Once we formulated the density profile of the extended accretion, apply the boundary conditions, and obtain the expression for $n\subH(\rho_d,0)$,
we can determined the disk scale length by evaluating Eq.~\ref{eq:diskdensity} in the plane at the disk edge, 
\begin{equation}
L_d = \frac{\rho_d}{\alpha_\rho} \ln \left\{ \frac{n_0}{n\subH(\rho_d,0)} \right\} \, .
\label{eq:L_d}
\end{equation}
We will apply the boundary conditions from the extended planar accretion in Section~\ref{sec:xpa-density} (see Eq.~\ref{eq:nHrho0}).

Integrating Eq.~\ref{eq:diskdensity} along the perpendicular to the disk, we have,
\begin{equation}
N\subH(\rho) = 2n_0 H_d \, e^{-(\rho/L_d)^{\alpha_\rho}} \, .
\label{eq:DNE}
\end{equation}
Evaluating this relation at $\rho = \rho_d$ provides an expression from which the disk scale height can be computed in terms of the free parameters $n_0$, $N\subH(\rho_d)$ and the disk scale length, $L_d$,
\begin{equation}
H_d = \frac{N\subH(\rho_d)}{2n_0} \, e^{(\rho_d/L_d)^{\alpha_\rho}} \, .
\label{eq:H_d}
\end{equation}
As described further below, $N\subH(\rho_d)$, is a free parameter that is based on theoretical predictions. Thus, the disk hydrogen density model is described by two free parameters, $n_0$ and $N\subH(\rho_d)$, and its scale length and scale height are determined by the boundary condition in the plane $z=0$ at the disk edge, $\rho_d$.

\begin{figure*}[htb]
\centering
\includegraphics[width=1.0\linewidth]{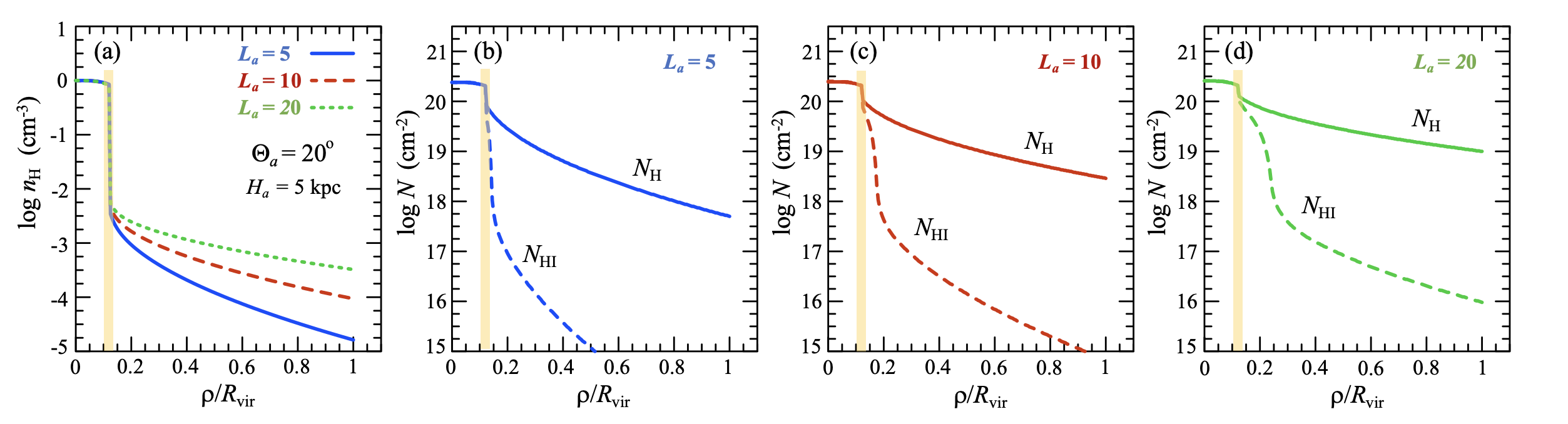}
\caption{\small  The idealized density profiles $n\subH(\rho,0)$, $N\subH(\rho)$, and $N\subHI(\rho)$ for a disk with $\rho_d=25$~kpc, $h_d=5$~kpc, $n_0=1$~cm$^{-3}$, and $N\subH(\rho_d) = 20.3$~cm$^{-2}$ interfaced with extended planar accretion having $\Theta_z = 20^{\circ}$ and $H_a=5$~kpc in a halo with $R_{\rm vir} = 200$~kpc. The accretion zone is shown as the shaded region, ranging from $\rho_{a,0}/R_{\rm vir} =0.10$ to $\rho_d/R_{\rm vir} =0.13$. (a) The hydrogen number density profiles, $n\subH(\rho,0)$ in the $z=0$ plane or three different extended accretion axial scale lengths, $L_a$.  (b,c,d) The perpendicular column density profiles, $N\subH(\rho)$ and $N\subHI(\rho)$, (b) for $L_a=5$~kpc, (c) for $L_a=10$~kpc, and (d) for $L_a=20$~kpc. These density and column density profiles are designed to capture the essential features of the theoretical predictions from simulations of \citet{hafen22} \citet{trapp22}, and \citet{stern23}; especially, see Fig.~14 of \citet[][]{trapp22}.}
\label{fig:xpadensity}
\vglue 0.15in
\end{figure*}

\subsection{Planar Accretion}
\label{sec:xpa-density}

Theoretically, extended planar accretion is predicted to be a significant component to the building of galaxy disks \citep[e.g.,][]{keres05, stewart11, stewart13, stewart17, stewart-proc17, vandevoort11, hafen22, trapp22, gurvich23, stern23, kocjan24}.  In simulations, \citet{stewart11, stewart13} reported the formation of a transient ``cold flow disk" structure around their galaxies. These massive extended planar structures were often warped, not rotationally supported, and comprise inflowing cold halo gas.  Most interestingly, these structures are kinematically aligned with the angular momentum of the central galaxy rather than with the angular momentum of the inflowing cold filamentary gas.  
Subsequent theoretical work using hydrodynamic cosmological simulations has confirmed that the infalling material spirals inward, and as doing so, tends to align with the plane of the galactic disk (see Fig.~2 of \citealp{hafen22} and Fig.~1 of \citealp{stern23}).  \citet[][see their Fig.~7]{trapp22} shows that this accretion fans inward over a flare angle ranging from $15^\circ$--$30^\circ$.  According to \citet[][]{hafen22}, \citet[][]{trapp22}, and \citet[][]{stern23} rotational support sets in within a narrow annulus at the galaxy disk edge. 

The Milky Way {\HI} disk is significantly warped but shows a coherent structure out to $R \sim 35$~kpc \citep{kalberla08}. Most galaxies {\HI} layers display flairs beyond the stellar disk \citep[e.g.][]{sancisi08, vanderKruit11}, though they exhibit a universal neutral gas surface density profile out to several $R_{25}$ \citep{bigiel12}.  Gas in the outer region of the Milky Way disk ($\rho \sim 40$--60~kpc) is well described as having an exponential radial scale length of 7.5~kpc \citep{kalberla08}.

As such, in terms of an idealized spatial structure, an extended planar accretion CGM component can be modeled as a flared extension to the galactic disk. Geometrically, this simple spatial structure should exhibit azimuthal symmetry and have a vertical height that increases with axial distance from the galaxy center. The void space of the hyperboloid of one sheet we adopted is well suited. In addition to having an azimuthal symmetry about the galaxy rotation axis, the hyperboloid skirt radius can be interpreted as the axial distance at which the accretion interfaces to the galactic disk. 

For the extended accretion structure, we consider a density distribution in which the density profile is symmetric above and below the $z=0$ plane,  
\begin{equation}
n\subH(\rho,z) = n\subH(\rho,0) \, {\rm sech}^2 (z/H_a) \, ,
\label{eq:density-accretion}
\end{equation}
valid for $\rho \geq \rho_d$ where $n\subH(\rho,0)$ is the hydrogen density profile as a function of axial radius $\rho$ in the $z=0$ plane, and where the free parameter $H_a$ is the vertical scale height of accreting gas.

In order to determine $n\subH(\rho,0)$, we normalize $n\subH(\rho_d,z)$ at the edge of the disk to the perpendicular column density $N\subH(\rho_d)$. For the column density distribution along the axial radius, we assume the form 
\begin{equation}
N\subH (\rho) = N\subH(\rho_d) 
\exp 
\left\{ - \left[
\frac{\rho-\rho_{a,0}}{L_a}
\right] ^{1/2} \right\} \, ,
\label{eq:NHrho1}
\end{equation}
valid for $\rho \geq \rho_d$, where the free parameter $L_a$ is the axial radial scale length of the extended planar gas.  The functional form $f(x) \propto e^{-\sqrt{x}}$ provides a rapid decline for small $x$ followed by a slower decline as $x$ increases; this emulates the findings of the simulations of \citet{hafen22} and \citet{trapp22}, who show a steep increase in density and column density as the disk edge is approached. Clearly, this distribution function is not unique and other forms can be imagined. Here, the square root of the exponent can be loosely identified with the average S\`ersic index of $n\sim 0.5$ for the parameterized expression for observed $N\subHI$ profiles along de-projected galaxy disks \citep{ianjamas18}.  Furthermore, it captures the ``sharp edge" of the Galactic warm ionized gas at the disk edge as modeled by \citet{marasco11} and adopted for our disk model (Eq.~\ref{eq:diskdensity}).  

Integrating Eq.~\ref{eq:density-accretion}, we write the column density integrated perpendicular to the disk plane
\begin{equation}
N\subH(\rho) = 2 \int_{0}^{H_\perp(\rho)} 
\!\!\!\!\!\!\!\!\!\!\!
n\subH(\rho,0) \, {\rm sech}^2 (z/H_a) \, dz \, ,
\end{equation}
where $H_\perp(\rho) = [{\rho^2-\rho^2_{a,0}}]^{1/2}\tan\Theta_a$ is the half-path length perpendicular to the accretion structure at axial radius $\rho$, and obtain,
\begin{equation}
N\subH(\rho) = 2 \, n\subH(\rho,0) H_a 
\, {\rm tanh} (
{H_\perp(\rho) }/{H_a} ) \, .
\label{eq:NHrho2}
\end{equation}
Equating Eqs.~\ref{eq:NHrho1} and \ref{eq:NHrho2} and rearranging, we obtain the hydrogen density profile in the plane,
\begin{equation}
n\subH(\rho,0) = \displaystyle
\frac{N\subH(\rho_d) \, \exp \left\{ - [ ( \rho-\rho_{a,0})/ L_a ]^{1/2} \right\}}
{2H_a \, {\rm tanh} ( H_\perp(\rho) /H_a)} \, ,
\label{eq:nHrho0}
\end{equation}
which we insert into Eq.~\ref{eq:density-accretion} to complete our expression for $n\subH(\rho,z)$ for the extended planar accretion.

\subsection{The Combined Disk Extra Planar Accretion}

Evaluating Eq.~\ref{eq:nHrho0} at the disk edge yields $n(\rho_d,0)$, which we insert into Eq.~\ref{eq:L_d} to obtain the disk scale length, $L_d$. We then can evaluate the disk scale height, $H_d$ from Eq.~\ref{eq:H_d}.

In summary, the two main free parameters of the combined disk combined extended accretion density profile are $n_0 = n\subH(0,0)$, the hydrogen density at the origin, and $N\subH(\rho_d)$, the perpendicular hydrogen column density at the disk edge.  The other two free parameters are the extended accretion axial scale length, $L_a$, and vertical scale height, $H_a$.  Recall that the disk axial scale length, $L_d$ (Eq.~\ref{eq:L_d}), and vertical scale height, $H_d$ (Eq.~\ref{eq:H_d}), are computed to match the disk and extended accretion density and column density at the disk edge.  

In Figure~\ref{fig:xpadensity}, we show selected disk plus extended accretion density profiles (in the galactic plane) and column density distributions (perpendicular to the structure). Guided by the simulations, we adopt fiducial values of $n_0 = 1$~cm$^{-3}$ (see Fig.~7 of \citealp{hafen22} and Fig.~4 of \citealp{stern23}) and $\log N\subH(\rho_d)=20.3$ \citep[see Fig.~14 of][]{trapp22}.  For $H_a = 5$~kpc, we illustrate the density profiles for $L_a=5$, 10, and 20~kpc.  These parameters yield a disk axial scale length on the order of $L_d \sim 40$~kpc and vertical scale height of $H_d \sim 40$~pc. 

The simulations suggest that the above model proposed here, which is characterized by a clearly truncated disk in terms of hydrogen column density, is best suited for hot accretion onto thin disks.  For thick disks, the accretion mechanisms are likely to be more chaotic \citep{hafen22, stern23}. Hot accretion in this case is characterized by gas with $T_a \sim 10^{5.5}$~K (see Fig.~3(a) of \citealp{hafen22} and Figs.~1 and 4 of \citealp{stern23}). The hot  accreting gas cools over a short $\sim\!300$--600~Myr period as it journeys into the what we are calling the ``accretion zone."  \citet{hafen22}, \cite{trapp22}, and \citet{stern23} all find that the gas temperature decreases by $\simeq\! 1.5$ dex to $T_d \simeq 10^4$~K and the hydrogen number density increases by $\sim\! 2$ dex to $10^{-1}$~cm$^{-3}$ as the accreting gas infuses into the disk zone; these rapid changes induce a rapid change toward lower ionization conditions at the edge of the gas disk.

\begin{figure}[h!bt]
\centering
\includegraphics[width=1.0\linewidth]{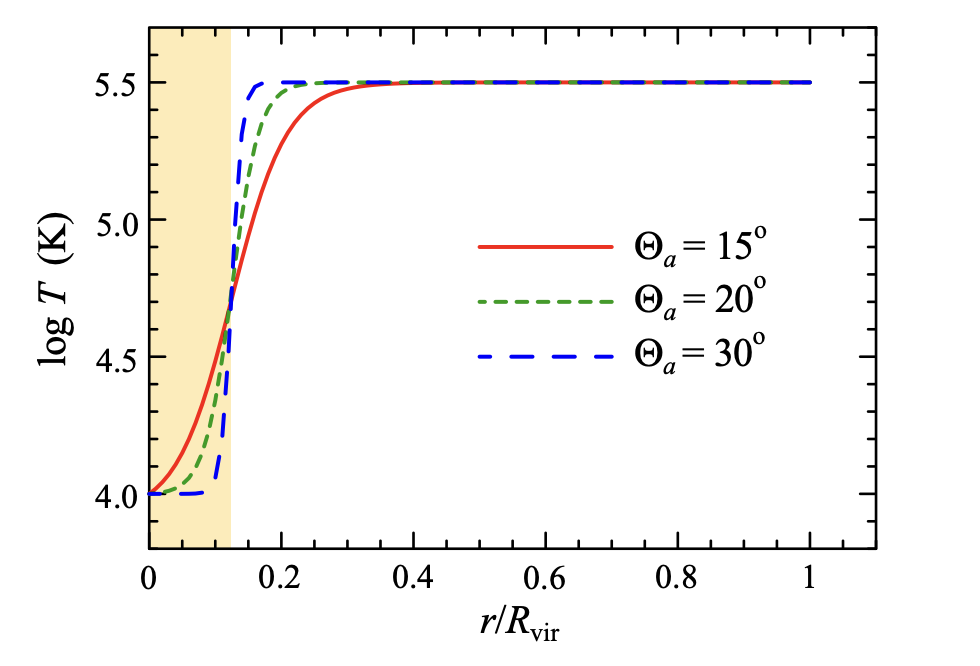}
\caption{\small  Temperature profiles with accretion $\tau_a=5.5$ and disk $\tau_d=4.0$ for extended planar accretion/disk for flare angles $\Theta_a=15^\circ$, $\Theta_a=20^\circ$, $\Theta_a=30^\circ$. These temperature profiles (see Eq.~\ref{eq:fermi-dirac-temp}) are designed to capture the essential features of the theoretical predictions from simulations of \citet{hafen22} \citet{trapp22}, and \citet{stern23}. The yellow shading represents the radial extent of the disk for $\rho_d = 25$~kpc and $R_{\rm vir} = 200$~kpc.}
\vglue 0.1in
\label{fig:xpatemperature}
\end{figure}

In Figure~\ref{fig:xpatemperature}, we show an idealized temperature profile for the planar accretion as a function of $r/R_{\rm vir}$.  The profiles are shown for $\Theta_a=15^\circ$, $20^\circ$, and $30^\circ$. We have adopted a modified Fermi-Dirac temperature profile ranging from $\tau_h = \log T_h$ to $\tau_d = \log T_d$,
\begin{equation}
 \log T(r) = \tau_a - (\tau_a\!-\!\tau_d) \left[
 \frac
 {1\!+\! \exp\left\{ - \rho_d / \Delta \rho\right\}}
 {1\!+\! \exp\left\{ (r\!-\!\rho_d)/\Delta \rho \right\}} \right] \, ,
\label{eq:fermi-dirac-temp}
\end{equation}
where $\Delta \rho = \rho_d-\rho_{a,0}$. Recall that $\rho_{a,0}$ is a function of $\Theta_a$ and is given by $\rho^2_{a,0} = \rho_d^2 - h_d^2 \cot^2 \Theta_a$.

Theory supports the notion that hot accretion models predict disk truncation similar to what we have emulated here with our proposed gas-phase distribution functions \citep[see][]{trapp22, hafen22}. That thinner disks have sharper cutoffs has been supported by observations for some time \citep[e.g.,][]{comeron12}. \citet{hafen22} also noted that hot accretion does not appear to operate as a mechanism for building thicker disks. One could imagine building gas-phase distribution functions that would emulate cold accretion as guided by theoretical works \citep[e.g.,][]{keres05, dekel06, brooks09, vandevoort11, noguchi18, kocjan24}.

\section{Absorption Profiles}
\label{sec:abslines}

Quasar absorption line profiles are typically modeled assuming pure absorption, $I(v)/I_0 = \exp \{ -\tau(v) \}$, where $I(v)/I_0$ is the relative flux and $\tau(v)$ is the optical depth at LOS velocity $v$. Since a given absorption line is specific to an electron transition in a given ion, we adopt notation in which the subscript $k$ represents the elemental species, $j$ represents the ionization stage, and $i$ represents the transition. Though the spatial-kinematics of the galaxy/CGM structures are specified as continuous density, temperature, and velocity fields, the absorption spectra must be discretized. This is required by the radiative transfer because changes in the radiation field occur over finite paths within the medium.  Optical depth for absorption quantifies the number of mean free paths surviving non-absorbed photons have traveled along a path length of LOS. 

We thus begin by dividing the LOS into finite intervals of fixed length, $\Delta t$. The center of the $m^{\rm th}$ LOS interval is denoted $t_m$ and will span fixed interval $\Delta t = t_m^+ - t_m^-$. In this interval, the mean hydrogen number density is $n\subH(t_m)$, the mean temperature is $T(t_m)$, and the mean LOS velocity is $v(t_m)$, which hereafter is referred to as $v_m$.  The mean abundance of species $k$ is $\epsilon_k(t_m)$ and the ionization fraction for ion $kj$ is $f_{kj}(t_m)$.  The mean ion number density in the LOS interval is then given by 
\begin{equation}
n_{kj}(t_m) = f_{kj}(t_m) \, \epsilon_k(t_m) \, n\subH(t_m) \, .  
\label{eq:nkjoft}
\end{equation}

For estimating the ionization fractions, we employed Cloudy \citep[version C22.02,][]{ferland98, ferland13, ferland17}. As previously mentioned, we built grids that are a function of $\log (n\subH/{\rm cm}^{-3}) \in (-5.0,+1.0)$ in steps of 0.25 dex and $\log (N\subHI/{\rm cm}^{-2}) \in (14.0,21.5)$ in 0.5 dex steps. We store the ionization fractions as a 2D array function of $n\subH$ and $N\subH$, where the latter is determined from $N\subH = f\subHI N\subHI$, where $f\subHI$ is the ionization fraction of neutral hydrogen. For the ionization models, we have assumed a \citet{haardt96, haardt12} ionizing spectrum and a 10\% solar metallicity. 

To locate the Cloudy grid point $(n\subH,N\subH)$ from which we pull the ionization fraction $f_{kj}$ for our ion of interest, we must estimate $n\subH$ and $N\subH$ at LOS location $t$.  We first determine the average $n\subH$ in the LOS segment computing the geometric mean,
\begin{equation}
  n\subH = \exp \left\{ \frac{1}{t_+ - t_-} \int_{t_-}^{t_+} \!\!\!\! \ln (n\subH(t)) dt \, , \right\}  \, .
\end{equation}
However, if $\Delta t$ is small enough, one can simple adopt $n\subH(\bar{t})$, where $\bar{t} = (t_-+t_+)/2$.  This is also the value of $n\subH(t_m)$ used in Eq.~\ref{eq:nkjoft}.

The value of $N\subH$ for selecting the appropriate Cloudy model in the $(n\subH,N\subH)$ grid is not the column density of the LOS segment, i.e., $N\subH \neq n\subH \Delta t$. For gas that is  embedded deep within the galaxy/CGM, the ionizing radiation will be shielded by the large {\HI} column in the structure \citep[][]{rahmati13, ploeckinger20}. Shielding softens the ionizing spectrum resulting in lower ionization conditions. To account for shielding, we determine the perpendicular hydrogen column density from the location ${\bf r}(t)$ to the nearest surface of the galaxy/CGM structure being probed by the LOS.  For locations inside the wind, we obtain
\begin{equation}
    N\subH = \int_{\rho(t)}^{\rho_w(z(t))_{\phantom{X}}} \!\!\!\!\!\!\!\!\! \!\!\!\!\!\!\!\!\!\! n\subH (\rho,z(t)) \, d\rho \, ,
\end{equation}
where ${\rho_w(z(t)) = [\rho^2_{w,0} + z^2(t)\tan^2 \Theta_w ]^{1/2}}$, and where $n\subH (\rho,z(t))$ is given by Eq.~\ref{eq:temperature-wind}. For locations inside the planar accretion we obtain,
\begin{equation}
N\subH =  n\subH(\rho(t),0) \! \int_{z(t)}^{H_\perp(\rho(t))_{\phantom{X_2}}} 
\!\!\!\!\!\!\!\!\!\!\! \!\!\!\!\!
  {\rm sech}^2 (z/H_a) \, dz \, ,
\end{equation}
where the density in the plane, $n\subH(\rho(t),0) $, is given by Eq.~\ref{eq:nHrho0}, $H_a$ is the density scale height, and the half-height of the accretion structure at axial distance $\rho(t)$ is $H_\perp(\rho(t)) = [{\rho^2(t)-\rho^2_{a,0}}]^{1/2}\tan\Theta_a$.
The ionization fraction for the ion of interest is then obtained using 2D linear interpolation of the four Cloudy grid points bracketing ($n\subH$,\, $N\subH$). 

The optical depth for transition $i$ of ion $kj$ in the m$^{\rm th}$ LOS interval of length $\Delta t$ is given by
\begin{equation}
\begin{array}{rcl}
\tau_{kji}(v,v_m) &\!\!=\!\!& \displaystyle \int_{t_m^{-}}^{t_m^{+}} \!\!\!\! n_{kj}(t_m) \, \sigma_{kji}(v,v_m)\, dt  \, , \\[16pt]
&\!\!=\!\!& \displaystyle N_{kj}(t_m) \, \sigma_{kji}(v,v_m)
\end{array}
\label{eq:tau-defined}
\end{equation}
where $N_{kj}(t_m) = n_{kj}(t_m) \Delta t$ is the mean column density of ion $kj$ following integration over the LOS interval, and where $\sigma_{kji}(v,v_m)$ is the total absorption cross section for transition $i$ of ion $kj$, which is a Voigt function peaking at the centroid $v=v_m$.  
The total absorption cross section evaluated for LOS interval $t_m$ as a function of $v$  can be written
\begin{equation}
\sigma  _{kji}(v,v_m) = 
\frac{\sqrt{\pi}e^2}{m_ec} \frac{f_{kji}\lambda_{kji}}{b_m} {\cal H}(x_m,y_m) \, ,
\label{eq:thevoigtxsec}
\end{equation}
where $\lambda_{kji}$ is the transition wavelength, $f_{kji}$ is the transition oscillator strength, and where ${\cal H}(x_m,y_m)$ is the Hjerting function, 
\begin{equation}
    {\cal H}(x_m,y_m) = \frac{y_m}{\pi} \int_{-\infty}^{\infty} \frac{e^{-z^2}\, dz}{(x_m - z)^2 + y_m^2} \, ,
\end{equation}
with peak at $x_m=0$ with full-width half maximum $2y_m$, where 
\begin{equation}
\begin{array}{rcl}
x_m &\!\!=\!\!& (v-v_m)/b_m  \, , \\[4pt]
y_m &\!\!=\!\!& \Gamma_{kji} \lambda_{kji}/(4\pi \, b_m) \, ,\\[4pt] 
b_m &\!\!=\!\!& \left[ (2k\subB/m_k)\, T(t_m) \right]^{1/2}
\end{array}
\end{equation}
where $\Gamma_{kji}$ is the transition damping constant and $b_m$ is the mean thermal Doppler parameter for an ion with mass $m_k$ in the LOS interval centered on LOS location $t_m$. Note that $x_m$ and $y_m$ are unitless. Several works have presented computationally expedient algorithms for computing the Hjerting function \citep[][]{humlicek79, tepper-garcia06, kochanov11, schreier18}. We employed the fast routine of \citet{humlicek79}.

Multiplying the function given by Eq.~\ref{eq:thevoigtxsec} by the column density, $N_{kj}(t_m)$, yields the optical depth profile, $\tau(v,v_m)$, as given by Eq.~\ref{eq:tau-defined}. The optical depth profile for the m$^{\rm th}$ LOS segment is a Voigt function with central peak at $v_m$. Stepping along the LOS one interval at a time, one computes the absorption spectrum for transition $i$ of ion $kj$ by summing the optical depths for each LOS segment at each $v$
\begin{equation}
\tau_{kji}(v) = \sum_{m} \tau_{kji}(v,v_m) \, ,
\label{eq:finaltau}
\end{equation}
and applying the 1D solution of radiative transfer for pure absorption, $[I(v)/I_0]_{kji} = \exp \{ -\tau_{kji}(v) \}$. 

Computationally, in order to obtain the optical depth as a function of velocity, $\sigma_{kji}(v,v_m)$ must be evaluated at discrete values of $v$. In practice, this can be done by stepping the velocity using some small interval $\Delta v$ ranging from some minimum value to some maximum value, say $-500 \leq v  \leq +500$~{\kms}. An appropriate value of $\Delta v$ might be the velocity width of a pixel for a commonly used spectrograph.\footnote{Be aware that the smaller $\Delta v$ is the smaller $\Delta t$ will need to be in order to resolve such small velocity differences along adjacent locations along the LOS (even though the SKAM is a continuous model, the sampling of the LOS is discretized.)}  For example, if there are $p$ pixels per resolution element for a spectrograph with resolution $R=\lambda/\Delta\lambda\subR = c /\Delta v\subR = c/(p \Delta v)$, where  $\Delta\lambda\subR$ and $\Delta v\subR$ are the FWHM of the instrument line spread function, then $\Delta v = c/(pR)$. For example, HIRES and UVES have $R=45,000$ and $p=3$, yielding $\Delta v = 2.2$~{\kms}. 

If spectra corresponding to a given astronomical spectrograph are desired, one can convolve the absorption spectrum with the instrumental line spread function, expressed  $[I(v)/I_0]_{kji} = \Phi(\Delta v) \ast \exp \{ -\tau_{kji}(v)\}$, where $\Delta v = R/c$  is the full-width half-maximum of the spectrograph resolution element and $R$ is the spectrograph resolution. One can then add noise if desired \citep[see][]{churchill15-direct}.

Different LOS locations can have the same LOS velocity; thus, absorption at a given value of $v$ can arise from spatially separate LOS segments from different spatially remote galaxy/CGM structures.  If we wish to compute the absorption spectrum for only a single individual galaxy/CGM structure, we can carry out Eq.~\ref{eq:finaltau} only over the LOS intervals that probe that structure. Doing so for each structure, we can generate a final absorption spectrum (by summing the optical depths of all structures) while also knowing what fraction of each flux decrement, i.e., $1-[I(v)/I_0]_{kji}$, comes from each structure (see Figure~\ref{fig:exampleSKAM} below).  This can be very powerful for dissecting absorption profile kinematics in terms of galaxy/CGM spatial-kinematics, LOS impact parameter, and galaxy orientation.

\section{Discussion}
\label{sec:discussion}

The overall SKAM formalism presented here provides a manageable tool for testing our hypotheses and theoretical predictions about the structure and dynamics of the CGM and about the various components of the baryon cycle. The utility of SKAMs lies in their intuitive simplification of the CGM and their versatility and flexibility.  One of the flexibilities is that the CGM can be highly simplified into idealized astrophysically motivated geometric component structures, each with their specific kinematics and baryon gas phases. A second flexibility is that any hypothesized geometric structure, velocity field, or gas phase distribution can be added, subtracted, or swapped out. In practice, implementing these flexibilities may be as simple as conditional execution and control flow of function calls from a package or packages of module libraries.   

Here, we discuss the SKAMs parameters described in this work. In Section~\ref{sec:examples}, we present a fiducial SKAM model and selected example absorption lines. Following our main conclusions, we describe the development of a SKAM GUI (see Section~\ref{sec:theGUI}) and then contemplate methods for incorporating multiphase gas models with clumpy distributions of cool dense clouds embedded in a warm/hot phase (see Section~\ref{sec:multiphase}).

\subsection{SKAM Parameterizations}

SKAMs require a number of free parameters that describe (1) the galaxy/GCM geometries, (2) their velocity fields, and (3) their gas phase distributions and conditions.  One can envision almost endless possibilities. It is desirable to keep the number of free parameters to a minimum, while diversifying the explorable parameter space. In \citetalias{churchill25-skamI}, we derived the geometric formalism for ``observing" SKAMs that requires only two free parameters, the coordinate rotation angles; we reviewed this formalism in Section~\ref{sec:thesetup}.  In \citetalias{churchill25-skamI}, we also presented the observational and theoretical background motivating selected geometric and kinematic representations of the CGM that we then developed. We summarized those spatial-kinematic models in Sections~\ref{sec:spatialmodels} and \ref{sec:kinematicmodels}.  In Table~\ref{tab:downtocases} we listed the parameters required for the proposed geometric structures and in Table~\ref{tab:kinematics-downtocases} we listed the parameters for the kinematic models we proposed. Here, we briefly review these parameters and discuss their applicability.

\subsubsection{Dark Matter Halo}

First and foremost are the parameters that yield the virial mass and virial radius of the dark-matter halo hosting the galaxy/CGM.  For the formalism presented here, which adopts NFW profiles \citep{NFW96}, three parameters are required: the maximum circular velocity, $V_c$, the halo overdensity $\Delta_c$, and the halo concentration parameter, ${\cal C}$. These also yield the halo scale radius $R_s = R_{\rm vir}/\mathcal{C}$ and the radius of maximum circular velocity  $R_{c} = 2.16528 R_s= (2.16528/\mathcal{C})R_{\rm vir}$.

\subsubsection{Galaxy/CGM Geometries}

The spherical CGM halo is defined by its radius, $R\subCGM = \eta\subCGM R_{\rm vir}$, where $\eta\subCGM$ is the free parameter.  The cylindrical disk is defined by its axial radius, $\rho_d$ and its height, $h_d$. The wind hyperboloid geometry is defined by the opening angle, $\Theta_w$, the base radius, $\rho_{w,0}$, and its radial extent, $\eta_w R_{\rm vir}$.  The accretion hyperboloid geometry is defined by the opening angle, $\Theta_a$, the accretion radius, $\rho_{a,0}$, and its radial extent, $\eta_a R_{\rm vir}$. The accretion and disk geometries can be tied together by creating an accretion zone with inner radius $\rho_{a,0} = \{\rho^2_d - h_d^2 \cot ^2 \Theta_a \}^{1/2}$ and outer radius $\rho_d$. This option reduced the free parameters for the accretion geometry to $\Theta_a$ and $\eta_a$.

\subsubsection{Galaxy/CGM Kinematics}

For our purposes, we will define the halo gas as the material residing within the spherical region of radius $R\subCGM$ that is not occupied by the disk, wind, or accretion structures (refer to Figure~\ref{fig:LOSskyview}(a)).  Halo gas kinematics can be modeled as a simple radial infall or outflow by specifying $V_r$ or $-V_r$, respectively, at all points inside the sphere. Streams moving along great circles can be modeled on polar orbits by specifying $V_\uptheta$. 
 
In reality, the halo is a kinematically complex region where turbulent and/or differently directed semi-coherent multiphase flows are quasi-randomized between outward-directed winds and IGM accretion \citep[see, for example][]{ramesh23}. Much of the infalling IGM gas is originally concentrated within large-scale filaments \citep[e.g.,][]{ceverino09, vandevoort11, danovich15, nelson18}. These infalling filaments meet resistance from stellar-driven outflows \citep[e.g.,][]{habe82, nelson19, peroux20, nguyen22, trapp22, faucher-giguere23, hidalgo-pineda25}.  The outflows tend to fan out as they rise above the disk \citep[see][]{rupke18, zhang18, nguyen22}, whereas the infalling gas tends to spiral inward and migrate into the galaxy plane \citep[e.g.,][]{keres05, stewart11, stewart17, stewart-proc17, hafen22, trapp22, stern23, kocjan24}. 

Gas at the wind/accretion interfaces can be stirred up, become turbulent, and then kinematically and spatially decouple from winds and accretion \citep[e.g.,][]{bisht25, kakoly25}. Furthermore, tidal streams and winds from satellite dwarf galaxies can further inject more gas mass, kinetic energy, and quasi-systematic kinematics, perhaps accounting for some proportion of the high-velocity cloud population \citep[e.g.,][]{york86, kacprzak10-pks1127, putman12, tpw-araa17, tripp22}. 

As such, the kinematics of the halo component is very challenging to characterize using simple velocity fields. This leads to the notion that it may be advantageous to omit the halo component kinematics from a SKAM, as this allows isolated and controlled exploration of disk/EPG, wind, and accretion kinematics. Since disk/EPG absorption will arise for only smaller impact parameters, for larger impact parameters, only wind and accretion gas flows will be giving rise to absorption. This may allow a simplification that leverages interpretation of observed  absorption lines. When comparing SKAM generated mock absorption lines to observed data for a given galaxy-absorber pair, one might then attribute unexplained absorption as arising in the halo component and then work backwards to constrain its kinematic properties.

For this work, we described three simple models of Disk/EPG kinematics. The first is flat rotation, which can be parameterized by a constant circular velocity. The halo maximum circular velocity, $V_c$, confined within the geometric cylinder representing the galaxy disk is an appropriate approximation. In practice, when we say we are confining the kinematics to the disk, we ae saying the velocity field is applied only when the LOS is probing the disk, the condition of which is $\rho(t) \leq \rho_d$ and $|z(t)| \leq h_d$.  In order to model lagging halo EPG kinematics, one can adopt the exponential model of \citet{steidel02} which has rotation curve $V_\upphi (\rho,0)$ in the disk plane and a declining rotation velocity above the plane following the exponential scale-height $H_v$ (see Eq.~\ref{eq:vsteidel02}). Alternatively, one can adopt the linear lagging halo model of \citet{bizyaev17}, defined by the parameters $V_\upphi (\rho,0)$, $\rho_{d,0}$, $|dV_\upphi/dz|$, and $dV_\upphi/d\rho$ (see Eqs.~\ref{eq:vrbizaev17} and \ref{eq:vphibizaev17}). Application of either lagging halo model will extend the disk/EPG material above the disk height, i.e., $|z| > h_d$; thus, one relaxes the condition $|z(t)| \leq h_d$ for the LOS. 

A facile model of wind kinematics for a simple cone geometry is constant radial outflow specified by the wind speed parameter $V_w$. We introduced a kinematic model for the wind that captures the hyperbolic nature of the outflowing velocity field. The parameter $V_w$ serves to define the flow rate of the wind for a constant wind speed along the streamlines. Wind material departs the galactic plane vertically and fans out on hyperbolic streamlines, becoming asymptotically radial as galactocentric distance increases. The streamlines are simply defined by the wind geometric structural parameters, $\rho_{w,0}$, $\Theta_w$ (see Eqs.~\ref{eq:thehyperveltotal}, \ref{eq:mywindmodel}, and \ref{eq:upsilon4wind}), so there are no added model parameters compared to pure radial kinematics.  

Spatially variable wind speeds can be modeled using  axial symmetric enhancement functions (see Section~\ref{sec:enhancefuncs} and Appendix~D of \citetalias{churchill25-skamI}). For example, the wind can be fastest in the core and slower on the surface. Enhanced wind kinematics take the general form $Y(x) = Y_1 + Y_2 f\subE(x)$, where $Y_1$ is the zero offset, $Y_2$ is the amplitude of the enhancement, and where the enhancement function is $f\subE(x) \in [0,1]$ with a peak value $f\subE(0)=1$, where $x$ is the spatial domain within the CGM geometric structure.  In the case of enhanced wind velocities, we have $V'_w(\rho,z) = V_w + \tilde{V}\subE f\subE(\rho/\rho_w(z))$. Functions such as those illustrated in Figure~\ref{fig:enhance} can be invoked to emulate theoretical predictions.  

In addition, gradual or rapid deceleration of the wind can be modeled using stall functions. Denoted ${\cal S}(r; \eta_s,\Delta \eta_s)$ and described and illustrated in \citetalias{churchill25-skamI}, they are multiplicative. The full model is written $V'_w(\rho,z) = {\cal S}(z; \eta_s,\Delta \eta_s)[V_w +\tilde{V}\subE f\subE(\rho/\rho_w(z)]$ with the parameter $\eta_s \in [0,1]$ setting the ``stall height,'' $\eta_s R_w$. At this height the outflow velocity has decreased to $\sim\! 50$\% of its launch velocity.  The parameter $\Delta\eta_s$ defines the ``stall ratio,'' $\Delta \eta_s/\eta_s$.   The smaller the stall ratio, the more rapid the ``stall.''  For $\Delta\eta_s/\eta_s \leq 0.2$, the wind velocity has decreased to $\sim\! 75$\% of its launch value at height $(\eta_s \!-\! \Delta\eta_s) R_w $, and it has decreased to $\sim\! 25$\% for $(\eta_s \!+\! \Delta\eta_s) R_r$. Very rapid ``stall outs" can be modeled with small stall ratios, $\Delta\eta_s/\eta_s \leq 0.05$.  Illustrations of selected enhanced wind kinematic models with stalling can be viewed in \citetalias{churchill25-skamI}. 

The accretion kinematics model we proposed is Keplerian infall from infinity.\footnote{In Appendix~E of \citetalias{churchill25-skamI}, we also worked out an alternative infall velocity field  known as the logarithmic spiral, which has been used to model spiral arm patterns \citep{bertin96}.} As written in Eqs.~\ref{eq:vaccretion}, \ref{eq:keplervels}, and \ref{eq:vaccretion-rhomag}, the Keplerian orbits are defined by their eccentricity with $e \geq 1$, the radius of periapsis, which is the accretion radius $\rho_{a,0}$, and the orbital kinetic energy, defined by matching the azimuthal velocity at the accretion radius to the rotation speed of the disk, i.e., $V_c$.  The $z$ component to the velocity field is added to these planar models in order to have the infalling material converge at the accretion radius.  This vertical component is given by $z/\rho$ times the axial component of the velocity field and thus does not require additional free parameters.

\subsubsection{Idealized Winds and Disk/Accretion}

For the wind, we presented a simple model that can account for density and temperature structure. The simplest density profile for $n\subH(\rho,z)$ is a constant, A first-order variation would allow the density to vary with height above the disk plane $n_w(z)$. Using enhancement functions, we can also vary the density as a function of axial distance from the wind core, $\rho=0$, to the wind surface, $\rho_w(z) = [ \rho^2_{w,0}+ z^2 \tan^2 \Theta_w ]^{1/2}$. In terms of an enhancement function, the form of this model is $\log n\subH(\rho,z) = \nu_1  + \nu_2 f\subE(\rho/\rho_w(z)$, where $\nu_1$ is the zero offset density, $\nu_2$ is the amplitude of the axial variation, and $f\subE(\rho/\rho_w(z))$ is symmetric about $\rho=0$. As shown in Eq.~\ref{eq:temperature-wind}, the temperature can be similarly handled, with height dependence $T(z)$, zero-point offset $\tau_1$, and axial variation with amplitude $\tau_2$. These functional forms are well-suited for modeling the core of the wind as a hot diffuse phase and the wind surface as a cooler denser phase, perhaps interpreted as a $\gamma$-ray/X-ray emitting wind like the Galactic Fermi/e-Rosita bubbles with a swept-up entrained ISM gas boundary \citep[e.g.][]{fox15, predehl20, mann23}.

Regarding the disk and the extended planar accretion, we have adopted the view that there should be continuity in the baryon gas phases across the accretion zone, defined as $\rho_{a,0} \leq \rho \leq \rho_d$. The axial extent of this zone is obtained matching the height of the disk, $h_d$, and the height of the accretion structure at the disk radius, $r_d$, yielding $\rho_{a,0} = [\rho_d^2 - h^2_d \cot^2 \Theta_a ]^{1/2}$.

We proposed a density-temperature distribution in the disk that is motivated by observations of {\HI} and intermediate metal lines in the Milky Way and local edge on galaxies. For the extended planar accretion, we assume a vertical distribution of $n\subH(\rho,z) = n\subH(\rho,0) \, {\rm sech}^2 (z/H_a)$ and a declining axial profile with exponential scale length $L_a$ for the vertically integrated hydrogen column density, $N\subH(\rho)$.   The main free parameters are the central density of the disk, $n\subH(0,0)$, the scale height of the accretion, $H_a$, and the vertically integrated hydrogen column density at the edge of the disk, $N\subH(\rho_d,d)$, and the accretion axial scale length, $L_d$. From these four parameters, we determine single smooth functional for the density spanning the plane of both the disk and the accretion, $n\subH(\rho,0)$, the disk axial scale length, $L_d$, and the disk vertical scale height, $H_d$, such that the baryon density distribution at the transition from disk to accretion is continuous and smooth in both the axial and vertical coordinates.  See Figure~\ref{fig:xpadensity} for example $n\subH(\rho,0)$ and $N\subH(\rho,0)$ axial profiles.

For the temperature distribution, we considered a simple function that yields a profile motivated by hot accretion, in which the accretion temperature drops to the disk temperature across the accretion zone. An example is shown in Figure~\ref{fig:xpatemperature} for accretion temperature of $T_a = 10^{5.5}$~K and a disk temperature of $T_d = 10^{4}$~K, as given by Eq.~\ref{eq:fermi-dirac-temp}.  Note that the temperature range of this function can be easily adjusted.

\subsection{Absorption Lines}

Multiple steps are required in order to generate absorption lines from a SKAM.  First, a metallicity and abundance pattern is assumed.  As with the velocity fields and baryonic gas phase distributions, the gas phase abundances can be unique to each galaxy/CGM structure and can vary spatially within them.  For example, in the wind, the core could reflect the yield of Type~II core collapse supernovae, whereas the wind surface might reflect the yield of later stages of stellar evolution or ISM gas-phase dust depletion patterns. Here, we have assumed a 10\% solar metallicity with solar photosphere abundances \citep{lodders19} for all spatial locations in all galaxy/CGM structures. The second step is to estimate the ionization fractions for ions that give rise to the absorption lines one wishes to generate. For example, in order to generate {\MgII}~$\lambda 2786$ and {\CIV}~$\lambda 1548$ absorption lines, we require the ionization fractions for the Mg$^+$ and C$^{+3}$ ions. Third, the optical depth must be determined as a function of LOS velocity, from which the spectrum can be generated assuming pure absorption.

Among all of the simplifications and approximations inherent to SKAMs, the critical ionization corrections are probably the most fraught with inaccuracy. Though ionization codes themselves are highly sophisticated and vetted, what is unknown is where in the galaxy/CGM structures gas is predominately photoionized or collisionally ionized. When photoionized, it is challenging to determine the degree to which the embedded gas is shielded from the background ionizing radiation. One approach is to estimate the solid-angle averaged transmitted spectral energy distribution of the ionizing radiation at LOS location ${\bf r}(t)$. Here, we adopted a simpler estimate and determined the attenuation from the closest surface of the galaxy/CGM structure. This could systematically underestimate the shielding and leading to systematically higher ionization conditions.

For collisional ionization, there is the issue of whether the ionization is in equilibrium. Both these uncertainties may be especially true for LOS locations near the core of the wind, where radiation from young hot may dominate over the cosmic ultraviolet background. Furthermore, the ionization balance from shocked plasmas, conductive interfaces,  turbulent interfaces, and cooling flows are challenging to constrain due to degeneracy in their predicted column density ratios \citep[e.g.,][]{dopita03, wakker12}. 

Some challenges of ionization modeling in SKAMs could be bypassed if, instead of modeling the baryons as a single gas phase at each point in the galaxy frame specified by $n\subH$, the density distribution of the ion of interest, $n_{kj}$, is itself specified. This removes the ionization fraction and the abundance fraction from the estimate of that ion's column density in the calculation of the optical depth in Eq.~\ref{eq:tau-defined}. However, if multiple ions of interest are to be modeled, then each will require that its own unique density distribution is fully specified.  A benefit of this approach is that ion density fields can be directly explored.  The downside is that the ratios of ion densities at any given location may not fall within the ranges known for astrophysical gases, though one can imagine implementing constraints based on theoretical expectations.

\section{A Fiducial SKAM}
\label{sec:examples}

\begin{figure*}[thb]
\centering
\vglue -0.05in
\includegraphics[width=0.65\linewidth]{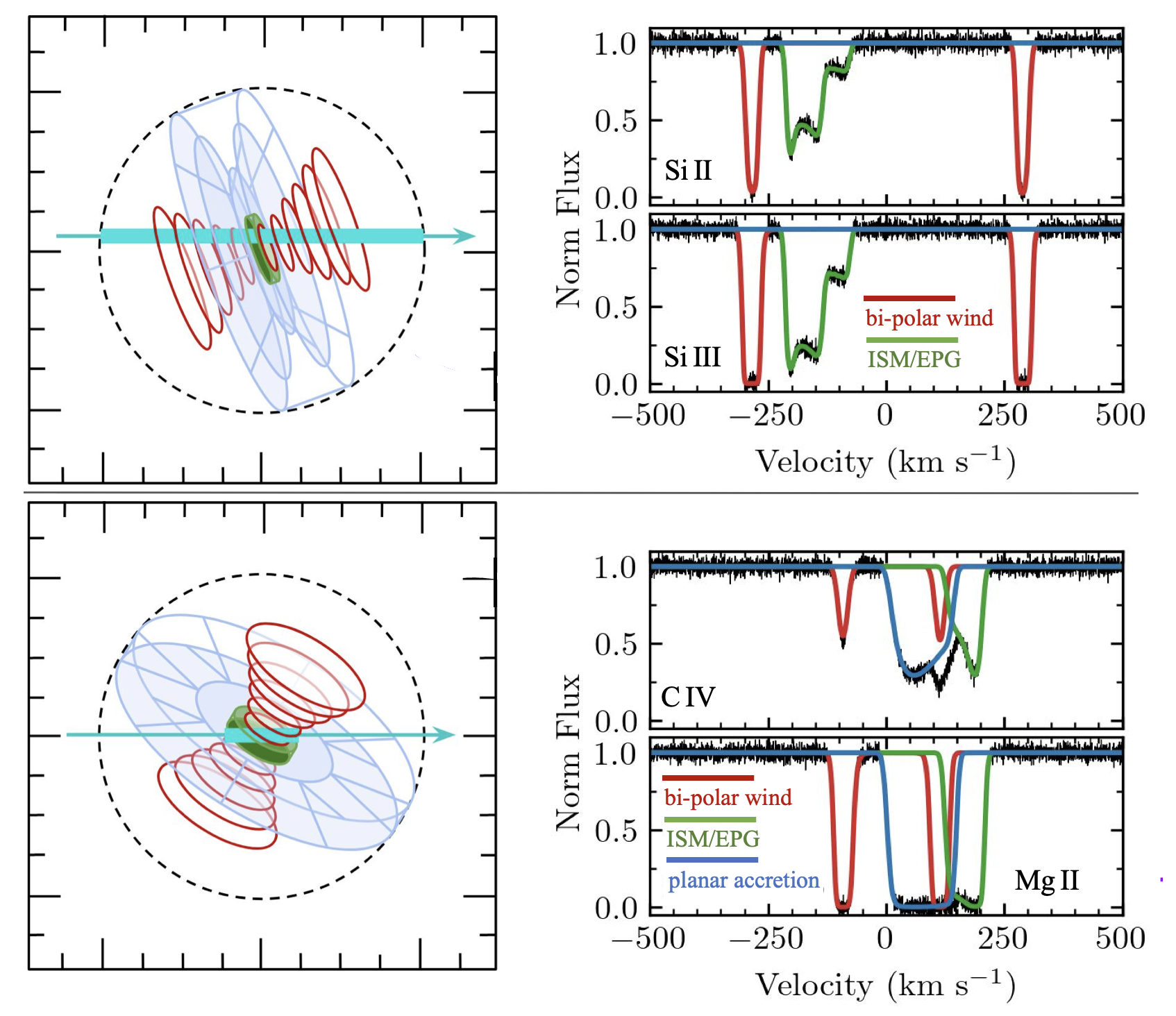}
\caption{\small  A SKAM as seen from two different observers. The free parameters for the spatial-kinematics and gas phases are listed in Table~\ref{tab:galpars}.  Note, only the disk, wind, and accretion structures are populated with gas; the spherical halo is not. See text for additional details and discussion. (left panels) The observer frame side view ($x_oz_o$ plane) of the SKAM, showing the disk (green), wind (red), accretion (blue), and spherical halo (black dashed curve). The LOS is the turquoise horizontal line, which is thicker where the LOS probes the SKAM structures. The observer is toward the left and the quasar toward the right. (right panels) Selected absorption lines color coded according to the galaxy/CGM structure from which they arise. (upper panels) The observer viewing parameters are $R_y(\beta) R_z(\alpha) = R_y(70^\circ) R_z(10^\circ)$, yielding $i = 22^{\circ}$ (nearly face-on orientation). The quasar has sky location $R_\perp = 25$~kpc and $\gamma = 45^{\circ}$, yielding $\Phi = 70^{\circ}$. The LOS passes inside the accretion radius and thus never intercepts the accreting material.  {\SiII}~$\lambda 1260$ and {\SiIII}~$\lambda 1206$ absorption from the disk and wind is shown to the right. (lower panels) The observer viewing parameters are $R_y(\beta) R_z(\alpha) = R_y(20^\circ) R_z(10^\circ)$, yielding $i = 70^{\circ}$ (approaching edge-on orientation). The quasar has sky location $R_\perp = 35$~kpc and $\gamma = 5^{\circ}$, yielding $\Phi = 35^{\circ}$. All three structures are probed but over a much shorter line of sight. {\CIV}~$\lambda 1548$ and {\MgII}~$\lambda 2796$ absorption from the disk, wind, and accretion is shown to the right.}
\label{fig:exampleSKAM}
\vglue 0.1in
\end{figure*}

We present a SKAM characterized by what we consider to be fiducial values of the free parameters describing each structural components and their kinematics and gas phase distributions. The fiducial parameters are listed in Table~\ref{tab:galpars} and the SKAM is illustrated in  Figure~\ref{fig:exampleSKAM}.

For the halo, we assume $V_c=220$~{\kms} for an NFW dark matter profile with a concentration parameter of $\mathcal{C} = 10$ and  mean overdensity $\Delta_c=200$.  These parameters yield a virial radius and mass similar to that of the Milky Way. The peak circular velocity occurs at $R_c = 44$~kpc.  We assume $\eta\subCGM = 1$, which yields $R\subCGM=R_{\rm vir}$.  We do not populate the halo (the region inside $R\subCGM$ that is not occupied by the galaxy disk, wind, and accretion structures) with baryons.  This provides an unimpeded examination of the latter three structures.

\begin{table}[htb]
\centering
\caption{Fiducial SKAM Parameters\label{tab:galpars}}
\begin{tabular}{lll}
\hline\\[-8pt]
Dark Matter/CGM Halo \\ \hline\\[-8pt]
circular velocity & $V_c$& 220~{\kms}\\
halo over density & $\Delta _c$& 200  \\
halo concentration  & ${\cal C}$ & 10 \\
CGM radial extent & $\eta\subCGM$ & 1.0  \\[2pt]
\hline \\[-8pt]
Disk/EPG \\ \hline\\[-8pt]
axial radius & $\rho_d$& $R_c$ \\
disk height & $h_d$ & 5~kpc \\
rotation velocity & $V_\upphi(\rho,z)$ & $V_c$\\
EPG model & \multicolumn{2}{l}{\citet{steidel02}}\\
velocity scale height & $H_{v}$ & 10~kpc \\
disk central density & $n\subH(0,0)$ & $0.03$ cm$^{-3}$\\
temperature, $T(r)$  & $\tau_a, \tau_d$ & 5.5, 4.0 $\log$(K) \\[2pt]
\hline \\[-8pt]
Accretion \\ \hline\\[-8pt]
opening angle & $\Theta_a$ & $25^\circ$ \\
accretion radius & $\rho_{a,0}$& $\sqrt{\rho_d^2\!-\! h_d^2 \cot^2 \Theta_a}$ \\
radial extent & $\eta_a$ & 1.0 \\
eccentricity & $e$ & 1.5  \\
periapsis velocity &  $V_\upphi(\rho_d,z)$ & $V_c$ \\
column density &  $\log N\subH(\rho_d)$ & $20.3$ $\log$(cm$^{-2}$) \\
$n\subH$ scale height & $H_a$ & 10~kpc\\
$N\subH$ scale length & $L_a$ & 5~kpc\\
temperature, $T(r)$  & $\tau_a, \tau_d$ & 5.5, 4.0 $\log$(K) \\[2pt]
\hline \\[-8pt]
Wind \\ \hline\\[-8pt]
opening angle & $\Theta_w$ & $\phantom{+}40^\circ$ \\
base radius & $\rho_{w,0}$& $\phantom{+}20$~kpc  \\
radial extent & $\eta_w$ & $\phantom{+}1.0$  \\
zero-point wind velocity & $V_w$ & $\phantom{+}200$~{\kms} \\
+ enhancement  & $V\subE$ & +150~{\kms} \\
zero-point $n\subH$ & $\nu_1$ & $-1.0$ $\log$(cm$^{-3}$)\\
+ enhancement & $\nu_2$ & $-3.0$ $\log$(cm$^{-3}$) \\
zero-point $T$ & $\tau_1$ & $\phantom{+}4.5$~$\log$(K)  \\
+ enhancement & $\tau_2$ & $+1.5$ $\log$(K) \\[2pt]
\hline \\
\end{tabular}
\end{table}

For the disk structure, we adopt a thin cylinder with height $h_d=5$~kpc and axial radius $\rho_d=R_c$.   For the kinematics, we adopt the ISM/EPG model of \citet{steidel02}, as given by Eq.~\ref{eq:VLOS-diskexp} with $V(\rho,z) = V_c$ and exponential velocity scale height $H_v= 10$~kpc.  Note that the LOS can probe the EPG at heights greater than $h_d$. For the central hydrogen density, we assume $n\subH = 0.03$~cm$^{-3}$. We adopt the temperature profile given by Eq.~\ref{eq:fermi-dirac-temp}, which yields $T \simeq 10^4$~K within the disk (for  $r< \rho_d$). By design, the gas phase distribution of the disk is coupled with that of the accretion.  

For the extended planar accretion component, the opening angle is set to $\Theta_a=25^\circ$. We adopt an accretion radius given by $\rho_{w,0}= [ R_c^2\!-\! h_d^2 \cot^2 \Theta_a ]^{1/2}$, which yields an accretion zone in the range $\rho_{w,0}$ to $\rho_d= R_c$. The radial extent is set to $R_a = R_{\rm vir}$ ($\eta_a = 1$).  For the accretion kinematics, we adopt a hyperbolic Keplerian trajectory with $e=1.5$  and enforce the velocity at the periapsis equal the maximum circular velocity, $V_c$, in the accretion zone.  For the tied disk/accretion gas phase profile, we adopt a vertically integrated column density at the accretion zone of $\log (N\subH(\rho_d)/{\rm cm}^{-2}) = 20.3$, a column density axial scale length of $L_a=5$~kpc, and a density vertical scale height of $H_a = 10$~kpc.  We adopt the temperature profile given by Eq.~\ref{eq:fermi-dirac-temp}, which yields $T_a \simeq 10^{5.5}$~K within the accretion structure (for  $r > \rho_d$) with transition to $T_d\simeq 10^4$~K across the accretion zone and into the disk.

For the wind, the opening angle is ${\Theta_w =40^\circ}$ and its radial extent is ${R_w=R_{\rm vir}}$ (${\eta_w = 1}$). The wind base radius is ${\rho_{w,0} = 20}$~kpc, 45\% of the disk axial radius.  We incorporated an enhanced wind velocity model given by
${V(\rho,z) = V_w + V\subE \cos^2 \{(\pi/2)(\rho/\rho_w(z))\} }$, with $V_w = 200$~{\kms} and $\Delta V\subE = +150$~{\kms}; the wind speed in the core is $V(\rho,0) = V_w + \Delta V_w  = 350$~{\kms} and on the surface is $V(\rho_w(z),z) = V_w = 200$~{\kms}. A stall function was not applied. For the gas phase, we used the enhanced models given by Eq.~\ref{eq:temperature-wind} with enhancement function $f\subE(\rho) = \cos^2 \{(\pi/2)(\rho/\rho_w(z))\}$. The zero-offset density and temperature are $\nu_1 = -1.0$ and $\tau_1 = 4.5$ with enhancement amplitudes $\nu_2 = -3.0$ and $\tau_2= +1.5$. Thus, the gas phase density varies from $n\subH = 10^{-4}$~{\rm cm}$^{-3}$ in the wind core to $n\subH = 10^{-1}$~{\rm cm}$^{-3}$ along the wind surface and the temperature varies from $T = 10^{6}$~{\rm K} in the wind core to $T = 3\times 10^{4}$~{\rm K} along the wind surface. The adopted parameters yield an idealized wind that is faster, lower density, and hotter in the core and slower, more dense, and cooler along the surface ``walls."

Figure~\ref{fig:exampleSKAM}, we present a schematic of two different LOS and orientations of this SKAM from the perspective in which the observer is toward the left and the quasar is toward the right (in the direction of the arrow). 

For the first LOS (the upper panels), the observer perspective is given by the rotations $R_y(\beta) R_z(\alpha) = R_y(70^\circ) R_z(10^\circ)$ with respect to the galaxy frame, which yields $i = 22^{\circ}$. The impact parameter is $R_\perp = 25$~kpc with sky position angle $\gamma = 45^{\circ}$, which yields $\Phi = 70^{\circ}$.  The wind is shown in red, the galaxy disk in green, and the accretion in blue.  In the right hand panel, {\SiII}~$\lambda 1260$ and {\SIII}~$\lambda 1206$ absorption lines are shown. The red profiles arise from the wind and the green profiles arise from the galaxy disk. A signal-to-noise ratio of 25 was added, shown as the black spectra.  The LOS passes inside the accretion radius and thus does not sample the accretion. Note that the wind gives rise to two symmetric absorption lines separated by $\sim 600$~{\kms}, whereas the galaxy disk yields a multi-component blended profile reflecting the disk rotation and EPG lagging halo. Profiles similar to these have been interpreted as supperbubbles \citep[e.g.,][]{bond01-bubbles, ellison03}.

For the second LOS (the lower panels), the observer perspective is given by the rotations $R_y(\beta) R_z(\alpha) = R_y(20^\circ) R_z(10^\circ)$ with respect to the galaxy frame, which yields $i = 70^{\circ}$. The impact parameter is $R_\perp = 35$~kpc with sky position angle $\gamma = 5^{\circ}$, which yields $\Phi = 34^{\circ}$.  In the right hand panel, {\CIV}~$\lambda 1548$ and {\MgII}~$\lambda 2796$ absorption lines are shown. The LOS probes all three structures. The disk is probed such that LOS velocity is effectively $V_c$. The wind gives rise to two symmetric absorption lines, but now separated by $\sim 100$~{\kms} due to the high inclination.  The core is too hot and low density to give rise to {\MgII} and {\CIV} absorption and the wind walls favor {\MgII} absorption over {\CIV}. Note that the accretion gives rise to a saturated {\MgII} absorption region over a broad $150$~{\kms} velocity spread.   In real world data, without insights from a SKAM, one might attribute this strong saturated absorption to a wind \citep[e.g.,][]{bond01, bouche07}. However, the admixture of multiple CGM components from the SKAM is consistent with the interpretation for the origin of ultra-strong {\MgII} absorbers favored by \citet{udhwani25}.

The SKAM shown in Figure~\ref{fig:exampleSKAM} can be considered a fairly run-of-the-mill, or fiducial model of the CGM. From the theoretical point of view, there is nothing built into it that contradicts the geometries, kinematics, nor the gas phases of the CGM
\citep[][]{oppenheimer08, vandevoort11, ford14, sharma14, tepper-garcia15, bustard16, stern16, oppenheimer18, zhang18, grand19,  ho-eagle20, stewart-proc17, nelson18, nelson19, fielding20, peroux20, appleby21, fielding22, hafen22, nguyen22, trapp22, stern23, tan23, kocjan24}. 
Thought the greatest uncertainty in these SKAM is the ionization physics, the resulting absorption strengths and kinematics of the generated {\MgII} and {\CIV} absorption lines are not atypical of those observed for real-world galaxy-absorber pairs 
\citep[][]{csv96, chen01, steidel02, kacprzak10, kacprzak11, kacprzak12, kacprzak21, kacprzak25, bouche12, gauthier12, borthakur13, bordoloi14-dwarfs, nielsen15, nielsen16, ho17, martin19, zabl19, zabl20, schroetter21, kumar24, lopez24, garza25}
This provides some confidence that SKAM models can have utility for interpreting observed absorption lines from the CGM. \\



\section{Conclusions}
\label{sec:conclusion}

We have developed a simple and flexible formalism for modeling absorption lines from the CGM of absorber-galaxy pairs. We call these spatial-kinematic absorption models (SKAMs). Spatial-kinematic only models and SKAMs have been used in several studies and have been instrumental in interpreting observed real-world galaxy-absorber pairs \citep[e.g.,][]{weisheit78, lanzetta92, steidel02, kacprzak10, bouche12, steidel02, kacprzak10, ho17, zabl20, beckett22, beckett23}. 

The SKAMs presented here are designed to be fully adjustable in order to facilitate exploration of how quasar absorption lines depend on a galaxy/CGM model of interest. Spatially, the CGM is modeled as having multiple components, each represented by a geometric solid. The number of CGM components and their representative solid geometries are arbitrary and can be tailored to the motivating science questions in the context of a given study. The kinematics (velocity fields) and the baryon gas phases (density and temperature distributions) are specified using analytical expressions. A unique velocity field and gas phase distribution can be specified for each individual CGM component.   

The background quasar impact parameter and sky position angle is fully adjustable. A coordinate system rotation allows the galaxy/CGM model to be viewed from an arbitrary perspective.  The orientation of the quasar line of sight (LOS) with respect to the orientation of the central galaxy maps uniquely to the pbserved galaxy inclination and quasar-galaxy azimuthal angle (measured relative to the projected major axis). 

Though we have reviewed the formalism of the geometric construction and kinematics of SKAMs, the details and derivations were presented in \citetalias{churchill25-skamI}. In this work, we described how a SKAM is populated with baryons, how ionization conditions can be treated, and how absorption lines are generated. The ionization modeling of the baryons is also entirely flexible.  Either the ion number densities themselves can be directly specified by the baryon gas-phase distribution functions, or ionization models can be applied to gas phase distributions written in terms of the hydrogen number density.  Though there are challenges for implementing realistic ionization modeling, we adopted the latter for the example SKAMs described in this work. 

SKAMs may have an array of possible avenues through which they contribute to a deeper understanding of the CGM from absorption line observations and to an improved understanding of the baryon cycle:

\begin{enumerate}

\item A teaching tool. A SKAM GUI can allow students and entry-level researchers to interactively manipulate SKAM parameters in a controlled fashion and instantly update a visual display. This display could show the observer perspective of the SKAM model on the sky plane, the LOS velocities of each galaxy/CGM component as a function of LOS position, and absorption lines spectra of selected transitions from ions of interest.  Such a tool, even using the most simple SKAMs, would yield instant insight and build intuition for recognizing connections between CGM structure, kinematics, baryon and ionization distributions, and observer viewing perspective. In particular, the behavior of multiphase absorption line systems could be explored.   

\item Understanding individual galaxy-absorber pairs: The quasar-galaxy impact parameter and the LOS-galaxy orientation can be configured to match a given observed galaxy-absorber pair for which a suite of absorption lines have been measured.  The galaxy/CGM spatial, kinematic, and gas phase distributions can be explored to qualitative or semi-quantitatively match the observed absorption profiles. The SKAM parameter space consistent with the data can be estimated, though ruling out regions of SKAM parameter space is just as important and more likely to be done with higher confidence.

\item Understanding group environments. If multiple galaxies are observed to potentially give rise to absorption and their relative LOS velocities, impact parameters, virial masses, and orientations are measured, then a multi-galaxy SKAM (using fiducial parameters for each) could be used to generate absorption for comparison with the observed absorption. Among other insights, unaccounted model absorption could be explained as possible intragroup gas. Additional details might include the individual galaxy star formation rates and adjustment of SKAM CGM wind parameters to  reflect the level of stellar activity in each galaxy. 

\item Statistical tests to constrain SKAM parameter space. Large surveys have documented the distribution of multiple absorption-derived quantities as a function of impact parameter. These include the equivalent width, column density, number of components, kinematic spread, component velocity clustering, galaxy-absorber LOS velocity difference, and covering fraction. For a given fixed set of SKAM parameters, one could construct a large dataset of galaxy-absorber pairs drawn from an unbiased distributions of impact parameters, inclinations, and quasar-galaxy azimuthal angles, and then directly compared SKAM generated absorption line quantities to the documented observed distributions. Drawing from a distribution of virial masses may be an important design element. 

\item A forward modeling tool. The SKAM parameter space could be wrapped in a Bayesian maximum-likelihood or a Markov Chain Monte Carlo formalism in which the likelihood function is the pixel-by-pixel $\chi^2$ or absolute residual between the SKAM generated absorption spectra and the observed spectra.  This may be possible only in cases of exquisite high-resolution spectra (as expected in the next generation of astronomical facilities).  In such cases, one could simply constrain the number densities of the absorbing ions directly and avoid complicated ionization calculations (unless metallicity and ionization are desire posterior quantities).    

\end{enumerate}

\subsection{A Budding GUI}
\label{sec:theGUI}

The scope of this work has been to develop and present a single construct for modeling absorption lines from the galaxy/CGM that can simultaneously and flexibly incorporate multiple spatial-kinematic components to the CGM.  A design element of SKAM as a tool is that any CGM component that can be envisioned can be implemented in the model.  Essentially, if one can imagine and formulate a given spatial structure, velocity field, and baryon gas phase distribution, they can build a SKAM. Although these idealized spatial-kinematic absorption models lack the higher degree of realism afforded by computational methods such as hydrodynamical simulations, they provide a valuable tool with which we can directly confront the distilled essence of our ideas about the nature of the CGM. Here, we have developed and presented what we are calling a fiducial SKAM and demonstrated how these selected idealized CGM components can be applied to generate reasonably realistic absorption lines. 

The SKAMs we have presented were built in a GUI that allows the manipulation of free parameters using a combination of control files, mouse/pad clicks, and key presses.  When the program is launched, a control file with SKAM structural parameters, ions of interest, quasar sky position, viewing orientation is read in and the specified SKAM and absorption lines are displayed. Various dashboards can be toggled, including (1) the multiple viewing perspectives of the SKAM in both the observer and galaxy frames, (2) the LOS velocities of the individual CGM components and the column densities, temperatures, and ionization fractions as a function of either LOS velocity or LOS position, and (3) the absorption line profiles.  The user can now interactively rotate the SKAM to change its orientation, adjust the position of the background quasar, and modify any of the structural, kinematic, or gas-phase parameters. 

\begin{figure*}[bht]
\centering
\vglue -0.1in
\includegraphics[width=0.95\linewidth]{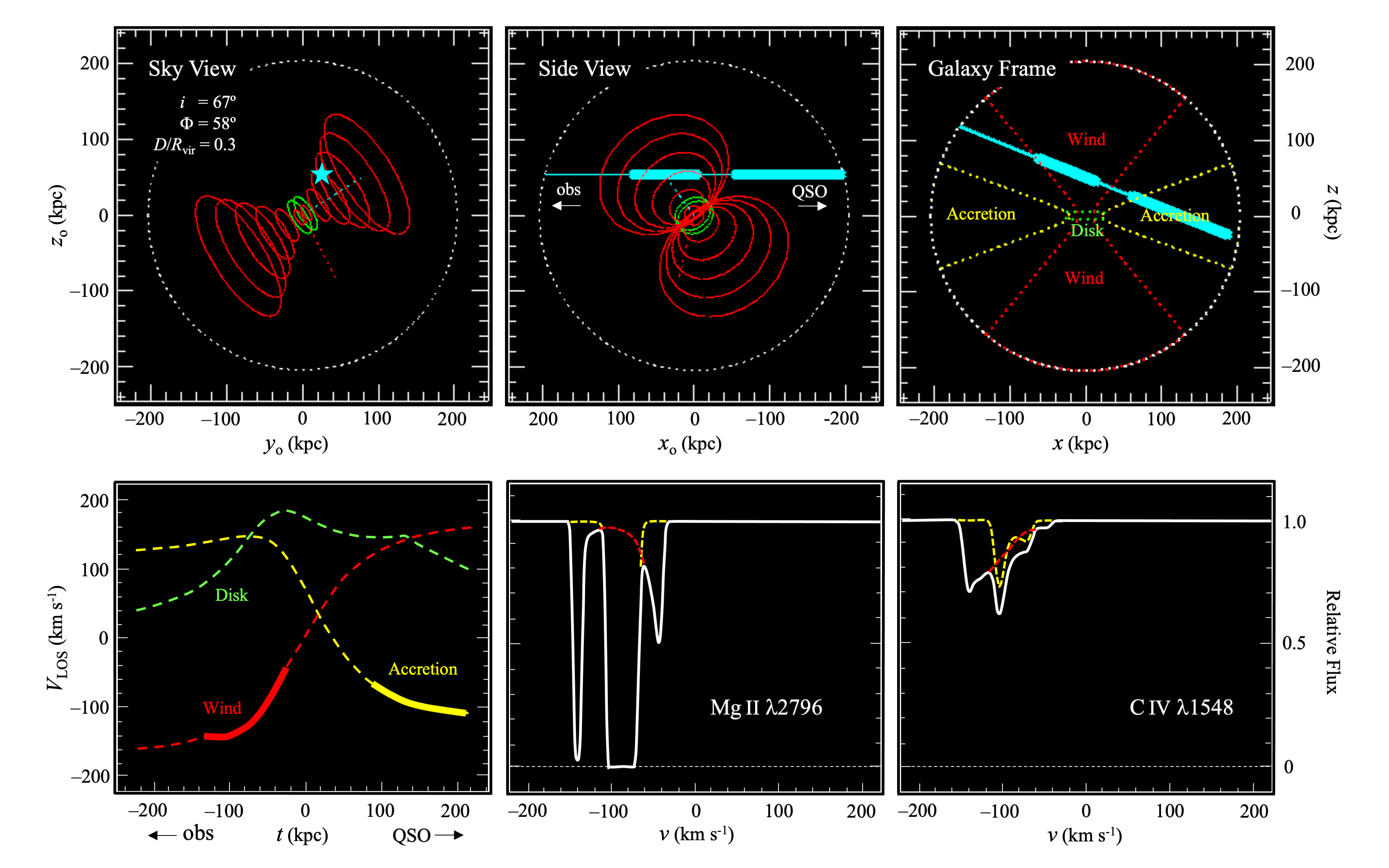}
\caption{\small  Mosaic of screenshots (with added annotations) of the dashboard from the author's GUI-based code.  
The observer perspective is given by the rotation $R_y(\beta) R_z(\alpha) = R_y(60^\circ) R_z(65^\circ)$ with respect to the galaxy frame, resulting in an observed inclination of $i = 67^\circ$.  The galaxy-quasar impact parameter is $R_\perp = 60$~kpc, which is $D/R_{\rm vir} = 0.3$ for the modeled dark matter halo. The quasar position angle is $\gamma = 65^\circ$, which yields $\Phi = 58^\circ$ from the observer's perspective. (upper left) The $y_oz_o$ plane. i.e., the observer frame  sky projection showing the disk (green), wind (red), and virial radius (dashed circle). The display of the accretion structure was toggled off for clarity of presentation. The quasar is represented by the turquoise star. (upper center) The $x_oz_o$ plane. i.e., the side view. The observer is toward the left ($x_o=\infty$) and the quasar toward the right ($x_o=-\infty$). Recall that ${\shat} = - {\ihato}$. Thick portions of the LOS show the LOS positions where the wind and accretion structures are probed. (upper right) The SKAM cross section on the $xz$ plane in the galaxy frame. The projection of the LOS is, including the thick regions where it probes the structures. (lower left) LOS velocity $V\subLOS$ as a function of LOS position $t$ showing the projected velocity fields for the disk, wind, and accretion. Dashed curves are ``ghost" projections of the projected velocities fields beyond the geometric solids and thick curves are where the LOS probes the wind and accretion. (lower center, right) {\MgII~$\lambda 2796$} and {\CIV}~$\lambda 1548$ absorption profiles for the assumed gas phases and velocity fields.} 
\vglue 0.1in
\label{fig:dashboard0}
\end{figure*}

The code, available at \href{https://github.com/CGM-World}{github.com/CGM-World}, is suitable for the writing of this paper.  It has not yet been refined to be a fully-fledged tool for teaching or research. It is written in Fortran~95 and uses the interactive Lick Mongo plotting package, which is a Fortran~77 object library embedded in the XVISTA reduction software package.\footnote{XVISTA is currently maintained by Jon Holtzman at New Mexico State University and is available at his \href{https://github.com/holtzmanjon/xvista/tree/master/lickmongo}{github.com page} or his \href{http://astro.nmsu.edu/~holtz/xvista}{university homepage}. }

A mosaic of screenshots of the GUI dashboard is shown in Figure~\ref{fig:dashboard0}.  The illustrated galaxy/CGM model uses the fiducial SKAM parameters presented in Table~\ref{tab:galpars}. In the upper panels, the virial radius is the white dashed circle, the disk is shown as a green cylinder, the wind as a red hyperboloid, and the accretion as a yellow hyperboloid.  The LOS is shown as a turquoise line and the background quasar is a turquoise star. In the ``Galaxy Frame" panel, the $xz$ plane of the SKAM is shown. The regions where the projected LOS probes the CGM wind and accretion structures are shown as the thick line portions (the disk is not probed). 

In the ``Sky View" panel, the observer's sky view (the $y_oz_o$ plane) is shown for rotation $R_y(\beta) R_z(\alpha) = R_y(60^\circ) R_z(65^\circ)$ with respect to the galaxy frame.  This rotation yields an observed inclination of $i = 67^\circ$ (a highly probable inclination).  The galaxy-quasar impact parameter is $R_\perp = 60$~kpc ($D/R_{\rm vir} = 0.3$) and the quasar position angle is $\gamma = 65^\circ$, which yields $\Phi = 58^\circ$. The quasar is probing the portion of the wind that is directed toward the observer. Note that the accretion structure is not drawn in for clarity.  In the ``Side View" panel, the observer is shown the view perpendicular to the sky plane, i.e., viewing the $x_oz_o$ plane. The upper portion of the wind is directed out of the plane and the lower portion is directed into the plane.

In the lower left panel of Figure~\ref{fig:dashboard0}, we show the LOS velocities as a function of LOS location.  Thick solid curves indicate where the LOS is probing each structure. The velocities of the absorption lines arising from each of these structures will reside in these velocity ranges, roughly $V\subLOS \in (-150,-40 )$~{\kms} for the wind and $V\subLOS \in (-110,-80)$~{\kms} for the accretion. However, the column densities of the various ions in the gas phases are not constant along the LOS, so the absorption kinematics may not span the full range of sampled velocities and may show component structure.  

In the lower center and right panels of Figure~\ref{fig:dashboard0}, we show the absorption profiles for  {\MgII}~$\lambda 2796$ and {\CIV}~$\lambda 1548$ transitions.  Due to the increasing density and declining temperature from the wind core to the wind walls, note that low-ionization {\MgII} absorption arises from LOS positions $t \simeq -120$~kpc and $t \simeq -20$~kpc corresponding to $V\subLOS \simeq -140$~{\kms} and $V\subLOS \simeq -40$~{\kms}, respectively. On the other hand, the absorption kinematics for the intermediate ion {\CIV} arising in the wind reveals the axial velocity and gas-phase gradients near the hotter, lower density wind core as the LOS obliquely crosses the wind axis. Similar mental mappings of accretion absorption can be inferred.  

A power of SKAM is its quick and easy parameter adjustments. It can be illuminating to explore the effects of changing the observer orientation, the location of the background QSO, or the velocity fields and/or baryon gas-phase distributions and experiencing how the absorption profiles adjust in response. 

\subsection{Considerations for Further Development}
\label{sec:multiphase}

As many uses as SKAMs may find, they do not represent physically self-consistent models of the CGM like those obtained using hydrodynamic simulations, such as processes that govern cloud growth, survival, and destruction via dynamic and global thermal instabilities and turbulent interfaces\footnote{We note that gas turbulence can also be modeled by introducing a turbulent velocity field, perhaps along the surfaces of the CGM structures.} \citep[e.g.,][]{maller04, faucher-giguere23, kakoly25}.  The spatial, kinematic, and baryon gas phase distributions are not based on fully self-consistent physical constraints such as the gravitational potential, hydrodynamics, stellar feedback, and energy and angular momentum relationships. Similarly, the gas-phases ``painted" into the idealized spatial-kinematic CGM components do not account for shocks or complex gas-phase mixing, nor are they constrained by mass flux conservation, dynamical instabilities, heating or cooling processes, nor the total baryonic gas mass budget as constrained by the cosmic baryon fraction, $f_b= \Omega_b/\Omega_m$.  

The gas ionization conditions are also idealized in that photoionization and collisional ionization equilibrium is assumed, though non-equilibrium collisional models can be incorporated with little effort. Furthermore, the modeling of the spaces not occupied by the wind and accretion structures, i.e., in the spherical halo, remain challenging as this region does not fit into any single idealized ``component" of the CGM, unless a simple static halos is assumed, or pure radial infall, for example. \cite{ho-eagle20}, note that, while simulations indicate low-ionization co-rotating gas extends to $R_\perp \geq R_{\rm vir}/2$, gas that resides at galactocentric distances beyond $R_{\rm vir}$ can mitigate the absorption line signatures of co-rotation for $R_\perp \geq R_{\rm vir}/4$.  This speaks to embedding the model in a broader IGM for additional realism.

Further development of the SKAM GUI would be fruitful as it has the potential not only as a teaching tool, but also as a scientific tool.  A key area for future development would be to incorporate multiphase gas.  We turn to that topic now.


The gas phases described in this work are simple analytical expression designed to emulate realistic astrophysical distributions; however, while these expressions are highly flexible and intuitive, they are essentially painted-in ``toy models" that are not physically coupled by an equation of state or governed by kinetic or thermal equilibrium principles.  The advantage of simple analytic expressions is that they are easily formulated and implemented, allowing expeditious exploration of various idealized density and temperature distributions. After all, SKAMs are, by design and spirit, tools for exploring our basic broad brush ideals about the real world. However, the incorporated baryon gas phases should be constrained by physical principles. For example, \citet{oppenheimer25} develop physics-based halo models suitable for mock X-ray, radio, and ultraviolet observations that assume a spherically symmetric distribution filled with a single-phase warm/hot gas \citep[also see][]{mo96}. 

For the SKAMs presented here, we have adopted single-phase models in which the gas phase can spatially vary in a smooth manner.  For a given galaxy/CGM structure, for example the wind, the gas can transition from a warm/hot diffuse phase to a cooler denser phase from one localized spatial location to another across the structure. The transition can be gradual or on a small spatial scale, but multiple phases do not spatially overlap. It is not entirely clear to what degree this type of smooth spatial model of the gas phases is consistent with the physical picture of the CGM inferred from recent developments in the Bayesian forward modeling of kinematically complex multi-ion absorption systems as multiphase multi-cloud gas complexes  \citep[see][]{sameer21, sameer22, sameer24}. Such an investigation would be an additional possible application of SKAMs. 

What is clear, however, is that the fiducial SKAM model presented here omits the fact that cool, higher density clumps, perhaps ``mist," may be embedded in and distributed throughout the warm/hot phase \citep[][]{hummels19, gronke20-mist, liang20}, or entrained in winds \citep[e.g.,][]{gronke20, sparre20, antipov25, ghosh25}. 

The belief that metal-line absorbers comprise multiple small mass ``clouds" has been prevalent for decades \citep[e.g.,][]{rees77, cowie77, begelman90}. First generation SKAMs, such as the spherical halo and galaxy disk models of \citet{charlton96} and \citet{charlton98}, employed Monte Carlo methods to populate the geometric structures with kiloparsec and sub-kiloparsec sized clouds having simple radial infall and rotation kinematics; these models were quite successful at matching the observed covering fractions and absorption profile morphologies of {\MgII} absorbers. Using lensed quasars, \citet{rauch02} determined upper limits on {\MgII} cloud sizes of 200~pc at $z \sim 3$ and  \citet{rigby02} argued that $z<1.4$ weak {\MgII} absorber sizes are on the order of 10~pc. Multi-component multiphase ionization modeling also yields cloud sizes as small as $\sim10$~pc for low-ionization absorption lines \citep{sameer21, sameer24}. Other studies have shown that {\CIV} absorbers are also expected to arise from small clouds (in subclumps or complexes of clumps), with sizes in the range of 300--1000 pc \citep[][]{rauch01, schaye07, sameer24}.

Various pressure-equalized two-phase spherical models that account for a hot virialized phase and cool cloud condensation and destruction balance have been developed \citep[e.g.,][]{maller04, faerman17, faerman20, faerman23}; we have drawn inspiration from these for designing our fiducial SKAM.  These halo models provide physics-based expressions for the cool-phase cloud masses, radii, lifetimes, velocities, and in some cases, predict the average number of clouds intercepted and their total integrated column density per LOS as a function of impact parameter. Inspired by these models, \cite{dutta24} described the multiphase CGM as a series of 2D density-temperature log-normal distibution functions. Though their model ``provides a simple analytical framework that is useful for describing important aspects of the multiphase CGM,'' the authors conceded that ``interpreting cold/warm phase diagnostics is not straightforward since these phases are patchy, with inherent variability in intercepting these clouds along arbitrary LOS.''

Guided by simulations and two-phase halo models, other approaches directly model and insert cool clouds and/or cloud complexes directly into a hot halo \citep[e.g.,][]{hummels24, bisht25, yang25}. The adopted radial profile of the number and spatial number density of cool clouds in the halo, the distribution of their absorption properties (densities, radii, masses, and column densities), and their velocity dispersion in the CGM are all constrained by or tested against observations (such as covering fractions, equivalent width radial profiles, etc.).

\citet{hummels24} built an open-source tool called {\sc cloudflex} that is suitable for generating mock absorption lines of the small-scale cool gas structures embedded in halos.  Free parameters include the power-law index for the distribution of cloud masses and for the minimum cloud mass, cloud size, volume filling factor, and the velocity distribution of cloud complexes. \citet{bisht25} and \citet{yang25} also developed halo models in which a clumpy cool CGM component is built of small clouds.  \citet{bisht25} focused on properties of cloud complexes comprising a ``mist of tiny cool cloudlets" dispersed in the warm/hot halo, whereas \citet{yang25} modeled the radial distribution of the cool gas density probability distribution using a classic $\beta$-model of \citet{cavaliere76}.  In all these models, clouds are distributed throughout the halo using Monte-Carlo methods and mock absorption lines are generated. The methods have demonstrated that most all cloud distribution parameters can be meaningfully constrained by statistical comparisons with observations.  

Using the hydrodynamic code AthenaPK, \citet{hidalgo-pineda25} studied the formation and survivability of cool clumps in high-velocity laminar outflowing winds.  They measured a power-law clump mass spectrum ($dN/dm \propto m^{-2}$) and concluded that turbulent mixing and radiative condensation is a key attribute of multiphase outflows. Although the volume filling factor of the cool clumps in the hot wind is very low ($\sim 10^{-3}$) the covering fraction is near unity, facts that, in combination with the ``plume" and ``shell" morphologies of the clumps, lead the authors to point out that this explains the ``misty" appearance observed in real-world outflows. 

Using the hydrodynamic code Enzo, the FOGGIE collaboration \cite[see][]{augustin25} resolved $T\sim 10^4$~K clumps down sizes of $\sim 250$~pc and found the clump mass distribution peaks at $m \sim 10^5$~M$_\odot$. This mass is consistent with the models of \citet{maller04}. The kinematics of these clumps tend be infalling and have slightly lower metallicities than the hot diffuse halo in which they are moving.   Interestingly, using FIRE cosmological simulations, \citet{kakoly25} found that, in lower-mass halos ($M_h < 10^{12}$~M$_\odot$), low-ionization mock absorption lines are consistent with arising from log-normal density distributions of a turbulent volume-filling phase rather than from small distributed cool ``clouds" embedded in a hot medium. This would support the modeling approach of \citet{dutta24}, at least for lower-mass halos. \citet{kakoly25} also concluded that the inner CGM of lower-mass galaxies are ``qualitatively distinct" from the halos of more massive galaxies, which may suggest that models of multiphase CGM may need to adopt different parameterized distribution functions  for lower-mass and higher-mass halos.

The upshot in the context of SKAMs, is that {\sc cloudflex} and other multiphase models that constrain cool cloud/clump properties can be important informants for SKAMs that would incorporate clumpy multiphase gas phase distributions; elements of {\sc cloudflex} and SKAM could be combined. Guided by multiphase models and simulations, Monte-Carlo insertion of cool clouds in the spherical halo, for example, could be implemented, though it may be computationally expensive to generate the mock absorption spectra.  An efficient method for scanning which clumps are intersected by the LOS at a given LOS position would need to be developed. Multiphase modeling represents one future direction for the continued development of SKAMs.

\section*{Acknowledgments}

Much gratitutude to Jane Charlton, Stephanie Ho, Glenn Kacprzak, Hasti Nateghi, Dylan Nelson, Nikki Nielsen, and Sameer for valuable comments and/or discussions that helped focus this project.  Gratitude to Nikki Nielsen for Figure~\ref{fig:exampleSKAM}. The author would be interested to consult and collaborate with individuals interested in the further development of the SKAM GUI, in particular the creation of a portable Python code or online web based GUI.


\bibliography{main}{}

\begin{thebibliography}{}
\expandafter\ifx\csname natexlab\endcsname\relax\def\natexlab#1{#1}\fi
\providecommand{\url}[1]{\href{#1}{#1}}
\providecommand{\dodoi}[1]{doi:~\href{http://doi.org/#1}{\nolinkurl{#1}}}
\providecommand{\doeprint}[1]{\href{http://ascl.net/#1}{\nolinkurl{http://ascl.net/#1}}}
\providecommand{\doarXiv}[1]{\href{https://arxiv.org/abs/#1}{\nolinkurl{https://arxiv.org/abs/#1}}}

\bibitem[{{Antipov} {et~al.}(2025){Antipov}, {Banda-Barrag{\'a}n}, {Birnboim}, {Federrath}, {Gnat}, \& {Br{\"u}ggen}}]{antipov25}
{Antipov}, A., {Banda-Barrag{\'a}n}, W.~E., {Birnboim}, Y., {et~al.} 2025, \mnras, 540, 3798, \dodoi{10.1093/mnras/staf949}

\bibitem[{{Appleby} {et~al.}(2021){Appleby}, {Dav{\'e}}, {Sorini}, {Storey-Fisher}, \& {Smith}}]{appleby21}
{Appleby}, S., {Dav{\'e}}, R., {Sorini}, D., {Storey-Fisher}, K., \& {Smith}, B. 2021, \mnras, 507, 2383, \dodoi{10.1093/mnras/stab2310}

\bibitem[{{Augustin} {et~al.}(2025){Augustin}, {Tumlinson}, {Peeples}, {O'Shea}, {Smith}, {Lochhaas}, {Wright}, {Acharyya}, {Werk}, {Lehner}, {Howk}, {Corlies}, {Simons}, \& {O'Meara}}]{augustin25}
{Augustin}, R., {Tumlinson}, J., {Peeples}, M.~S., {et~al.} 2025, arXiv e-prints, arXiv:2501.06551, \dodoi{10.48550/arXiv.2501.06551}

\bibitem[{{Beckett} {et~al.}(2022){Beckett}, {Morris}, {Fumagalli}, {Tejos}, {Jannuzi}, \& {Cantalupo}}]{beckett22}
{Beckett}, A., {Morris}, S.~L., {Fumagalli}, M., {et~al.} 2022, \mnras, 517, 1020, \dodoi{10.1093/mnras/stac2630}

\bibitem[{{Beckett} {et~al.}(2023){Beckett}, {Morris}, {Fumagalli}, {Tejos}, {Jannuzi}, \& {Cantalupo}}]{beckett23}
---. 2023, \mnras, 521, 1113, \dodoi{10.1093/mnras/stad596}

\bibitem[{{Begelman} \& {McKee}(1990)}]{begelman90}
{Begelman}, M.~C., \& {McKee}, C.~F. 1990, \apj, 358, 375, \dodoi{10.1086/168994}

\bibitem[{{Berkhuijsen} {et~al.}(2006){Berkhuijsen}, {Mitra}, \& {Mueller}}]{berkhuijsen06}
{Berkhuijsen}, E.~M., {Mitra}, D., \& {Mueller}, P. 2006, Astronomische Nachrichten, 327, 82, \dodoi{10.1002/asna.200510488}

\bibitem[{{Bertin} \& {Lin}(1996)}]{bertin96}
{Bertin}, G., \& {Lin}, C.~C. 1996, {Spiral Structure in Galaxies a Density Wave Theory} (Cambridge, MA: MIT Press)

\bibitem[{{Bigiel} \& {Blitz}(2012)}]{bigiel12}
{Bigiel}, F., \& {Blitz}, L. 2012, \apj, 756, 183, \dodoi{10.1088/0004-637X/756/2/183}

\bibitem[{{Bisht} {et~al.}(2025){Bisht}, {Sharma}, {Dutta}, \& {Nath}}]{bisht25}
{Bisht}, M.~S., {Sharma}, P., {Dutta}, A., \& {Nath}, B.~B. 2025, \mnras, 542, 1573, \dodoi{10.1093/mnras/staf1319}

\bibitem[{{Bizyaev} {et~al.}(2017){Bizyaev}, {Walterbos}, {Yoachim}, {Riffel}, {Fern{\'a}ndez-Trincado}, {Pan}, {Diamond-Stanic}, {Jones}, {Thomas}, {Cleary}, \& {Brinkmann}}]{bizyaev17}
{Bizyaev}, D., {Walterbos}, R.~A.~M., {Yoachim}, P., {et~al.} 2017, \apj, 839, 87, \dodoi{10.3847/1538-4357/aa6979}

\bibitem[{{Bond} {et~al.}(2001{\natexlab{a}}){Bond}, {Churchill}, {Charlton}, \& {Vogt}}]{bond01-bubbles}
{Bond}, N.~A., {Churchill}, C.~W., {Charlton}, J.~C., \& {Vogt}, S.~S. 2001{\natexlab{a}}, \apj, 557, 761, \dodoi{10.1086/321689}

\bibitem[{{Bond} {et~al.}(2001{\natexlab{b}}){Bond}, {Churchill}, {Charlton}, \& {Vogt}}]{bond01}
---. 2001{\natexlab{b}}, \apj, 562, 641, \dodoi{10.1086/323876}

\bibitem[{{Bordoloi} {et~al.}(2014{\natexlab{a}}){Bordoloi}, {Lilly}, {Kacprzak}, \& {Churchill}}]{bordoloi14}
{Bordoloi}, R., {Lilly}, S.~J., {Kacprzak}, G.~G., \& {Churchill}, C.~W. 2014{\natexlab{a}}, \apj, 784, 108, \dodoi{10.1088/0004-637X/784/2/108}

\bibitem[{{Bordoloi} {et~al.}(2014{\natexlab{b}}){Bordoloi}, {Tumlinson}, {Werk}, {Oppenheimer}, {Peeples}, {Prochaska}, {Tripp}, {Katz}, {Dav{\'e}}, {Fox}, {Thom}, {Ford}, {Weinberg}, {Burchett}, \& {Kollmeier}}]{bordoloi14-dwarfs}
{Bordoloi}, R., {Tumlinson}, J., {Werk}, J.~K., {et~al.} 2014{\natexlab{b}}, \apj, 796, 136, \dodoi{10.1088/0004-637X/796/2/136}

\bibitem[{{Borthakur} {et~al.}(2013){Borthakur}, {Heckman}, {Strickland}, {Wild}, \& {Schiminovich}}]{borthakur13}
{Borthakur}, S., {Heckman}, T., {Strickland}, D., {Wild}, V., \& {Schiminovich}, D. 2013, \apj, 768, 18, \dodoi{10.1088/0004-637X/768/1/18}

\bibitem[{{Bouch{\'e}} {et~al.}(2012){Bouch{\'e}}, {Hohensee}, {Vargas}, {Kacprzak}, {Martin}, {Cooke}, \& {Churchill}}]{bouche12}
{Bouch{\'e}}, N., {Hohensee}, W., {Vargas}, R., {et~al.} 2012, \mnras, 426, 801, \dodoi{10.1111/j.1365-2966.2012.21114.x}

\bibitem[{{Bouch{\'e}} {et~al.}(2007){Bouch{\'e}}, {Murphy}, {P{\'e}roux}, {Csabai}, \& {Wild}}]{bouche07}
{Bouch{\'e}}, N., {Murphy}, M.~T., {P{\'e}roux}, C., {Csabai}, I., \& {Wild}, V. 2007, \nar, 51, 131, \dodoi{10.1016/j.newar.2006.11.050}

\bibitem[{{Brook} {et~al.}(2008){Brook}, {Governato}, {Quinn}, {Wadsley}, {Brooks}, {Willman}, {Stilp}, \& {Jonsson}}]{brook08}
{Brook}, C.~B., {Governato}, F., {Quinn}, T., {et~al.} 2008, \apj, 689, 678, \dodoi{10.1086/591489}

\bibitem[{{Brooks} {et~al.}(2009){Brooks}, {Governato}, {Quinn}, {Brook}, \& {Wadsley}}]{brooks09}
{Brooks}, A.~M., {Governato}, F., {Quinn}, T., {Brook}, C.~B., \& {Wadsley}, J. 2009, \apj, 694, 396, \dodoi{10.1088/0004-637X/694/1/396}

\bibitem[{{Bustard} {et~al.}(2016){Bustard}, {Zweibel}, \& {D'Onghia}}]{bustard16}
{Bustard}, C., {Zweibel}, E.~G., \& {D'Onghia}, E. 2016, \apj, 819, 29, \dodoi{10.3847/0004-637X/819/1/29}

\bibitem[{{Carr} \& {Scarlata}(2022)}]{carr22}
{Carr}, C., \& {Scarlata}, C. 2022, \apj, 939, 47, \dodoi{10.3847/1538-4357/ac93fa}

\bibitem[{{Casavecchia} {et~al.}(2023){Casavecchia}, {Banda-Barrag{\'a}n}, {Br{\"u}ggen}, \& {Brighenti}}]{casavecchia23}
{Casavecchia}, B., {Banda-Barrag{\'a}n}, W.~E., {Br{\"u}ggen}, M., \& {Brighenti}, F. 2023, IAU Symposium, 362, 56, \dodoi{10.1017/S1743921322001211}

\bibitem[{{Cavaliere} \& {Fusco-Femiano}(1976)}]{cavaliere76}
{Cavaliere}, A., \& {Fusco-Femiano}, R. 1976, \aap, 49, 137

\bibitem[{{Ceverino} \& {Klypin}(2009)}]{ceverino09}
{Ceverino}, D., \& {Klypin}, A. 2009, \apj, 695, 292, \dodoi{10.1088/0004-637X/695/1/292}

\bibitem[{{Charlton} \& {Churchill}(1996)}]{charlton96}
{Charlton}, J.~C., \& {Churchill}, C.~W. 1996, \apj, 465, 631, \dodoi{10.1086/177448}

\bibitem[{{Charlton} \& {Churchill}(1998)}]{charlton98}
---. 1998, \apj, 499, 181, \dodoi{10.1086/305632}

\bibitem[{{Chen} {et~al.}(2014){Chen}, {Gauthier}, {Sharon}, {Johnson}, {Nair}, \& {Liang}}]{chen14}
{Chen}, H.-W., {Gauthier}, J.-R., {Sharon}, K., {et~al.} 2014, \mnras, 438, 1435, \dodoi{10.1093/mnras/stt2288}

\bibitem[{{Chen} {et~al.}(2001){Chen}, {Lanzetta}, \& {Webb}}]{chen01}
{Chen}, H.-W., {Lanzetta}, K.~M., \& {Webb}, J.~K. 2001, \apj, 556, 158, \dodoi{10.1086/321537}

\bibitem[{{Chevalier} \& {Clegg}(1985)}]{chevalier85}
{Chevalier}, R.~A., \& {Clegg}, A.~W. 1985, \nat, 317, 44, \dodoi{10.1038/317044a0}

\bibitem[{{Churchill}(2025{\natexlab{a}})}]{churchill25-skamI}
{Churchill}, C.~W. 2025{\natexlab{a}}, Paper I, concurrently submitted to OAJ

\bibitem[{{Churchill}(2025{\natexlab{b}})}]{churchill25-book1}
---. 2025{\natexlab{b}}, {Quasar Absorption Lines. Volume 1. Introduction, Discoveries, and Methods} (Cambridge: Cambridge university Press)

\bibitem[{{Churchill} {et~al.}(1996){Churchill}, {Steidel}, \& {Vogt}}]{csv96}
{Churchill}, C.~W., {Steidel}, C.~C., \& {Vogt}, S.~S. 1996, \apj, 471, 164, \dodoi{10.1086/177960}

\bibitem[{{Churchill} {et~al.}(2015){Churchill}, {Vander Vliet}, {Trujillo-Gomez}, {Kacprzak}, \& {Klypin}}]{churchill15-direct}
{Churchill}, C.~W., {Vander Vliet}, J.~R., {Trujillo-Gomez}, S., {Kacprzak}, G.~G., \& {Klypin}, A. 2015, \apj, 802, 10, \dodoi{10.1088/0004-637X/802/1/10}

\bibitem[{{Comer{\'o}n} {et~al.}(2012){Comer{\'o}n}, {Elmegreen}, {Salo}, {Laurikainen}, {Athanassoula}, {Bosma}, {Knapen}, {Gadotti}, {Sheth}, {Hinz}, {Regan}, {Gil de Paz}, {Mu{\~n}oz-Mateos}, {Men{\'e}ndez-Delmestre}, {Seibert}, {Kim}, {Mizusawa}, {Laine}, {Ho}, \& {Holwerda}}]{comeron12}
{Comer{\'o}n}, S., {Elmegreen}, B.~G., {Salo}, H., {et~al.} 2012, \apj, 759, 98, \dodoi{10.1088/0004-637X/759/2/98}

\bibitem[{{Cowie} \& {McKee}(1977)}]{cowie77}
{Cowie}, L.~L., \& {McKee}, C.~F. 1977, \apj, 211, 135, \dodoi{10.1086/154911}

\bibitem[{{Danovich} {et~al.}(2015){Danovich}, {Dekel}, {Hahn}, {Ceverino}, \& {Primack}}]{danovich15}
{Danovich}, M., {Dekel}, A., {Hahn}, O., {Ceverino}, D., \& {Primack}, J. 2015, \mnras, 449, 2087, \dodoi{10.1093/mnras/stv270}

\bibitem[{{Dav{\'e}} {et~al.}(2011){Dav{\'e}}, {Oppenheimer}, \& {Finlator}}]{dave11}
{Dav{\'e}}, R., {Oppenheimer}, B.~D., \& {Finlator}, K. 2011, \mnras, 415, 11, \dodoi{10.1111/j.1365-2966.2011.18680.x}

\bibitem[{{de Grijs} \& {Peletier}(1997)}]{deGrijs97}
{de Grijs}, R., \& {Peletier}, R.~F. 1997, \aap, 320, L21, \dodoi{10.48550/arXiv.astro-ph/9702215}

\bibitem[{{Dekel} \& {Birnboim}(2006)}]{dekel06}
{Dekel}, A., \& {Birnboim}, Y. 2006, \mnras, 368, 2, \dodoi{10.1111/j.1365-2966.2006.10145.x}

\bibitem[{{Diamond-Stanic} {et~al.}(2016){Diamond-Stanic}, {Coil}, {Moustakas}, {Tremonti}, {Sell}, {Mendez}, {Hickox}, \& {Rudnick}}]{diamond16}
{Diamond-Stanic}, A.~M., {Coil}, A.~L., {Moustakas}, J., {et~al.} 2016, \apj, 824, 24, \dodoi{10.3847/0004-637X/824/1/24}

\bibitem[{{Dickey} \& {Lockman}(1990)}]{dickey-lockman90}
{Dickey}, J.~M., \& {Lockman}, F.~J. 1990, \araa, 28, 215, \dodoi{10.1146/annurev.aa.28.090190.001243}

\bibitem[{{D'Onghia} \& {Fox}(2016)}]{d-onghia116}
{D'Onghia}, E., \& {Fox}, A.~J. 2016, \araa, 54, 363, \dodoi{10.1146/annurev-astro-081915-023251}

\bibitem[{{Dopita} \& {Sutherland}(2003)}]{dopita03}
{Dopita}, M.~A., \& {Sutherland}, R.~S. 2003, {Astrophysics of the Diffuse Universe} (Berlin, New York: Springer), \dodoi{10.1007/978-3-662-05866-4}

\bibitem[{{Dutta} {et~al.}(2024){Dutta}, {Bisht}, {Sharma}, {Ghosh}, {Roy}, \& {Nath}}]{dutta24}
{Dutta}, A., {Bisht}, M.~S., {Sharma}, P., {et~al.} 2024, \mnras, 531, 5117, \dodoi{10.1093/mnras/stae977}

\bibitem[{{Ellison} {et~al.}(2003){Ellison}, {Mall{\'e}n-Ornelas}, \& {Sawicki}}]{ellison03}
{Ellison}, S.~L., {Mall{\'e}n-Ornelas}, G., \& {Sawicki}, M. 2003, \apj, 589, 709, \dodoi{10.1086/374660}

\bibitem[{{Esmerian} {et~al.}(2021){Esmerian}, {Kravtsov}, {Hafen}, {Faucher-Gigu{\`e}re}, {Quataert}, {Stern}, {Kere{\v{s}}}, \& {Wetzel}}]{esmerian21}
{Esmerian}, C.~J., {Kravtsov}, A.~V., {Hafen}, Z., {et~al.} 2021, \mnras, 505, 1841, \dodoi{10.1093/mnras/stab1281}

\bibitem[{{Faerman} {et~al.}(2017){Faerman}, {Sternberg}, \& {McKee}}]{faerman17}
{Faerman}, Y., {Sternberg}, A., \& {McKee}, C.~F. 2017, \apj, 835, 52, \dodoi{10.3847/1538-4357/835/1/52}

\bibitem[{{Faerman} {et~al.}(2020){Faerman}, {Sternberg}, \& {McKee}}]{faerman20}
---. 2020, \apj, 893, 82, \dodoi{10.3847/1538-4357/ab7ffc}

\bibitem[{{Faerman} \& {Werk}(2023)}]{faerman23}
{Faerman}, Y., \& {Werk}, J.~K. 2023, arXiv e-prints, arXiv:2302.00692, \dodoi{10.48550/arXiv.2302.00692}

\bibitem[{{Fang} {et~al.}(2013){Fang}, {Bullock}, \& {Boylan-Kolchin}}]{fang13}
{Fang}, T., {Bullock}, J., \& {Boylan-Kolchin}, M. 2013, \apj, 762, 20, \dodoi{10.1088/0004-637X/762/1/20}

\bibitem[{{Faucher-Gigu{\`e}re}(2020)}]{faucher-giguere20}
{Faucher-Gigu{\`e}re}, C.-A. 2020, \mnras, 493, 1614, \dodoi{10.1093/mnras/staa302}

\bibitem[{{Faucher-Giguere} \& {Oh}(2023)}]{faucher-giguere23}
{Faucher-Giguere}, C.-A., \& {Oh}, S.~P. 2023, arXiv e-prints, arXiv:2301.10253, \dodoi{10.48550/arXiv.2301.10253}

\bibitem[{{Ferland} {et~al.}(1998){Ferland}, {Korista}, {Verner}, {Ferguson}, {Kingdon}, \& {Verner}}]{ferland98}
{Ferland}, G.~J., {Korista}, K.~T., {Verner}, D.~A., {et~al.} 1998, \pasp, 110, 761, \dodoi{10.1086/316190}

\bibitem[{{Ferland} {et~al.}(2013){Ferland}, {Porter}, {van Hoof}, {Williams}, {Abel}, {Lykins}, {Shaw}, {Henney}, \& {Stancil}}]{ferland13}
{Ferland}, G.~J., {Porter}, R.~L., {van Hoof}, P.~A.~M., {et~al.} 2013, \rmxaa, 49, 137, \dodoi{10.48550/arXiv.1302.4485}

\bibitem[{{Ferland} {et~al.}(2017){Ferland}, {Chatzikos}, {Guzm{\'a}n}, {Lykins}, {van Hoof}, {Williams}, {Abel}, {Badnell}, {Keenan}, {Porter}, \& {Stancil}}]{ferland17}
{Ferland}, G.~J., {Chatzikos}, M., {Guzm{\'a}n}, F., {et~al.} 2017, \rmxaa, 53, 385, \dodoi{10.48550/arXiv.1705.10877}

\bibitem[{{Ferrara}(1996)}]{ferrara96}
{Ferrara}, A. 1996, in Unsolved Problems of the Milky Way, ed. L.~{Blitz} \& P.~J. {Teuben}, Vol. 169, 479

\bibitem[{{Fielding} \& {Bryan}(2022)}]{fielding22}
{Fielding}, D.~B., \& {Bryan}, G.~L. 2022, \apj, 924, 82, \dodoi{10.3847/1538-4357/ac2f41}

\bibitem[{{Fielding} {et~al.}(2020){Fielding}, {Tonnesen}, {DeFelippis}, {Li}, {Su}, {Bryan}, {Kim}, {Forbes}, {Somerville}, {Battaglia}, {Schneider}, {Li}, {Choi}, {Hayward}, \& {Hernquist}}]{fielding20}
{Fielding}, D.~B., {Tonnesen}, S., {DeFelippis}, D., {et~al.} 2020, \apj, 903, 32, \dodoi{10.3847/1538-4357/abbc6d}

\bibitem[{{Finke} {et~al.}(2022){Finke}, {Ajello}, {Dom{\'\i}nguez}, {Desai}, {Hartmann}, {Paliya}, \& {Saldana-Lopez}}]{finke22}
{Finke}, J.~D., {Ajello}, M., {Dom{\'\i}nguez}, A., {et~al.} 2022, \apj, 941, 33, \dodoi{10.3847/1538-4357/ac9843}

\bibitem[{{Finke} {et~al.}(2010){Finke}, {Razzaque}, \& {Dermer}}]{finke10}
{Finke}, J.~D., {Razzaque}, S., \& {Dermer}, C.~D. 2010, \apj, 712, 238, \dodoi{10.1088/0004-637X/712/1/238}

\bibitem[{{Ford} {et~al.}(2014){Ford}, {Dav{\'e}}, {Oppenheimer}, {Katz}, {Kollmeier}, {Thompson}, \& {Weinberg}}]{ford14}
{Ford}, A.~B., {Dav{\'e}}, R., {Oppenheimer}, B.~D., {et~al.} 2014, \mnras, 444, 1260, \dodoi{10.1093/mnras/stu1418}

\bibitem[{{Fox} {et~al.}(2015){Fox}, {Bordoloi}, {Savage}, {Lockman}, {Jenkins}, {Wakker}, {Bland-Hawthorn}, {Hernandez}, {Kim}, {Benjamin}, {Bowen}, \& {Tumlinson}}]{fox15}
{Fox}, A.~J., {Bordoloi}, R., {Savage}, B.~D., {et~al.} 2015, \apjl, 799, L7, \dodoi{10.1088/2041-8205/799/1/L7}

\bibitem[{{French} \& {Wakker}(2020)}]{french20}
{French}, D.~M., \& {Wakker}, B.~P. 2020, \apj, 897, 151, \dodoi{10.3847/1538-4357/ab9905}

\bibitem[{{Gaensler} {et~al.}(2008){Gaensler}, {Madsen}, {Chatterjee}, \& {Mao}}]{gaensler08}
{Gaensler}, B.~M., {Madsen}, G.~J., {Chatterjee}, S., \& {Mao}, S.~A. 2008, \pasa, 25, 184, \dodoi{10.1071/AS08004}

\bibitem[{{Garza} {et~al.}(2025){Garza}, {Werk}, {Berg}, {Faerman}, {Oppenheimer}, {Bordoloi}, \& {Ellison}}]{garza25}
{Garza}, S.~L., {Werk}, J.~K., {Berg}, T. A.~M., {et~al.} 2025, \apjl, 978, L12, \dodoi{10.3847/2041-8213/ad9c69}

\bibitem[{{Gauthier} \& {Chen}(2012)}]{gauthier12}
{Gauthier}, J.-R., \& {Chen}, H.-W. 2012, \mnras, 424, 1952, \dodoi{10.1111/j.1365-2966.2012.21327.x}

\bibitem[{{Ghosh} {et~al.}(2025){Ghosh}, {Gronke}, {Sharma}, \& {Dutta}}]{ghosh25}
{Ghosh}, R., {Gronke}, M., {Sharma}, P., \& {Dutta}, A. 2025, arXiv e-prints, arXiv:2510.03552, \dodoi{10.48550/arXiv.2510.03552}

\bibitem[{{Grand} {et~al.}(2019){Grand}, {van de Voort}, {Zjupa}, {Fragkoudi}, {G{\'o}mez}, {Kauffmann}, {Marinacci}, {Pakmor}, {Springel}, \& {White}}]{grand19}
{Grand}, R. J.~J., {van de Voort}, F., {Zjupa}, J., {et~al.} 2019, \mnras, 490, 4786, \dodoi{10.1093/mnras/stz2928}

\bibitem[{{Gronke} \& {Oh}(2020{\natexlab{a}})}]{gronke20}
{Gronke}, M., \& {Oh}, S.~P. 2020{\natexlab{a}}, \mnras, 492, 1970, \dodoi{10.1093/mnras/stz3332}

\bibitem[{{Gronke} \& {Oh}(2020{\natexlab{b}})}]{gronke20-mist}
---. 2020{\natexlab{b}}, \mnras, 494, L27, \dodoi{10.1093/mnrasl/slaa033}

\bibitem[{{Gurvich} {et~al.}(2023){Gurvich}, {Stern}, {Faucher-Gigu{\`e}re}, {Hopkins}, {Wetzel}, {Moreno}, {Hayward}, {Richings}, \& {Hafen}}]{gurvich23}
{Gurvich}, A.~B., {Stern}, J., {Faucher-Gigu{\`e}re}, C.-A., {et~al.} 2023, \mnras, 519, 2598, \dodoi{10.1093/mnras/stac3712}

\bibitem[{{Haardt} \& {Madau}(1996)}]{haardt96}
{Haardt}, F., \& {Madau}, P. 1996, \apj, 461, 20, \dodoi{10.1086/177035}

\bibitem[{{Haardt} \& {Madau}(2012)}]{haardt12}
---. 2012, \apj, 746, 125, \dodoi{10.1088/0004-637X/746/2/125}

\bibitem[{{Habe} \& {Ikeuchi}(1982)}]{habe82}
{Habe}, A., \& {Ikeuchi}, S. 1982, Progress of Theoretical Physics, 68, 1131, \dodoi{10.1143/PTP.68.1131}

\bibitem[{{Hafen} {et~al.}(2022){Hafen}, {Stern}, {Bullock}, {Gurvich}, {Yu}, {Faucher-Gigu{\`e}re}, {Fielding}, {Angl{\'e}s-Alc{\'a}zar}, {Quataert}, {Wetzel}, {Starkenburg}, {Boylan-Kolchin}, {Moreno}, {Feldmann}, {El-Badry}, {Chan}, {Trapp}, {Kere{\v{s}}}, \& {Hopkins}}]{hafen22}
{Hafen}, Z., {Stern}, J., {Bullock}, J., {et~al.} 2022, \mnras, 514, 5056, \dodoi{10.1093/mnras/stac1603}

\bibitem[{{Haffner} {et~al.}(2009){Haffner}, {Dettmar}, {Beckman}, {Wood}, {Slavin}, {Giammanco}, {Madsen}, {Zurita}, \& {Reynolds}}]{haffner09}
{Haffner}, L.~M., {Dettmar}, R.~J., {Beckman}, J.~E., {et~al.} 2009, Reviews of Modern Physics, 81, 969, \dodoi{10.1103/RevModPhys.81.969}

\bibitem[{{Hidalgo-Pineda} {et~al.}(2025){Hidalgo-Pineda}, {Gronke}, \& {Grete}}]{hidalgo-pineda25}
{Hidalgo-Pineda}, F., {Gronke}, M., \& {Grete}, P. 2025, arXiv e-prints, arXiv:2510.14829, \dodoi{10.48550/arXiv.2510.14829}

\bibitem[{{Ho} \& {Martin}(2020)}]{ho20}
{Ho}, S.~H., \& {Martin}, C.~L. 2020, \apj, 888, 14, \dodoi{10.3847/1538-4357/ab58cd}

\bibitem[{{Ho} {et~al.}(2017){Ho}, {Martin}, {Kacprzak}, \& {Churchill}}]{ho17}
{Ho}, S.~H., {Martin}, C.~L., {Kacprzak}, G.~G., \& {Churchill}, C.~W. 2017, \apj, 835, 267, \dodoi{10.3847/1538-4357/835/2/267}

\bibitem[{{Ho} {et~al.}(2020){Ho}, {Martin}, \& {Schaye}}]{ho-eagle20}
{Ho}, S.~H., {Martin}, C.~L., \& {Schaye}, J. 2020, \apj, 904, 76, \dodoi{10.3847/1538-4357/abbe88}

\bibitem[{{Hodges-Kluck} {et~al.}(2016){Hodges-Kluck}, {Miller}, \& {Bregman}}]{hodges-kluck16}
{Hodges-Kluck}, E.~J., {Miller}, M.~J., \& {Bregman}, J.~N. 2016, \apj, 822, 21, \dodoi{10.3847/0004-637X/822/1/21}

\bibitem[{{Huang} {et~al.}(2022){Huang}, {Katz}, {Cottle}, {Scannapieco}, {Dav{\'e}}, \& {Weinberg}}]{huang22}
{Huang}, S., {Katz}, N., {Cottle}, J., {et~al.} 2022, \mnras, 509, 6091, \dodoi{10.1093/mnras/stab3363}

\bibitem[{{Huang} {et~al.}(2020){Huang}, {Katz}, {Scannapieco}, {Cottle}, {Dav{\'e}}, {Weinberg}, {Peeples}, \& {Br{\"u}ggen}}]{huang20}
{Huang}, S., {Katz}, N., {Scannapieco}, E., {et~al.} 2020, \mnras, 497, 2586, \dodoi{10.1093/mnras/staa1978}

\bibitem[{{Huml{\'\i}cek}(1979)}]{humlicek79}
{Huml{\'\i}cek}, J. 1979, \jqsrt, 21, 309, \dodoi{10.1016/0022-4073(79)90062-1}

\bibitem[{{Hummels} {et~al.}(2024){Hummels}, {Rubin}, {Schneider}, \& {Fielding}}]{hummels24}
{Hummels}, C.~B., {Rubin}, K. H.~R., {Schneider}, E.~E., \& {Fielding}, D.~B. 2024, \apj, 972, 148, \dodoi{10.3847/1538-4357/ad5965}

\bibitem[{{Hummels} {et~al.}(2019){Hummels}, {Smith}, {Hopkins}, {O'Shea}, {Silvia}, {Werk}, {Lehner}, {Wise}, {Collins}, \& {Butsky}}]{hummels19}
{Hummels}, C.~B., {Smith}, B.~D., {Hopkins}, P.~F., {et~al.} 2019, \apj, 882, 156, \dodoi{10.3847/1538-4357/ab378f}

\bibitem[{{Ianjamasimanana} {et~al.}(2018){Ianjamasimanana}, {Walter}, {de Blok}, {Heald}, \& {Brinks}}]{ianjamas18}
{Ianjamasimanana}, R., {Walter}, F., {de Blok}, W.~J.~G., {Heald}, G.~H., \& {Brinks}, E. 2018, \aj, 155, 233, \dodoi{10.3847/1538-3881/aabbaa}

\bibitem[{{Kacprzak} {et~al.}(2011){Kacprzak}, {Churchill}, {Barton}, \& {Cooke}}]{kacprzak11}
{Kacprzak}, G.~G., {Churchill}, C.~W., {Barton}, E.~J., \& {Cooke}, J. 2011, \apj, 733, 105, \dodoi{10.1088/0004-637X/733/2/105}

\bibitem[{{Kacprzak} {et~al.}(2010{\natexlab{a}}){Kacprzak}, {Churchill}, {Ceverino}, {Steidel}, {Klypin}, \& {Murphy}}]{kacprzak10}
{Kacprzak}, G.~G., {Churchill}, C.~W., {Ceverino}, D., {et~al.} 2010{\natexlab{a}}, \apj, 711, 533, \dodoi{10.1088/0004-637X/711/2/533}

\bibitem[{{Kacprzak} {et~al.}(2012){Kacprzak}, {Churchill}, \& {Nielsen}}]{kacprzak12}
{Kacprzak}, G.~G., {Churchill}, C.~W., \& {Nielsen}, N.~M. 2012, \apjl, 760, L7, \dodoi{10.1088/2041-8205/760/1/L7}

\bibitem[{{Kacprzak} {et~al.}(2010{\natexlab{b}}){Kacprzak}, {Murphy}, \& {Churchill}}]{kacprzak10-pks1127}
{Kacprzak}, G.~G., {Murphy}, M.~T., \& {Churchill}, C.~W. 2010{\natexlab{b}}, \mnras, 406, 445, \dodoi{10.1111/j.1365-2966.2010.16667.x}

\bibitem[{{Kacprzak} {et~al.}(2021){Kacprzak}, {Nielsen}, {Nateghi}, {Churchill}, {Pointon}, {Nanayakkara}, {Muzahid}, \& {Charlton}}]{kacprzak21}
{Kacprzak}, G.~G., {Nielsen}, N.~M., {Nateghi}, H., {et~al.} 2021, \mnras, 500, 2289, \dodoi{10.1093/mnras/staa3461}

\bibitem[{{Kacprzak} {et~al.}(2019){Kacprzak}, {Vander Vliet}, {Nielsen}, {Muzahid}, {Pointon}, {Churchill}, {Ceverino}, {Arraki}, {Klypin}, {Charlton}, \& {Lewis}}]{kacprzak19}
{Kacprzak}, G.~G., {Vander Vliet}, J.~R., {Nielsen}, N.~M., {et~al.} 2019, \apj, 870, 137, \dodoi{10.3847/1538-4357/aaf1a6}

\bibitem[{{Kacprzak} {et~al.}(2025){Kacprzak}, {Oppenheimer}, {Nielsen}, {Fern{\'a}ndez-Figueroa}, {Murphy}, {Allen}, {Barone}, {Sameer}, {Churchill}, {Burchett}, {Gupta}, {Charlton}, \& {Platukis}}]{kacprzak25}
{Kacprzak}, G.~G., {Oppenheimer}, B., {Nielsen}, N., {et~al.} 2025, \pasa, 42, e128, \dodoi{10.1017/pasa.2025.10091}

\bibitem[{{Kakoly} {et~al.}(2025){Kakoly}, {Stern}, {Faucher-Gigu{\`e}re}, {Fielding}, {Goldner}, {Sun}, \& {Hummels}}]{kakoly25}
{Kakoly}, A., {Stern}, J., {Faucher-Gigu{\`e}re}, C.-A., {et~al.} 2025, arXiv e-prints, arXiv:2504.17001, \dodoi{10.48550/arXiv.2504.17001}

\bibitem[{{Kalberla} \& {Dedes}(2008)}]{kalberla08}
{Kalberla}, P.~M.~W., \& {Dedes}, L. 2008, \aap, 487, 951, \dodoi{10.1051/0004-6361:20079240}

\bibitem[{{Kere{\v{s}}} {et~al.}(2005){Kere{\v{s}}}, {Katz}, {Weinberg}, \& {Dav{\'e}}}]{keres05}
{Kere{\v{s}}}, D., {Katz}, N., {Weinberg}, D.~H., \& {Dav{\'e}}, R. 2005, \mnras, 363, 2, \dodoi{10.1111/j.1365-2966.2005.09451.x}

\bibitem[{{Khaire} {et~al.}(2019){Khaire}, {Walther}, {Hennawi}, {O{\~n}orbe}, {Luki{\'c}}, {}, {Prochaska}, {Tripp}, {Burchett}, \& {Rodriguez}}]{ks19}
{Khaire}, V., {Walther}, M., {Hennawi}, J.~F., {et~al.} 2019, \mnras, 486, 769, \dodoi{10.1093/mnras/stz344}

\bibitem[{{Kochanov}(2011)}]{kochanov11}
{Kochanov}, V.~P. 2011, \jqsrt, 112, 2762, \dodoi{10.1016/j.jqsrt.2011.08.006}

\bibitem[{{Kocjan} {et~al.}(2024){Kocjan}, {Cadiou}, {Agertz}, \& {Pontzen}}]{kocjan24}
{Kocjan}, Z., {Cadiou}, C., {Agertz}, O., \& {Pontzen}, A. 2024, \mnras, 534, 918, \dodoi{10.1093/mnras/stae2128}

\bibitem[{{Kumar} {et~al.}(2024){Kumar}, {Chen}, {Qu}, {Chen}, {Zahedy}, {Johnson}, {Muzahid}, \& {Cantalupo}}]{kumar24}
{Kumar}, S., {Chen}, H.-W., {Qu}, Z., {et~al.} 2024, The Open Journal of Astrophysics, 7, 94, \dodoi{10.33232/001c.124930}

\bibitem[{{Lan} \& {Mo}(2018)}]{lan18}
{Lan}, T.-W., \& {Mo}, H. 2018, \apj, 866, 36, \dodoi{10.3847/1538-4357/aadc08}

\bibitem[{{Langer} {et~al.}(2014){Langer}, {Pineda}, \& {Velusamy}}]{langer14}
{Langer}, W.~D., {Pineda}, J.~L., \& {Velusamy}, T. 2014, \aap, 564, A101, \dodoi{10.1051/0004-6361/201323281}

\bibitem[{{Langer} {et~al.}(2021){Langer}, {Pineda}, {Goldsmith}, {Chambers}, {Riquelme}, {Anderson}, {Luisi}, {Justen}, \& {Buchbender}}]{langer21}
{Langer}, W.~D., {Pineda}, J.~L., {Goldsmith}, P.~F., {et~al.} 2021, \aap, 651, A59, \dodoi{10.1051/0004-6361/202040223}

\bibitem[{{Lanzetta} \& {Bowen}(1992)}]{lanzetta92}
{Lanzetta}, K.~M., \& {Bowen}, D.~V. 1992, \apj, 391, 48, \dodoi{10.1086/171325}

\bibitem[{{Leitherer} {et~al.}(1999){Leitherer}, {Schaerer}, {Goldader}, {Delgado}, {Robert}, {Kune}, {de Mello}, {Devost}, \& {Heckman}}]{leitherer99}
{Leitherer}, C., {Schaerer}, D., {Goldader}, J.~D., {et~al.} 1999, \apjs, 123, 3, \dodoi{10.1086/313233}

\bibitem[{{Liang} \& {Remming}(2020)}]{liang20}
{Liang}, C.~J., \& {Remming}, I. 2020, \mnras, 491, 5056, \dodoi{10.1093/mnras/stz3403}

\bibitem[{{Lilly} {et~al.}(2013){Lilly}, {Carollo}, {Pipino}, {Renzini}, \& {Peng}}]{lilly13}
{Lilly}, S.~J., {Carollo}, C.~M., {Pipino}, A., {Renzini}, A., \& {Peng}, Y. 2013, \apj, 772, 119, \dodoi{10.1088/0004-637X/772/2/119}

\bibitem[{{Lodders}(2019)}]{lodders19}
{Lodders}, K. 2019, arXiv e-prints, arXiv:1912.00844, \dodoi{10.48550/arXiv.1912.00844}

\bibitem[{{Lopez} {et~al.}(2024){Lopez}, {Afruni}, {Zamora}, {Tejos}, {Ledoux}, {Hernandez}, {Berg}, {Cortes}, {Urbina}, {Johnston}, {Barrientos}, {Bayliss}, {Cuellar}, {Krogager}, {Noterdaeme}, \& {Solimano}}]{lopez24}
{Lopez}, S., {Afruni}, A., {Zamora}, D., {et~al.} 2024, \aap, 691, A356, \dodoi{10.1051/0004-6361/202451200}

\bibitem[{{Macci{\`o}} {et~al.}(2006){Macci{\`o}}, {Moore}, \& {Stadel}}]{maccio06}
{Macci{\`o}}, A.~V., {Moore}, B., \& {Stadel}, J. 2006, \apjl, 636, L25, \dodoi{10.1086/499778}

\bibitem[{{Maller} \& {Bullock}(2004)}]{maller04}
{Maller}, A.~H., \& {Bullock}, J.~S. 2004, \mnras, 355, 694, \dodoi{10.1111/j.1365-2966.2004.08349.x}

\bibitem[{{Mann}(2023)}]{mann23}
{Mann}, A. 2023, Proceedings of the National Academy of Science, 120, e2318720120, \dodoi{10.1073/pnas.2318720120}

\bibitem[{{Marasco} \& {Fraternali}(2011)}]{marasco11}
{Marasco}, A., \& {Fraternali}, F. 2011, \aap, 525, A134, \dodoi{10.1051/0004-6361/201015508}

\bibitem[{{Martin} {et~al.}(2019){Martin}, {Ho}, {Kacprzak}, \& {Churchill}}]{martin19}
{Martin}, C.~L., {Ho}, S.~H., {Kacprzak}, G.~G., \& {Churchill}, C.~W. 2019, \apj, 878, 84, \dodoi{10.3847/1538-4357/ab18ac}

\bibitem[{{Mathews} \& {Prochaska}(2017)}]{mathews17}
{Mathews}, W.~G., \& {Prochaska}, J.~X. 2017, \apjl, 846, L24, \dodoi{10.3847/2041-8213/aa8861}

\bibitem[{{Mo} \& {Miralda-Escude}(1996)}]{mo96}
{Mo}, H.~J., \& {Miralda-Escude}, J. 1996, \apj, 469, 589, \dodoi{10.1086/177808}

\bibitem[{{Naab} \& {Ostriker}(2017)}]{naab17}
{Naab}, T., \& {Ostriker}, J.~P. 2017, \araa, 55, 59, \dodoi{10.1146/annurev-astro-081913-040019}

\bibitem[{{Nateghi} {et~al.}(2021){Nateghi}, {Kacprzak}, {Nielsen}, {Muzahid}, {Churchill}, {Pointon}, \& {Charlton}}]{nateghi21}
{Nateghi}, H., {Kacprzak}, G.~G., {Nielsen}, N.~M., {et~al.} 2021, \mnras, 500, 3987, \dodoi{10.1093/mnras/staa3534}

\bibitem[{{Navarro} {et~al.}(1996){Navarro}, {Frenk}, \& {White}}]{NFW96}
{Navarro}, J.~F., {Frenk}, C.~S., \& {White}, S. D.~M. 1996, \apj, 462, 563, \dodoi{10.1086/177173}

\bibitem[{{Nelson} {et~al.}(2018){Nelson}, {Kauffmann}, {Pillepich}, {Genel}, {Springel}, {Pakmor}, {Hernquist}, {Weinberger}, {Torrey}, {Vogelsberger}, \& {Marinacci}}]{nelson18}
{Nelson}, D., {Kauffmann}, G., {Pillepich}, A., {et~al.} 2018, \mnras, 477, 450, \dodoi{10.1093/mnras/sty656}

\bibitem[{{Nelson} {et~al.}(2019){Nelson}, {Pillepich}, {Springel}, {Pakmor}, {Weinberger}, {Genel}, {Torrey}, {Vogelsberger}, {Marinacci}, \& {Hernquist}}]{nelson19}
{Nelson}, D., {Pillepich}, A., {Springel}, V., {et~al.} 2019, \mnras, 490, 3234, \dodoi{10.1093/mnras/stz2306}

\bibitem[{{Nguyen} \& {Thompson}(2022)}]{nguyen22}
{Nguyen}, D.~D., \& {Thompson}, T.~A. 2022, \apjl, 935, L24, \dodoi{10.3847/2041-8213/ac86c3}

\bibitem[{{Nielsen} {et~al.}(2015){Nielsen}, {Churchill}, {Kacprzak}, {Murphy}, \& {Evans}}]{nielsen15}
{Nielsen}, N.~M., {Churchill}, C.~W., {Kacprzak}, G.~G., {Murphy}, M.~T., \& {Evans}, J.~L. 2015, \apj, 812, 83, \dodoi{10.1088/0004-637X/812/1/83}

\bibitem[{{Nielsen} {et~al.}(2016){Nielsen}, {Churchill}, {Kacprzak}, {Murphy}, \& {Evans}}]{nielsen16}
---. 2016, \apj, 818, 171, \dodoi{10.3847/0004-637X/818/2/171}

\bibitem[{{Nielsen} {et~al.}(2024){Nielsen}, {Fisher}, {Kacprzak}, {Chisholm}, {Martin}, {Reichardt Chu}, {Sandstrom}, \& {Rickards Vaught}}]{nielsen24}
{Nielsen}, N.~M., {Fisher}, D.~B., {Kacprzak}, G.~G., {et~al.} 2024, Nature Astronomy, 8, 1602, \dodoi{10.1038/s41550-024-02365-x}

\bibitem[{{Noguchi}(2018)}]{noguchi18}
{Noguchi}, M. 2018, \apj, 853, 67, \dodoi{10.3847/1538-4357/aaa484}

\bibitem[{{Oosterloo} {et~al.}(2007){Oosterloo}, {Fraternali}, \& {Sancisi}}]{oosterloo07}
{Oosterloo}, T., {Fraternali}, F., \& {Sancisi}, R. 2007, \aj, 134, 1019, \dodoi{10.1086/520332}

\bibitem[{{Oppenheimer} \& {Dav{\'e}}(2008)}]{oppenheimer08}
{Oppenheimer}, B.~D., \& {Dav{\'e}}, R. 2008, \mnras, 387, 577, \dodoi{10.1111/j.1365-2966.2008.13280.x}

\bibitem[{{Oppenheimer} {et~al.}(2018){Oppenheimer}, {Schaye}, {Crain}, {Werk}, \& {Richings}}]{oppenheimer18}
{Oppenheimer}, B.~D., {Schaye}, J., {Crain}, R.~A., {Werk}, J.~K., \& {Richings}, A.~J. 2018, \mnras, 481, 835, \dodoi{10.1093/mnras/sty2281}

\bibitem[{{Oppenheimer} {et~al.}(2025){Oppenheimer}, {Voit}, {Bah{\'e}}, {Battaglia}, {Bregman}, {Burchett}, {Eckert}, {Faerman}, {Gibson}, {Hummels}, {Medlock}, {Nagai}, {Putman}, {Qu}, {Sun}, {Werk}, \& {Zhang}}]{oppenheimer25}
{Oppenheimer}, B.~D., {Voit}, G.~M., {Bah{\'e}}, Y.~M., {et~al.} 2025, \mnras, 543, 2649, \dodoi{10.1093/mnras/staf1581}

\bibitem[{{Padoan} {et~al.}(2001){Padoan}, {Kim}, {Goodman}, \& {Staveley-Smith}}]{padoan01}
{Padoan}, P., {Kim}, S., {Goodman}, A., \& {Staveley-Smith}, L. 2001, \apjl, 555, L33, \dodoi{10.1086/321735}

\bibitem[{{P{\'e}roux} \& {Howk}(2020)}]{peroux-howk20}
{P{\'e}roux}, C., \& {Howk}, J.~C. 2020, \araa, 58, 363, \dodoi{10.1146/annurev-astro-021820-120014}

\bibitem[{{P{\'e}roux} {et~al.}(2020){P{\'e}roux}, {Nelson}, {van de Voort}, {Pillepich}, {Marinacci}, {Vogelsberger}, \& {Hernquist}}]{peroux20}
{P{\'e}roux}, C., {Nelson}, D., {van de Voort}, F., {et~al.} 2020, \mnras, 499, 2462, \dodoi{10.1093/mnras/staa2888}

\bibitem[{{Pezzulli} {et~al.}(2017){Pezzulli}, {Fraternali}, \& {Binney}}]{pezzulli17}
{Pezzulli}, G., {Fraternali}, F., \& {Binney}, J. 2017, \mnras, 467, 311, \dodoi{10.1093/mnras/stx029}

\bibitem[{{Ploeckinger} \& {Schaye}(2020)}]{ploeckinger20}
{Ploeckinger}, S., \& {Schaye}, J. 2020, \mnras, 497, 4857, \dodoi{10.1093/mnras/staa2172}

\bibitem[{{Predehl} {et~al.}(2020){Predehl}, {Sunyaev}, {Becker}, {Brunner}, {Burenin}, {Bykov}, {Cherepashchuk}, {Chugai}, {Churazov}, {Doroshenko}, {Eismont}, {Freyberg}, {Gilfanov}, {Haberl}, {Khabibullin}, {Krivonos}, {Maitra}, {Medvedev}, {Merloni}, {Nandra}, {Nazarov}, {Pavlinsky}, {Ponti}, {Sanders}, {Sasaki}, {Sazonov}, {Strong}, \& {Wilms}}]{predehl20}
{Predehl}, P., {Sunyaev}, R.~A., {Becker}, W., {et~al.} 2020, \nat, 588, 227, \dodoi{10.1038/s41586-020-2979-0}

\bibitem[{{Prochaska} \& {Wolfe}(1998)}]{prochaska98}
{Prochaska}, J.~X., \& {Wolfe}, A.~M. 1998, \apj, 507, 113, \dodoi{10.1086/306325}

\bibitem[{{Putman} {et~al.}(2012){Putman}, {Peek}, \& {Joung}}]{putman12}
{Putman}, M.~E., {Peek}, J.~E.~G., \& {Joung}, M.~R. 2012, \araa, 50, 491, \dodoi{10.1146/annurev-astro-081811-125612}

\bibitem[{{Qu} \& {Bregman}(2019)}]{qu19}
{Qu}, Z., \& {Bregman}, J.~N. 2019, \apj, 880, 89, \dodoi{10.3847/1538-4357/ab2a0b}

\bibitem[{{Qu} {et~al.}(2020){Qu}, {Bregman}, {Hodges-Kluck}, {Li}, \& {Lindley}}]{qu20}
{Qu}, Z., {Bregman}, J.~N., {Hodges-Kluck}, E., {Li}, J.-T., \& {Lindley}, R. 2020, \apj, 894, 142, \dodoi{10.3847/1538-4357/ab774e}

\bibitem[{{Rahmati} {et~al.}(2013){Rahmati}, {Pawlik}, {Rai{\v{c}}evi{\'c}}, \& {Schaye}}]{rahmati13}
{Rahmati}, A., {Pawlik}, A.~H., {Rai{\v{c}}evi{\'c}}, M., \& {Schaye}, J. 2013, \mnras, 430, 2427, \dodoi{10.1093/mnras/stt066}

\bibitem[{{Ramesh} {et~al.}(2023){Ramesh}, {Nelson}, \& {Pillepich}}]{ramesh23}
{Ramesh}, R., {Nelson}, D., \& {Pillepich}, A. 2023, \mnras, 518, 5754, \dodoi{10.1093/mnras/stac3524}

\bibitem[{{Rauch} {et~al.}(2001){Rauch}, {Sargent}, \& {Barlow}}]{rauch01}
{Rauch}, M., {Sargent}, W. L.~W., \& {Barlow}, T.~A. 2001, \apj, 554, 823, \dodoi{10.1086/321402}

\bibitem[{{Rauch} {et~al.}(2002){Rauch}, {Sargent}, {Barlow}, \& {Simcoe}}]{rauch02}
{Rauch}, M., {Sargent}, W. L.~W., {Barlow}, T.~A., \& {Simcoe}, R.~A. 2002, \apj, 576, 45, \dodoi{10.1086/341267}

\bibitem[{{Rees} \& {Ostriker}(1977)}]{rees77}
{Rees}, M.~J., \& {Ostriker}, J.~P. 1977, \mnras, 179, 541, \dodoi{10.1093/mnras/179.4.541}

\bibitem[{{Rigby} {et~al.}(2002){Rigby}, {Charlton}, \& {Churchill}}]{rigby02}
{Rigby}, J.~R., {Charlton}, J.~C., \& {Churchill}, C.~W. 2002, \apj, 565, 743, \dodoi{10.1086/324723}

\bibitem[{{Rupke}(2018)}]{rupke18}
{Rupke}, D. 2018, Galaxies, 6, 138, \dodoi{10.3390/galaxies6040138}

\bibitem[{{Sameer} {et~al.}(2021){Sameer}, {Charlton}, {Norris}, {Gebhardt}, {Churchill}, {Kacprzak}, {Muzahid}, {Narayanan}, {Nielsen}, {Richter}, \& {Wakker}}]{sameer21}
{Sameer}, {Charlton}, J.~C., {Norris}, J.~M., {et~al.} 2021, \mnras, 501, 2112, \dodoi{10.1093/mnras/staa3754}

\bibitem[{{Sameer} {et~al.}(2022){Sameer}, {Charlton}, {Kacprzak}, {Narayanan}, {Sankar}, {Richter}, {Wakker}, {Nielsen}, \& {Churchill}}]{sameer22}
{Sameer}, {Charlton}, J.~C., {Kacprzak}, G.~G., {et~al.} 2022, \mnras, 510, 5796, \dodoi{10.1093/mnras/stac052}

\bibitem[{{Sameer} {et~al.}(2024){Sameer}, {Charlton}, {Wakker}, {Kacprzak}, {Nielsen}, {Churchill}, {Richter}, {Muzahid}, {Ho}, {Nateghi}, {Rosenwasser}, {Narayanan}, \& {Ganguly}}]{sameer24}
{Sameer}, {Charlton}, J.~C., {Wakker}, B.~P., {et~al.} 2024, \mnras, 530, 3827, \dodoi{10.1093/mnras/stae962}

\bibitem[{{Sancisi} {et~al.}(2008){Sancisi}, {Fraternali}, {Oosterloo}, \& {van der Hulst}}]{sancisi08}
{Sancisi}, R., {Fraternali}, F., {Oosterloo}, T., \& {van der Hulst}, T. 2008, \aapr, 15, 189, \dodoi{10.1007/s00159-008-0010-0}

\bibitem[{{Savage} \& {Wakker}(2009)}]{savage09}
{Savage}, B.~D., \& {Wakker}, B.~P. 2009, \apj, 702, 1472, \dodoi{10.1088/0004-637X/702/2/1472}

\bibitem[{{Schaye} {et~al.}(2007){Schaye}, {Carswell}, \& {Kim}}]{schaye07}
{Schaye}, J., {Carswell}, R.~F., \& {Kim}, T.-S. 2007, \mnras, 379, 1169, \dodoi{10.1111/j.1365-2966.2007.12005.x}

\bibitem[{{Schreier}(2018)}]{schreier18}
{Schreier}, F. 2018, \mnras, 479, 3068, \dodoi{10.1093/mnras/sty1680}

\bibitem[{{Schroetter} {et~al.}(2019){Schroetter}, {Bouch{\'e}}, {Zabl}, {Contini}, {Wendt}, {Schaye}, {Mitchell}, {Muzahid}, {Marino}, {Bacon}, {Lilly}, {Richard}, \& {Wisotzki}}]{schroetter19}
{Schroetter}, I., {Bouch{\'e}}, N.~F., {Zabl}, J., {et~al.} 2019, \mnras, 490, 4368, \dodoi{10.1093/mnras/stz2822}

\bibitem[{{Schroetter} {et~al.}(2021){Schroetter}, {Bouch{\'e}}, {Zabl}, {Rahmani}, {Wendt}, {Muzahid}, {Contini}, {Schaye}, {Schmidt}, \& {Wisotzki}}]{schroetter21}
---. 2021, \mnras, 506, 1355, \dodoi{10.1093/mnras/stab1447}

\bibitem[{{Sharma} {et~al.}(2014){Sharma}, {Nath}, {Chattopadhyay}, \& {Shchekinov}}]{sharma14}
{Sharma}, M., {Nath}, B.~B., {Chattopadhyay}, I., \& {Shchekinov}, Y. 2014, \mnras, 441, 431, \dodoi{10.1093/mnras/stu497}

\bibitem[{{Sormani} {et~al.}(2018){Sormani}, {Sobacchi}, {Pezzulli}, {Binney}, \& {Klessen}}]{sormani18}
{Sormani}, M.~C., {Sobacchi}, E., {Pezzulli}, G., {Binney}, J., \& {Klessen}, R.~S. 2018, \mnras, 481, 3370, \dodoi{10.1093/mnras/sty2500}

\bibitem[{{Sparre} {et~al.}(2020){Sparre}, {Pfrommer}, \& {Ehlert}}]{sparre20}
{Sparre}, M., {Pfrommer}, C., \& {Ehlert}, K. 2020, \mnras, 499, 4261, \dodoi{10.1093/mnras/staa3177}

\bibitem[{{Spavone} {et~al.}(2010){Spavone}, {Iodice}, {Arnaboldi}, {Gerhard}, {Saglia}, \& {Longo}}]{spavone10}
{Spavone}, M., {Iodice}, E., {Arnaboldi}, M., {et~al.} 2010, \apj, 714, 1081, \dodoi{10.1088/0004-637X/714/2/1081}

\bibitem[{{Steidel} {et~al.}(2002){Steidel}, {Kollmeier}, {Shapley}, {Churchill}, {Dickinson}, \& {Pettini}}]{steidel02}
{Steidel}, C.~C., {Kollmeier}, J.~A., {Shapley}, A.~E., {et~al.} 2002, \apj, 570, 526, \dodoi{10.1086/339792}

\bibitem[{{Stern} {et~al.}(2018){Stern}, {Faucher-Gigu{\`e}re}, {Hennawi}, {Hafen}, {Johnson}, \& {Fielding}}]{stern18}
{Stern}, J., {Faucher-Gigu{\`e}re}, C.-A., {Hennawi}, J.~F., {et~al.} 2018, \apj, 865, 91, \dodoi{10.3847/1538-4357/aac884}

\bibitem[{{Stern} {et~al.}(2019){Stern}, {Fielding}, {Faucher-Gigu{\`e}re}, \& {Quataert}}]{stern19}
{Stern}, J., {Fielding}, D., {Faucher-Gigu{\`e}re}, C.-A., \& {Quataert}, E. 2019, \mnras, 488, 2549, \dodoi{10.1093/mnras/stz1859}

\bibitem[{{Stern} {et~al.}(2024){Stern}, {Fielding}, {Hafen}, {Su}, {Naor}, {Faucher-Gigu{\`e}re}, {Quataert}, \& {Bullock}}]{stern23}
{Stern}, J., {Fielding}, D., {Hafen}, Z., {et~al.} 2024, \mnras, 530, 1711, \dodoi{10.1093/mnras/stae824}

\bibitem[{{Stern} {et~al.}(2016){Stern}, {Hennawi}, {Prochaska}, \& {Werk}}]{stern16}
{Stern}, J., {Hennawi}, J.~F., {Prochaska}, J.~X., \& {Werk}, J.~K. 2016, \apj, 830, 87, \dodoi{10.3847/0004-637X/830/2/87}

\bibitem[{{Stewart}(2017)}]{stewart-proc17}
{Stewart}, K.~R. 2017, in Astrophysics and Space Science Library, Vol. 430, Gas Accretion onto Galaxies, ed. A.~{Fox} \& R.~{Dav{\'e}}, 249, \dodoi{10.1007/978-3-319-52512-9_11}

\bibitem[{{Stewart} {et~al.}(2013){Stewart}, {Brooks}, {Bullock}, {Maller}, {Diemand}, {Wadsley}, \& {Moustakas}}]{stewart13}
{Stewart}, K.~R., {Brooks}, A.~M., {Bullock}, J.~S., {et~al.} 2013, \apj, 769, 74, \dodoi{10.1088/0004-637X/769/1/74}

\bibitem[{{Stewart} {et~al.}(2011){Stewart}, {Kaufmann}, {Bullock}, {Barton}, {Maller}, {Diemand}, \& {Wadsley}}]{stewart11}
{Stewart}, K.~R., {Kaufmann}, T., {Bullock}, J.~S., {et~al.} 2011, \apj, 738, 39, \dodoi{10.1088/0004-637X/738/1/39}

\bibitem[{{Stewart} {et~al.}(2017){Stewart}, {Maller}, {O{\~n}orbe}, {Bullock}, {Joung}, {Devriendt}, {Ceverino}, {Kere{\v{s}}}, {Hopkins}, \& {Faucher-Gigu{\`e}re}}]{stewart17}
{Stewart}, K.~R., {Maller}, A.~H., {O{\~n}orbe}, J., {et~al.} 2017, \apj, 843, 47, \dodoi{10.3847/1538-4357/aa6dff}

\bibitem[{{Tan} \& {Oh}(2021)}]{tan21}
{Tan}, B., \& {Oh}, S.~P. 2021, \mnras, 508, L37, \dodoi{10.1093/mnrasl/slab100}

\bibitem[{{Tan} {et~al.}(2023){Tan}, {Oh}, \& {Gronke}}]{tan23}
{Tan}, B., {Oh}, S.~P., \& {Gronke}, M. 2023, \mnras, 520, 2571, \dodoi{10.1093/mnras/stad236}

\bibitem[{{Tepper-Garc{\'\i}a}(2006)}]{tepper-garcia06}
{Tepper-Garc{\'\i}a}, T. 2006, \mnras, 369, 2025, \dodoi{10.1111/j.1365-2966.2006.10450.x}

\bibitem[{{Tepper-Garc{\'\i}a} {et~al.}(2015){Tepper-Garc{\'\i}a}, {Bland-Hawthorn}, \& {Sutherland}}]{tepper-garcia15}
{Tepper-Garc{\'\i}a}, T., {Bland-Hawthorn}, J., \& {Sutherland}, R.~S. 2015, \apj, 813, 94, \dodoi{10.1088/0004-637X/813/2/94}

\bibitem[{{Trapp} {et~al.}(2022){Trapp}, {Kere{\v{s}}}, {Chan}, {Escala}, {Hummels}, {Hopkins}, {Faucher-Gigu{\`e}re}, {Murray}, {Quataert}, \& {Wetzel}}]{trapp22}
{Trapp}, C.~W., {Kere{\v{s}}}, D., {Chan}, T.~K., {et~al.} 2022, \mnras, 509, 4149, \dodoi{10.1093/mnras/stab3251}

\bibitem[{{Tripp}(2022)}]{tripp22}
{Tripp}, T.~M. 2022, \mnras, 511, 1714, \dodoi{10.1093/mnras/stac044}

\bibitem[{{Tumlinson} {et~al.}(2017){Tumlinson}, {Peeples}, \& {Werk}}]{tpw-araa17}
{Tumlinson}, J., {Peeples}, M.~S., \& {Werk}, J.~K. 2017, \araa, 55, 389, \dodoi{10.1146/annurev-astro-091916-055240}

\bibitem[{{Udhwani} {et~al.}(2025){Udhwani}, {Sameer}, {Narayanan}, {Muzahid}, {Charlton}, \& {Cantalupo}}]{udhwani25}
{Udhwani}, P., {Sameer}, {Narayanan}, A., {et~al.} 2025, arXiv e-prints, arXiv:2510.00729, \dodoi{10.48550/arXiv.2510.00729}

\bibitem[{{van de Voort} {et~al.}(2011){van de Voort}, {Schaye}, {Booth}, {Haas}, \& {Dalla Vecchia}}]{vandevoort11}
{van de Voort}, F., {Schaye}, J., {Booth}, C.~M., {Haas}, M.~R., \& {Dalla Vecchia}, C. 2011, \mnras, 414, 2458, \dodoi{10.1111/j.1365-2966.2011.18565.x}

\bibitem[{{van der Kruit} \& {Freeman}(2011)}]{vanderKruit11}
{van der Kruit}, P.~C., \& {Freeman}, K.~C. 2011, \araa, 49, 301, \dodoi{10.1146/annurev-astro-083109-153241}

\bibitem[{{Veilleux} {et~al.}(2020){Veilleux}, {Maiolino}, {Bolatto}, \& {Aalto}}]{veilleux20}
{Veilleux}, S., {Maiolino}, R., {Bolatto}, A.~D., \& {Aalto}, S. 2020, \aapr, 28, 2, \dodoi{10.1007/s00159-019-0121-9}

\bibitem[{{Wakker} {et~al.}(2012){Wakker}, {Savage}, {Fox}, {Benjamin}, \& {Shapiro}}]{wakker12}
{Wakker}, B.~P., {Savage}, B.~D., {Fox}, A.~J., {Benjamin}, R.~A., \& {Shapiro}, P.~R. 2012, \apj, 749, 157, \dodoi{10.1088/0004-637X/749/2/157}

\bibitem[{{Weisheit}(1978)}]{weisheit78}
{Weisheit}, J.~C. 1978, \apj, 219, 829, \dodoi{10.1086/155844}

\bibitem[{{Wright} {et~al.}(2024){Wright}, {Somerville}, {Lagos}, {Schaller}, {Dav{\'e}}, {Angl{\'e}s-Alc{\'a}zar}, \& {Genel}}]{wright24}
{Wright}, R.~J., {Somerville}, R.~S., {Lagos}, C. d.~P., {et~al.} 2024, \mnras, 532, 3417, \dodoi{10.1093/mnras/stae1688}

\bibitem[{{Xu} {et~al.}(2023){Xu}, {Heckman}, {Henry}, {Berg}, {Chisholm}, {James}, {Martin}, {Stark}, {Hayes}, {Arellano-Cordova}, {Carr}, {Huberty}, {Mingozzi}, {Scarlata}, \& {Sugahara}}]{xu23-classy6}
{Xu}, X., {Heckman}, T., {Henry}, A., {et~al.} 2023, arXiv e-prints, arXiv:2301.11498, \dodoi{10.48550/arXiv.2301.11498}

\bibitem[{{Yang} {et~al.}(2025){Yang}, {Qu}, {Bregman}, \& {Ji}}]{yang25}
{Yang}, H., {Qu}, Z., {Bregman}, J.~N., \& {Ji}, L. 2025, \mnras, 538, 1871, \dodoi{10.1093/mnras/staf407}

\bibitem[{{York} {et~al.}(1986){York}, {Dopita}, {Green}, \& {Bechtold}}]{york86}
{York}, D.~G., {Dopita}, M., {Green}, R., \& {Bechtold}, J. 1986, \apj, 311, 610, \dodoi{10.1086/164800}

\bibitem[{{Zabl} {et~al.}(2019){Zabl}, {Bouch{\'e}}, {Schroetter}, {Wendt}, {Finley}, {Schaye}, {Conseil}, {Contini}, {Marino}, {Mitchell}, {Muzahid}, {Pezzulli}, \& {Wisotzki}}]{zabl19}
{Zabl}, J., {Bouch{\'e}}, N.~F., {Schroetter}, I., {et~al.} 2019, \mnras, 485, 1961, \dodoi{10.1093/mnras/stz392}

\bibitem[{{Zabl} {et~al.}(2020){Zabl}, {Bouch{\'e}}, {Schroetter}, {Wendt}, {Contini}, {Schaye}, {Marino}, {Muzahid}, {Pezzulli}, {Verhamme}, \& {Wisotzki}}]{zabl20}
---. 2020, \mnras, 492, 4576, \dodoi{10.1093/mnras/stz3607}

\bibitem[{{Zhang}(2018)}]{zhang18}
{Zhang}, D. 2018, Galaxies, 6, 114, \dodoi{10.3390/galaxies6040114}

\bibitem[{{Zheng} {et~al.}(2019){Zheng}, {Peek}, {Putman}, \& {Werk}}]{zheng19}
{Zheng}, Y., {Peek}, J.~E.~G., {Putman}, M.~E., \& {Werk}, J.~K. 2019, \apj, 871, 35, \dodoi{10.3847/1538-4357/aaf6eb}

\end{thebibliography}
\bibliographystyle{aasjournal}



\end{document}